\documentclass[traditabstract]{aa}
\usepackage{graphicx,subfigure,amsmath}
\usepackage{txfonts}
\usepackage{natbib}
\usepackage{ams}

\begin{document}

\title{{\it Herschel}/PACS photometry of transiting-planet host stars with candidate warm debris disks \thanks{Herschel is an ESA space observatory with science instruments provided by European-led Principal Investigator consortia and with important participation from NASA.}}

\author{Bruno Mer\'in \inst{1}
  \and David R. Ardila \inst{2}
  \and \'Alvaro Ribas \inst{3,1}
  \and Herv\'e Bouy \inst{3}
  \and Geoffrey Bryden \inst{4}
  \and Karl Stapelfeldt \inst{5}
  \and Deborah Padgett \inst{6,7}
}

\institute{European Space Astronomy Centre (ESA), P.O. Box 78, 28691 Villanueva de la Ca\~nada, Spain
  \and NASA Herschel Science Center, California Institute of Technology, Mail Code 100-22, Pasadena, CA 91125, USA
  \and  Centro de Astrobiolog\'ia, INTA-CSIC, P.O. Box - Apdo. de correos 78, Villanueva de la Ca\~nada Madrid 28691, Spain
  \and Jet Propulsion Laboratory, California Institute of Technology, Pasadena, CA 91109, USA 
  \and NASA Goddard Space Flight Center, Exoplanets and Stellar Astrophysics, Code 667, Greenbelt, MD 20771, USA 
  \and Spitzer Science Center, California Institute of Technology, Pasadena, CA 91125, USA 
  \and Current address: NASA Goddard Space Flight Center, Code 667, Greenbelt, MD 20771, USA 
}

\date{\today}

\abstract {
Dust in debris disks is produced by colliding or evaporating planetesimals, remnants of the planet formation process. Warm dust disks, known by their emission at $\le$ 24 $\mu$m, are rare (4\% of FGK main sequence stars) and especially interesting because they trace material in the region likely to host terrestrial planets, where the dust has a very short dynamical lifetime. Statistical analyses of the source counts of excesses as found with the mid-IR Wide Field Infrared Survey Explorer (WISE) suggest that warm-dust candidates found for the {\it Kepler} transiting-planet host-star candidates can be explained by extragalactic or galactic background emission aligned by chance with the target stars. These statistical analyses do not exclude the possibility that a given WISE excess could be due to a transient dust population associated with the target. Here we report {\it Herschel}/PACS 100 and 160 micron follow-up observations of a sample of {\it Kepler} and non-{\it Kepler} transiting-planet candidates' host stars, with candidate WISE warm debris disks,  aimed at detecting a possible cold debris disk in any of them. No clear detections were found in any one of the objects at either wavelength. Our upper limits confirm that most objects in the sample do not have a massive debris disk like that in $\beta$ Pic. We also show that the planet-hosting star WASP-33 does not have a debris disk comparable to the one around $\eta$ Crv. Although the data cannot be used to rule out rare warm disks around the {\it Kepler} planet-hosting candidates, the lack of detections and the characteristics of neighboring emission found at far-IR wavelengths support an earlier result suggesting that most of the WISE-selected IR excesses around Kepler candidate host stars are likely due to either chance alignment with background IR-bright galaxies and/or to interstellar emission.
}

\maketitle

\section{INTRODUCTION}

The existence of debris disks around stars has been known since the first far-infrared (far-IR) observations from space were carried out. The first debris disk was found around Vega by the Infrared Astronomical Satellite (IRAS) \citep{Aumann1984}, but hundreds of stars were revealed to host them soon after \citep[e.g.,][]{Oudmaijer1992,Mannings1998}. These disks are thought to be analogues of the Kuiper belt \citep{Moro2008}, comprised of planetesimals, which are the remnant of the planet formation process. Their detection is due to the presence of dust in these belts produced by collisions among or evaporation of planetesimals \citep[e.g.,][]{Wyatt2008}. Typical debris disks have estimated radii of some hundred AU and brightness temperatures around $\sim$\,50\,K. This results in a peak in their spectral energy distribution (SED) around 70-100\,$\mu$m. They are a normal feature in the universe, with an estimated frequency of $\sim$16 to 20\,\% around main sequence FGK stars, from \textit{Spitzer} and \textit{Herschel} studies, respectively \citep{Trilling2008,Eiroa2013}.

A fraction of debris disks emit significantly at shorter wavelengths, suggesting higher dust temperatures and hence smaller separations of the dust from the host star. Compared to the previously mentioned frequency of debris disks in main sequence FGK stars, only $\sim$\,4\,\%  of solar-type stars were found to show excesses at 24\,$\mu$m by the \textit{Spitzer} Space Telescope \citep{Trilling2008}. In most of the cases, this is produced by the Wien edge of longer wavelength dust emission (e.g., a highly massive debris disk), although some stars have been proved to harbor legitimate warm inner debris disks \citep{Siegler2007,Lawler2009}.

There is no single explanation for these kinds of warm excesses for objects older than $\sim$ 100\,Myrs. \citet{Melis2010} shows that for these ages, it is still possible to find strong mid-IR excesses due to the destruction of planetary embryos, resulting in a dusty environment in the inner regions of the host star. For older systems warm dust close to the star cannot be continuously replenished. For example, the steady state asteroid-belt planetesimal grinding model by \citet{Wyatt2007b} is not capable of explaining these warm excesses. Therefore, they are likely to be produced in transient events such as planetesimals collisions owing to gravitational instabilities that cause bodies in the outer disk to be pushed into their asteroid belt \citep{Absil2006,Wyatt2008}. Another possible explanation involves perturbation of close-in planets in the asteroid belt \citep[see][]{Wyatt2008}.

In a recent study, \citet{Ribas2012} (hereafter R12), we searched for warm IR excesses around the 997 stars released as candidate planetary host stars by the {\it Kepler} project in February 2011.  See also \cite{Lawler2012} for a similar work on the same topic. The sample of stars observed by {\it Kepler} is comprised of over 150,000 stars determined by the Kepler Input Catalog \citep[KIC,][]{KeplerInputCatalog,KICselection}, harboring mostly main sequence G type field dwarfs. \citet{Morton2011} suggest that 90-95\,\% of these candidates are real planetary systems, although the fraction of false positives and contamination is still under a strong debate \citep[see, e.g.,][]{Mann2012,Adams2013,Lillo-Box2014}. This catalog was combined with the Wide-field Infrared Survey Explorer (WISE) Preliminary Release Catalog (PRC)\footnote{http://wise2.ipac.caltech.edu/docs/release/prelim/}, which covered $\sim$\,57\,\% of the sky and half of the Kepler field of view at 3.4, 4.6, 12, and 22\,$\mu$m (hereafter W1, W2, W3, and W4). Using the W3 and W4 bands, R12 selected 13 objects in half the Kepler field as warm dust excess candidates.

On  February 2012 the \textit{Kepler mission} released a new sample of 2\,321 planetary systems candidates around 1\,790 stars\footnote{http://archive.stsci.edu/kepler/planet\_candidates.html}. On March 2012, the WISE All-Sky Catalog (ASC) was released\footnote{http://wise2.ipac.caltech.edu/docs/release/allsky/}, covering the complete {\it Kepler} field and providing an unprecedented opportunity to study the warm disks' incidence in an homogeneous sample and obtaining more significant statistical values.

Here we present an update of the work done in R12 using the new Kepler candidates and the WISE All-Sky Catalog, plus follow-up observations at 100 and 160 $\mu$m of a refined sample done with the PACS instrument \citep{Poglitsch2010} onboard the {\it Herschel Space Observatory} \citep{Pilbratt2010}. The aim of this work is to potentially detect a large IR excess at longer wavelengths indicating a large debris disk in one of the new candidate planet-hosting stars or, alternatively, to clarify the possible contamination of these mid-IR excesses by extra galactic sources, as recently proposed by Kennedy \& Wyatt (2012). Non-detections with {\it Herschel} would suggest a lack of a cold-component in the debris disk but  cannot rule out the presence of a warm debris disk only emitting at mid-IR wavelengths. Recent evidence of multi-component debris disks \citep{Su2013,Absil2013} indicates a potential for discovering the cold component of such disks around any of these candidates.

This work is organized as follows. Section \S~\ref{sample} describes the definition of the new sample to be followed up with {\it Herschel} based on the latest {\it Kepler} sample of planetary system candidates, transiting planet systems, and the WISE ASC. \S~\ref{observations} describes the {\it Herschel}/PACS observations of the sample and the data reduction. \S~\ref{results} shows the images, describes the flux measurements, and analyzes the resulting SEDs of the targets. The discussion and final conclusions are given in \S~\ref{discussion}.


\section{CANDIDATE SELECTION}
\label{sample}

\subsection{Catalog matching}

We followed the same approach as R12, by searching the 1\,790 Kepler planetary systems' candidate stars and previously known transiting systems\footnote{http://var2.astro.cz/ETD/index.php} as described in \cite{Poddany2010} for counterparts to the WISE catalog. We considered a search radius of 1\,\arcsec\, as a good agreement between the WISE astrometric accuracy ($\sim$\,0.2\,\, \arcsec, see the Explanatory Supplement to the WISE All-Sky Data Release) and a conservative cut in order not to lose potential candidates. A total of 1\,728 matches was obtained with this process.

The catalog was then cleaned in the following way. We required photometry from the 2 Micron All Sky Survey \cite[2MASS,][]{Skrutskie2006} with SN $\ge$\,7 and no  problems with artifacts or halos. We also selected only those sources marked as uncontaminated by extragalactic sources in the catalog. Objects with magnitudes below the saturation limit in any WISE bands are rejected, as well as those classified as extended. As a last prerequisite, we selected only those objects with SN $\ge$\,3 at W3 band in order to try to make sure that the detections are real. This whole process led to selecting 844 candidates.


\subsection{Target selection}

Following the procedure in R12, we defined the excess at 12 or 22\,$\mu$m as $\chi_\lambda \equiv (F_{\rm WISE}-F_{\rm phot}) / \sigma_{tot} \ge 2 $, where $F_{\rm WISE}$ is the detected, dereddened flux at W3 or W4 band, $F_{\rm phot}$ is the corresponding synthetic flux value obtained from photospheric modeling, and $\sigma_{tot}$ represents the total uncertainty computed as $\sigma_{tot}=\sqrt{\sigma^{2}_{\rm obs} + \sigma^{2}_{\rm cal}}$, where $\sigma_{\rm obs}$, the measurement uncertainty in the corresponding band, and $\sigma_{\rm cal}$ the absolute calibration uncertainty of the WISE instrument (see the Explanatory Supplement to the WISE All-Sky Data Release). We selected those objects with $\chi_\lambda > 2$ (corresponding to a 95\,\% confidence level) at one or both W3 and W4 bands (when detected), and obtained 293 objects.

\subsection{WISE image inspection}

The WISE images of the remaining candidates were visually inspected in order to assure they are real detections. The study in R12 showed that a significant number of the sources in the WISE PRC suffered from photometric problems or were spurious detections. We followed the same approach, rejecting sources based on three criteria (see R12 for different examples):

\begin{itemize}
\item Artifacts: The source is contaminated by halos or extended emissions.
\item Offsets: The centroid position of the emission peak of the source at the band where excess is found is different than at W1 or W2 bands. 
\item "Extended" PSFs or no detections: The photocenter at the band where the excess is found does not appear as a single PSF. This may be because the source is extended, because additional sources are present nearby at the same flux level or because the source is faint enough that significant noise can contribute to the detection. 
\end{itemize}

Table \ref{tab:sample} lists the final 19 objects that passed all selection criteria above and their corresponding $\chi_\lambda$ values after the process above. All but one of the 19 selected sources show strong 12 $\mu$m excess, suggesting that they are due to warm disks rather than the Wien tail of colder disks. The second column in the table gives the {\it Kepler} Object of Interest (KOI) identifications of the objects where available. A few objects that were observed by us with PACS were not identified by {\it Kepler} as KOIs at the time of our {\it Herschel} observations, but were observed as a control sample since they fulfilled the WISE excess and image inspection selection criteria, and the catalog of KOIs could grow with time to include any of the objects in the KIC catalog when more light-curve data were analyzed jointly.

Table \ref{tab:prevsample} gives the objects selected in R12 but with the updated $\chi_\lambda$ values and image inspection notes from the final WISE All-Sky Data Release. Half of the objects previously identified with the WISE PRC as having candidate warm excesses are discarded with the higher quality WISE ASC data, while the other half (namely KIC3547091, KIC3732821, KIC6422367, KIC6665695, KIC6924203, and KIC9214942) still fulfill our photometric excess criterion. In these objects, image inspection allowed us to rule them out as valid candidates for our study. Table \ref{tab:prevsample} describes the issues detected during the image inspection. Some of these issues were also identified in the WISE ASC quality flags for bands W3 and W4, but only manual image inspection allowed us to identify them as bogus warm-excess objects.
\begin{table*}
\begin{center}
\renewcommand{\footnoterule}{}  
\begin{tabular}{l l c c c c c c c c}
\hline
Object  & Alt. name & RA (J2000) & DEC (J2000) &  $\chi_{12}$ & $\chi_{22}$ & Herschel obsids & F$_{100 \mu m}$ & F$_{160 \mu m}$ & Separation$^\dagger$  \\
            & &   &  &    &   & 13422.. & (mJy) & (mJy) & ($''$) \\
\hline
CoRoT-10 & & 19:24:15.29 & +00:44:45.99 & 4.12 & \ldots & 49290, 91 & $<$ 5.8 & $<$ 23.8 & 12.3 \\
HAT-P-28  & & 00:52:00.18 & +34:43:42.22 & 2.67 & \ldots & 59254, 55 &  $<$ 3.1 & $<$ 5.8 & 6.1 \\
WASP-33 & & 02:26:51.06 & +37:33:01.73 & -0.14 & 2.11 & 49290, 91 & $<$ 2.8 & $<$ 5.0 & 27.1 \\
KIC2162635 & KOI-1032 & 19:27:54.62 & +37:31:57.00 & 9.34 & 3.63 & 57407, 08 & $<$ 6.33 & $<$ 17.6 & 28.1 \\
KIC3835670$^a$ & KOI-149 & 19:06:31.22 & +38:56:44.23 & 6.30 & \ldots & 56956, 57 & $<$ 4.7 & $<$ 8.7 & 3.7  \\
KIC4478168 & KOI-626 &19:40:46.41 & +39:32:23.00 & 3.62 & \ldots &  57409, 10 & $<$ 3.5 & $<$ 7.1 & 25.0 \\
KIC4918309  & KOI-1582 &19:20:30.87 & +40:01:18.61 & 3.77 & \ldots & 59364, 65 & $<$ 2.7 & $<$ 8.7 & 3.4 \\
KIC7097965  & & 18:57:57.68  & +42:38:53.77 & 2.65 & \ldots & 56212, 13 & $<$ 2.7 & $<$ 7.5 & 18.8 \\
KIC8766650 & & 19:46:11.42 & +44:56:30.71 & 7.73 & \ldots & 53513, 14 & $<$ 4.0 & $<$ 8.8 & 38.9 \\
KIC8962094 & KOI-700 & 19:39:53.65 & +45:12:49.32 & 7.73 & 2.94 & 57740, 41 & $<$ 8.8 & $<$ 24.9 & 28.2 \\
KIC9703198 & KOI-469 & 19:14:33.08 & +46:25:17.17 & 4.56 & \ldots & 57742, 43 & $<$ 3.5 & $<$ 7.9 & 13.1 \\
KIC9884104 & KOI-718 & 19:14:57.34 & +46:45:45.35 & 3.12 & \ldots & 59238, 39 & $<$ 2.4 & $<$ 15.4 & 18.4 \\
KIC9965439 & KOI-722 & 19:49:02.19 & +46:50:35.50 & 2.71 & \ldots & 59240, 41 & $<$ 3.6 & $<$ 7.3 &  16.6 \\
KIC10386922 & KOI-289 & 18:51:46.94 & +47:34:29.71 & 2.46 & \ldots & 57744, 45 & $<$ 3.2 & $<$ 8.8 & 19.6  \\
KIC10873260 & KOI-535 & 19:45:32.53 & +48:14:00.25 & 5.24 & \ldots & 59242, 43 & $<$ 5.3 & $<$14.0 & 20.5 \\
KIC11134879$^b$ & KOI-480 &19:21:45.02 & +48:47:30.83 & 3.54 & \ldots & 56215, 16 & $<$ 8.3 & $<$ 11.7 & 10.8 \\
KIC11624249$^b$  & KOI-356 & 19:50:56.73 & +49:38:13.62  & 5.04 & \ldots & 59244, 45 & $<$ 6.5 & $<$ 15.2 & 16.5 \\
KIC11673802 & & 19:49:26.24 & +49:47:51.21 & 5.95 & 3.70 & 57970, 71 & $<$ 4.2 & $<$ 20.8 & 24.6 \\
KIC11774991 & & 19:49:10.21 & +49:58:54.23 & 3.15 & \ldots & 57968, 69 & $<$ 3.1 & $<$ 8.1 & 31.8 \\
\hline
\end{tabular}
\end{center}
$^a$ {\it Herschel} reset of the Spacecraft Velocity Vector (corrected in HCSS v.10.1). \\
$^b$ WISE ASC flag w4flg indicates source confusion at W4 band. \\
$^\dagger$ Separation between intended target position and the closest point source centroid as measured in the 100 $\mu$m image. 
\caption{Final sample fulfilling all selection criteria, observing log and results
from the {\it Herschel}/PACS follow-up observations. }
\label{tab:sample}
\end{table*}

\begin{table*}
\begin{center}
\renewcommand{\footnoterule}{}  
\begin{tabular}{l c c c c l}
\hline
Object  & RA (J2000) & DEC (J2000) &  $\chi_{12}$ & $\chi_{22}$ & Notes from image inspection \\
             &   &  &    &   & \\
\hline
WASP-46 & 21:14:56.86 & -55:52:18.4 & 1.86 & \ldots & \\	
KIC2853093	& 19:26:19.00 & +38:02:09.0 & \ldots & \ldots & \\	
KIC3547091$^a$	& 19:28:18.84 & +38:37:53.3 & 8.74 & \ldots & Contamination by nearby source at W3 \\	
KIC3732821	& 19:07:40.11 & +38:52:20.0 & 2.13 & \ldots & Extended non-point source at W3 \\		
KIC4545187	& 19:04:38.89 & +39:40:40.8 & 0.83 & \ldots & \\
KIC6422367	& 18:53:53.07 & +41:52:23.6  & 7.05 & 1.92 & Offset in W3 \\	
KIC6665695	& 18:48:01.11 & +42:10:35.5 & 0.49 & 	2.25 & Extended non-point source at W4 \\	
KIC6924203	& 18:49:19.88 & +42:27:49.8 & 3.25 & 	\ldots & Extended non-point source at W3 \\
KIC8414716	& 18:57:43.32 & +44:28:49.9 & 1.93 & \ldots & \\	
KIC9008220	& 19:04:36.48 & +45:19:57.2 & 1.19 & \ldots & \\		
KIC9214942	& 19:21:55.19 & +45:36:02.5 & 1.19 & 2.05 & Extended non-point source at W4 \\	
\hline
\end{tabular}
\end{center}
$^a$ $\chi_{12}$ estimated by hand due to apparent excesses at W1 and W2 that yield standard R12 method inapplicable. \\
\caption{Updated $\chi_{12}$ and $\chi_{22}$ values for the previous sample from R12 using the WISE ASC, together with notes from the image inspection. These are no longer warm-excess candidates but are given here for reference to R12.}
\label{tab:prevsample}
\end{table*}

\section{OBSERVATIONS, DATA REDUCTION, AND RESULTS}
\label{observations}

Each of the 19 resulting targets were observed at 100 and 160 $\mu$m using the PACS mini scan-map AOT, in the recommended two cross-scan directions (within the observing program OT2\_dardila\_2). The observations contained ten scan legs, three-minutes long and were repeated six times, in two concatenated AORs at angles 70 and 110 degrees with respect to the PACS detector. The total time per source was 3611 sec, achieving a 1 $\sigma$ sensitivity of $\sim$ 1 mJy at 100 $\mu$m and 2-5 mJy at 160 $\mu$m. The {\it Herschel} observation IDs (obsids) are given in Table \ref{tab:sample} for each of the targets. Objects KIC4143755 and KIC7222086, both of which did not pass the $\chi_{\lambda} > 2$ criteria in an a posteriori analysis with the WISE ASC, were also observed with obsid pairs 1342259362 and 1342259363 and 1342257738 and 1342257739, respectively, within this program.  Exactly the same analysis as for the targets satisfying all selection criteria was applied to them, and they were kept in the rest of the work as part of the control sample.

The images were processed with HIPE version 11.0 \citep{Ott2010} using the PACS point source i-pipe and the latest PACS photometer flux calibration (calTree version 48, see the PACS calibration documents\footnote{See PACS page at http://www.cosmos.esa.int/web/herschel/home} for more information). The same results were also obtained with the same pipeline in HIPE 10.0 and with calTree version 45. This is the default high-pass filtering pipeline with the standard 15- and 25-pixel high-pass filter widths for the 100 and 160 $\mu$m images, respectively. To maximize point-source sensitivity at the center of the images, the user pipeline explicitly masks the source positions with a 20\arcsec\, mask before the final third pass of the high-pass filter takes place. The resulting pixel sizes are 2\arcsec x 2\arcsec and 3\arcsec x 3\arcsec for the 100 and 160 $\mu$m images, respectively. All resulting images are shown in figures \ref{fig:images1} and  \ref{fig:images2}. The linear inverted grayscale goes from zero to 3$\sigma$, where the $\sigma$ is calculated as the standard deviation of the 50 x 50 central pixels in each image, where the coverage is best and relatively constant. 

\begin{figure*}
  \centering
\includegraphics[width=4.5cm]{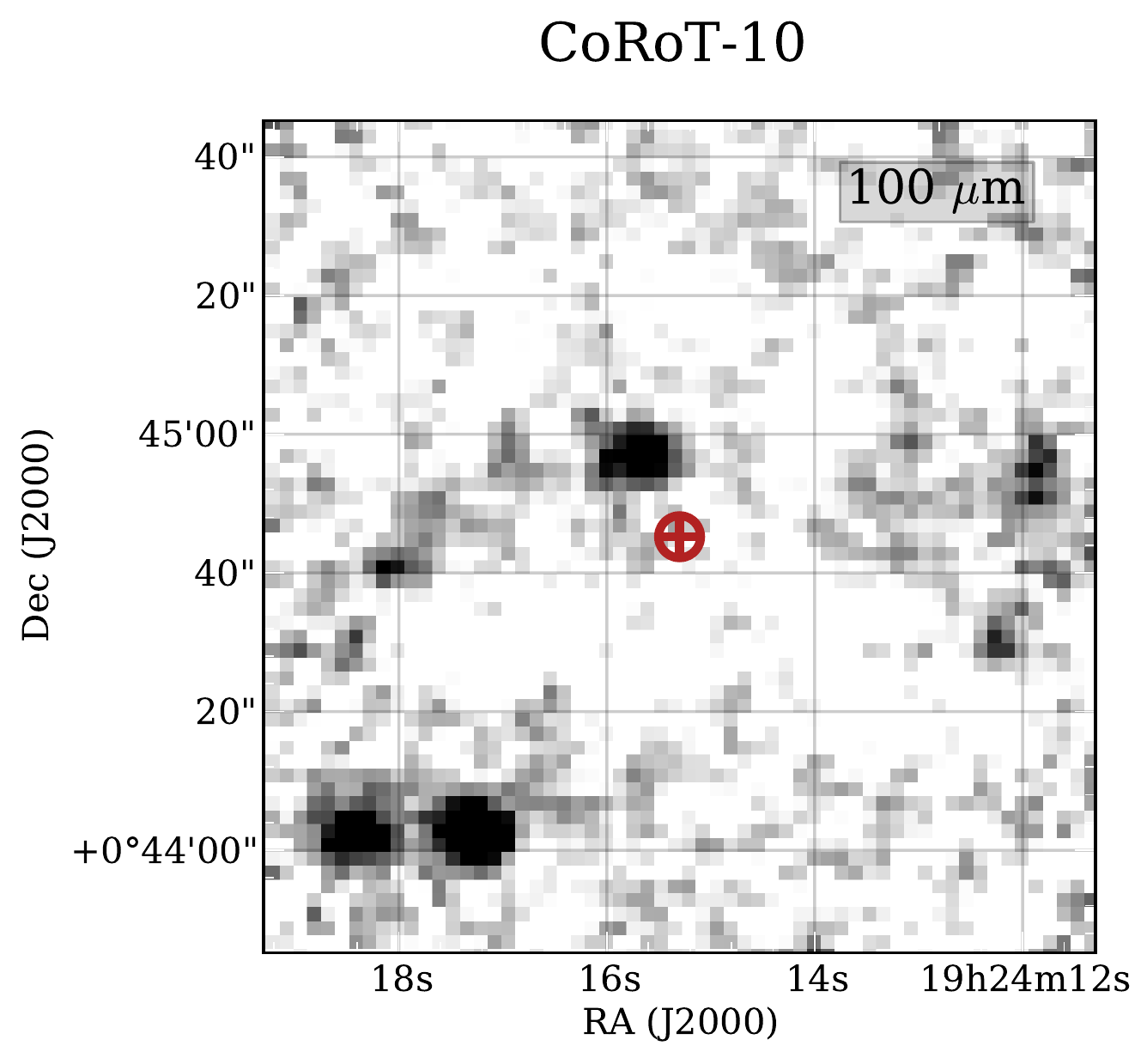}
\includegraphics[width=4.5cm]{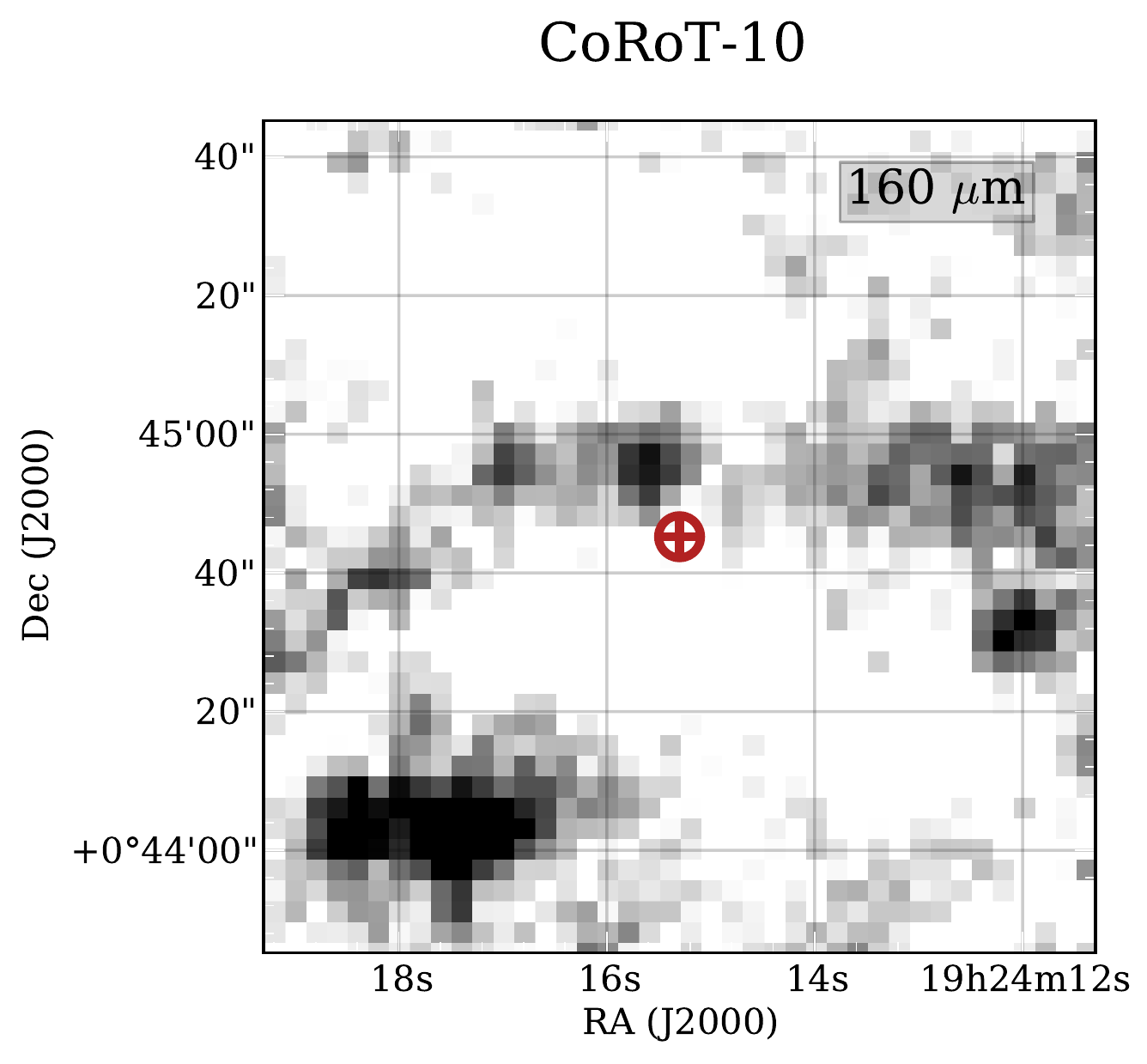}
\includegraphics[width=4.5cm]{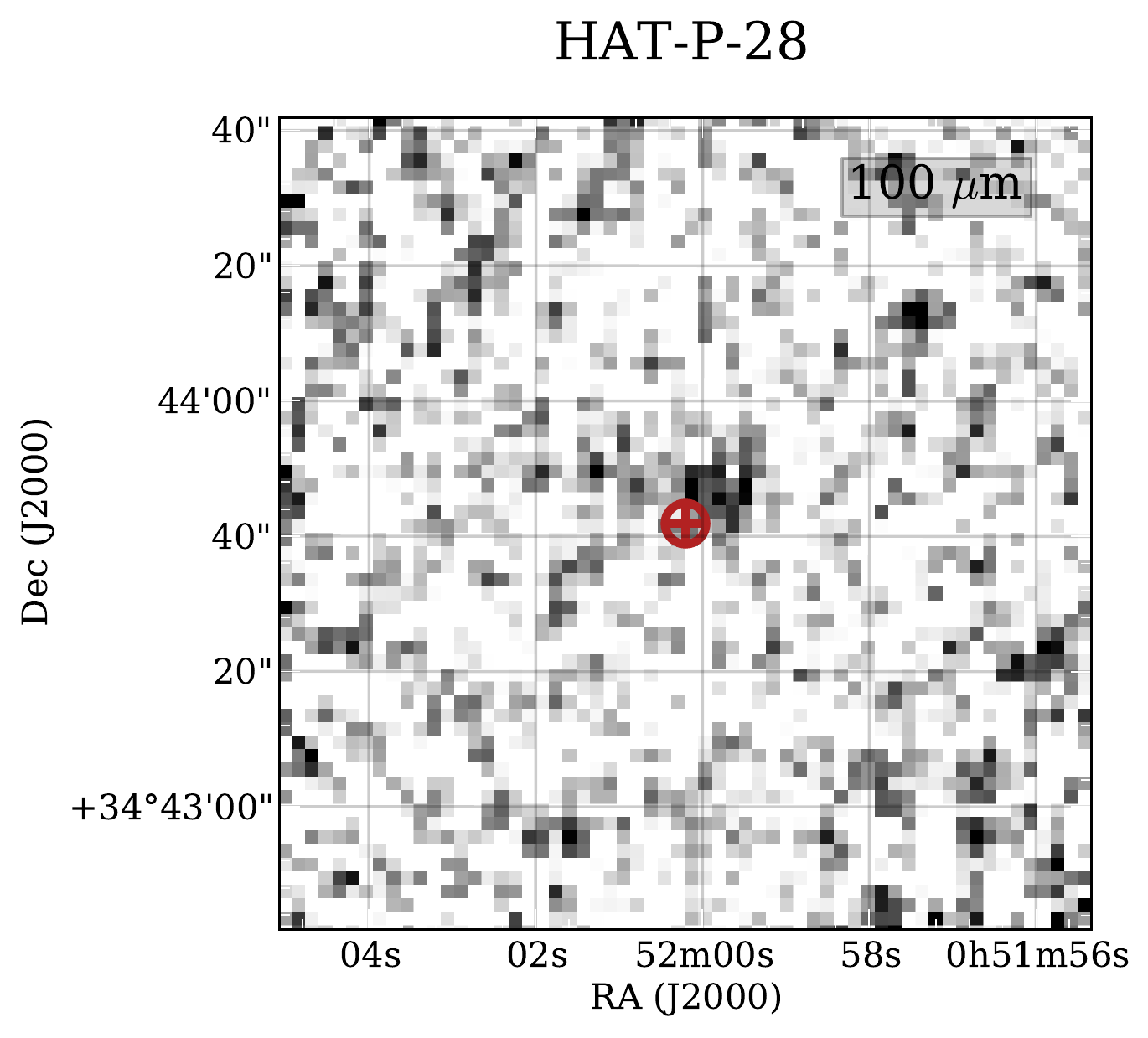}
\includegraphics[width=4.5cm]{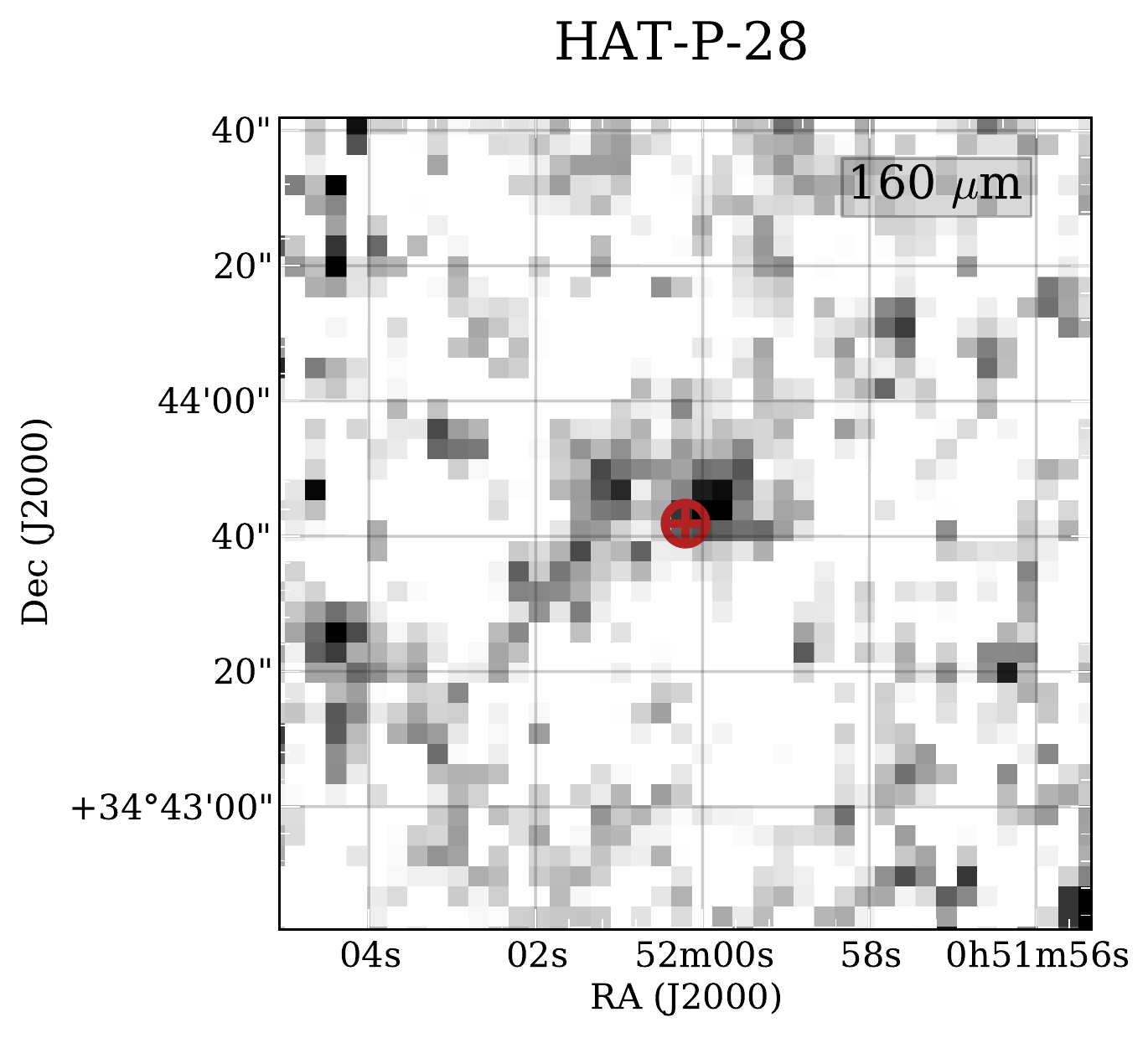}
\includegraphics[width=4.5cm]{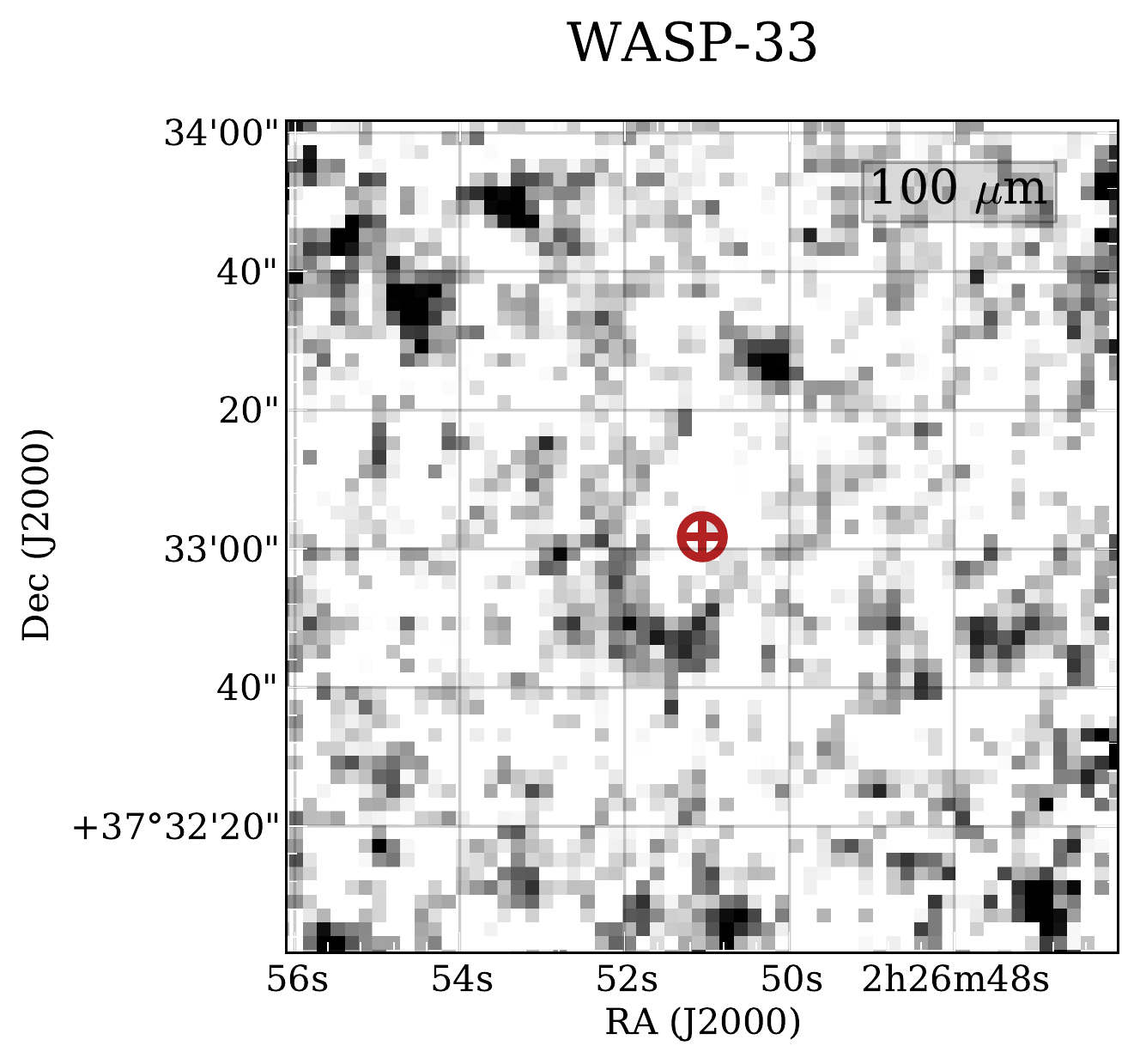}
\includegraphics[width=4.5cm]{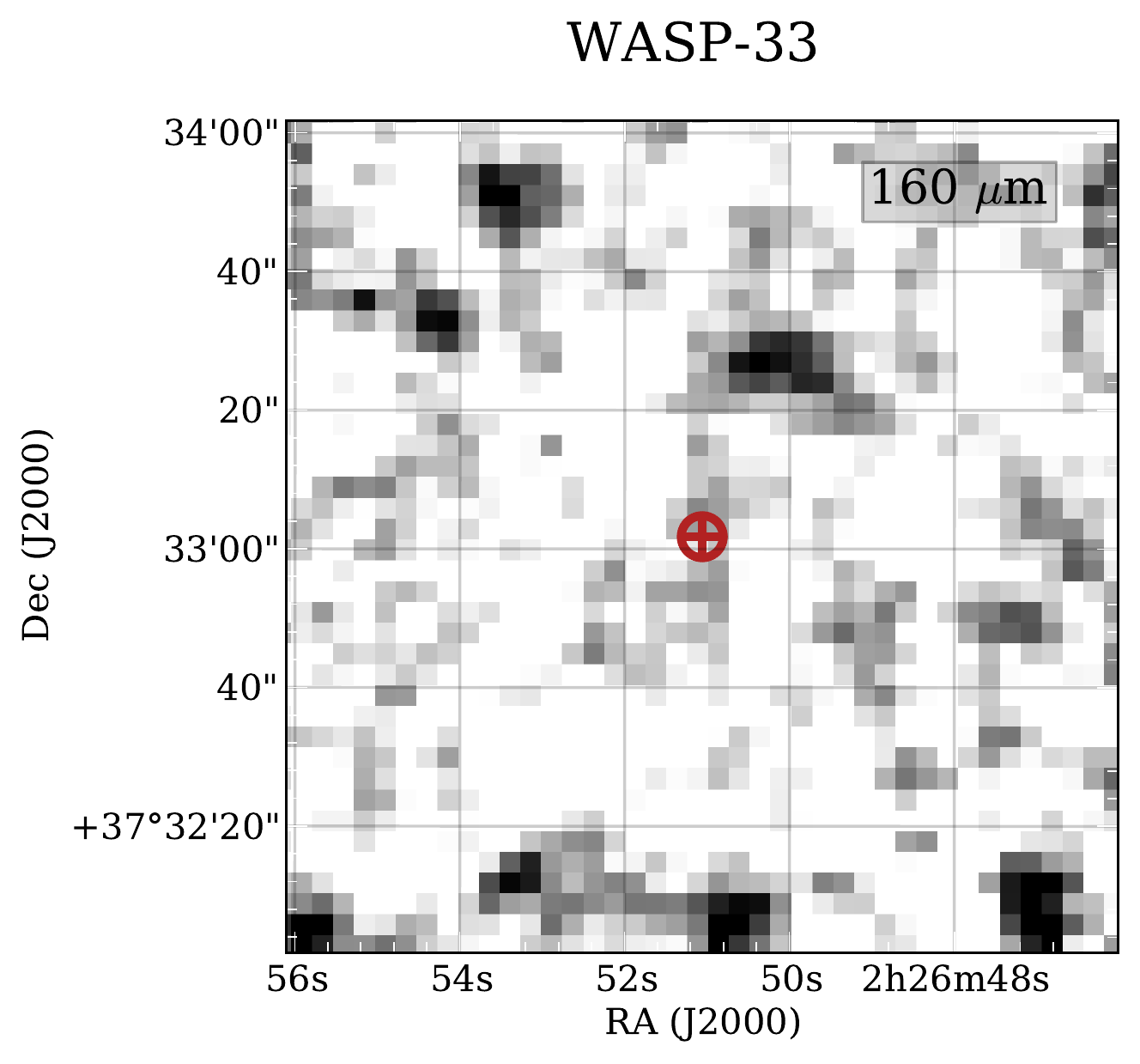}
\includegraphics[width=4.5cm]{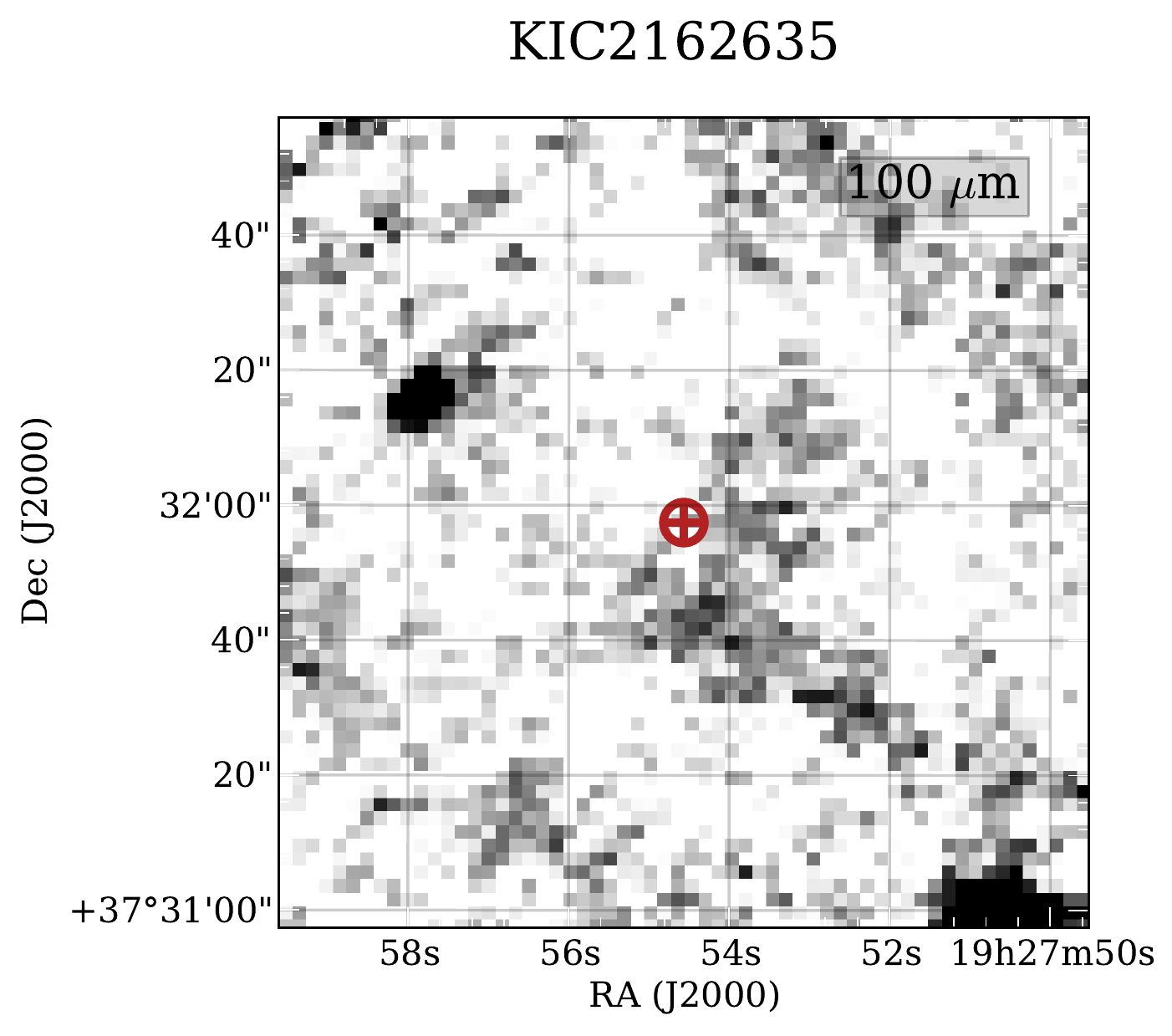}
\includegraphics[width=4.5cm]{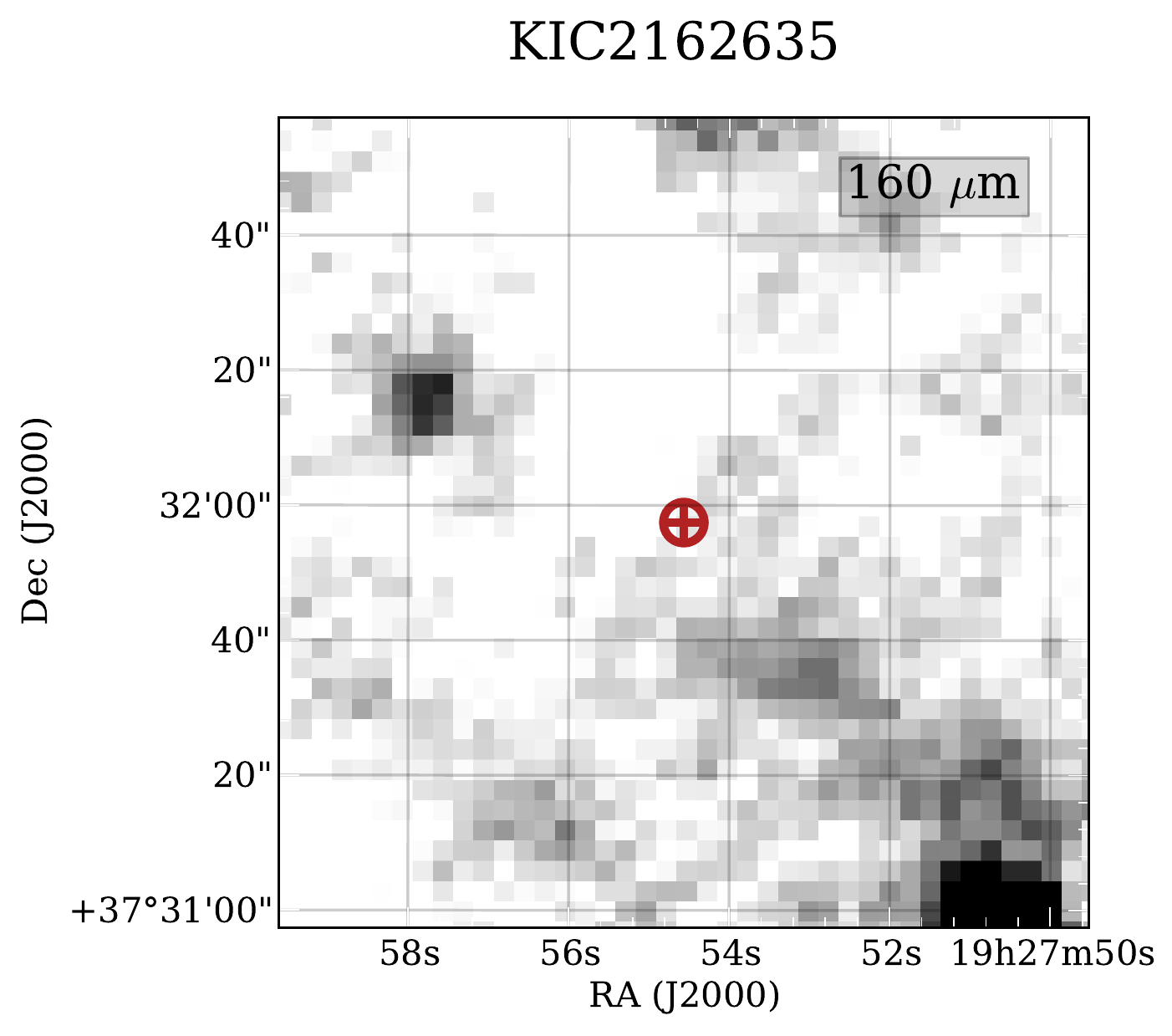}
\includegraphics[width=4.5cm]{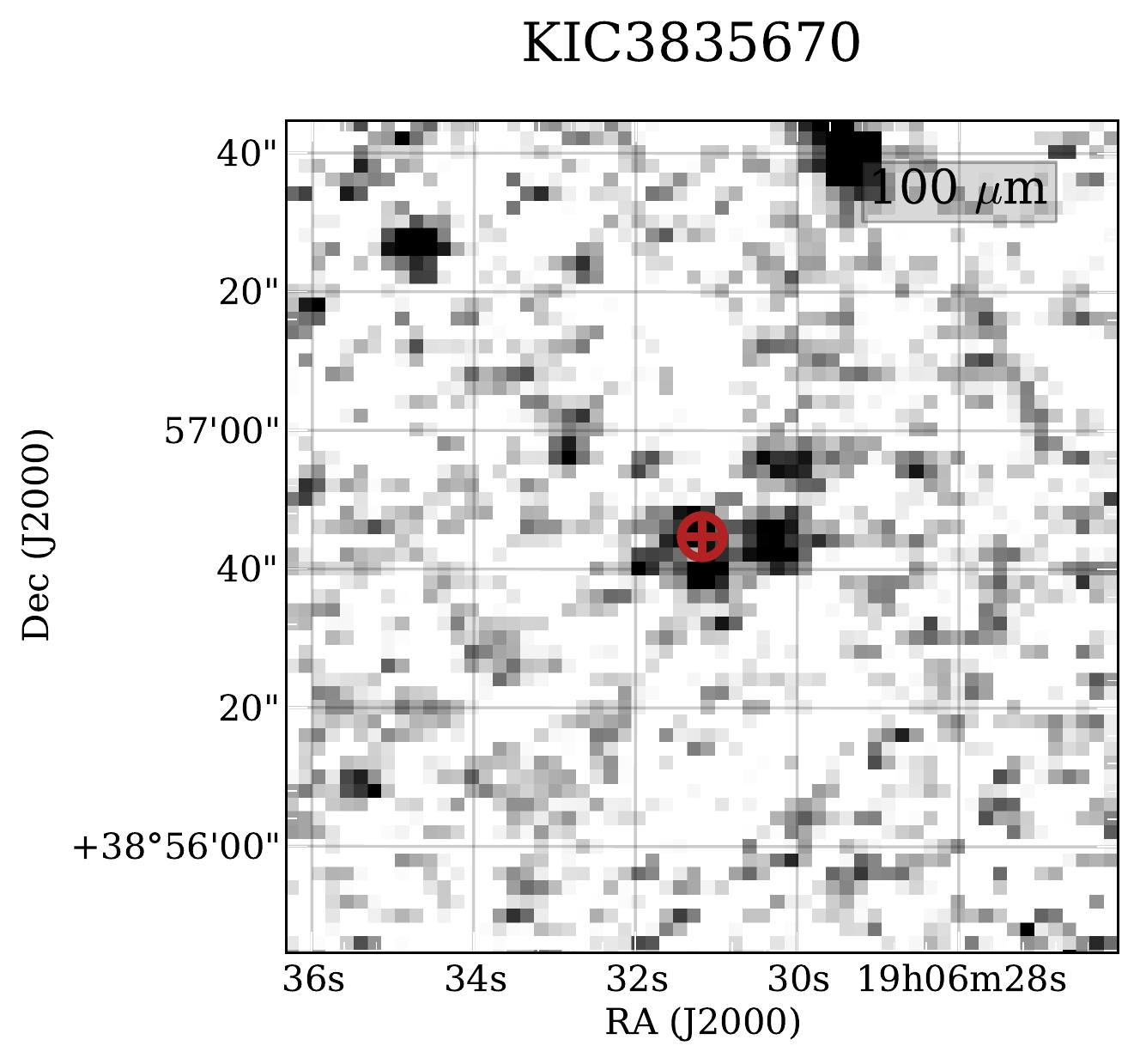}
\includegraphics[width=4.5cm]{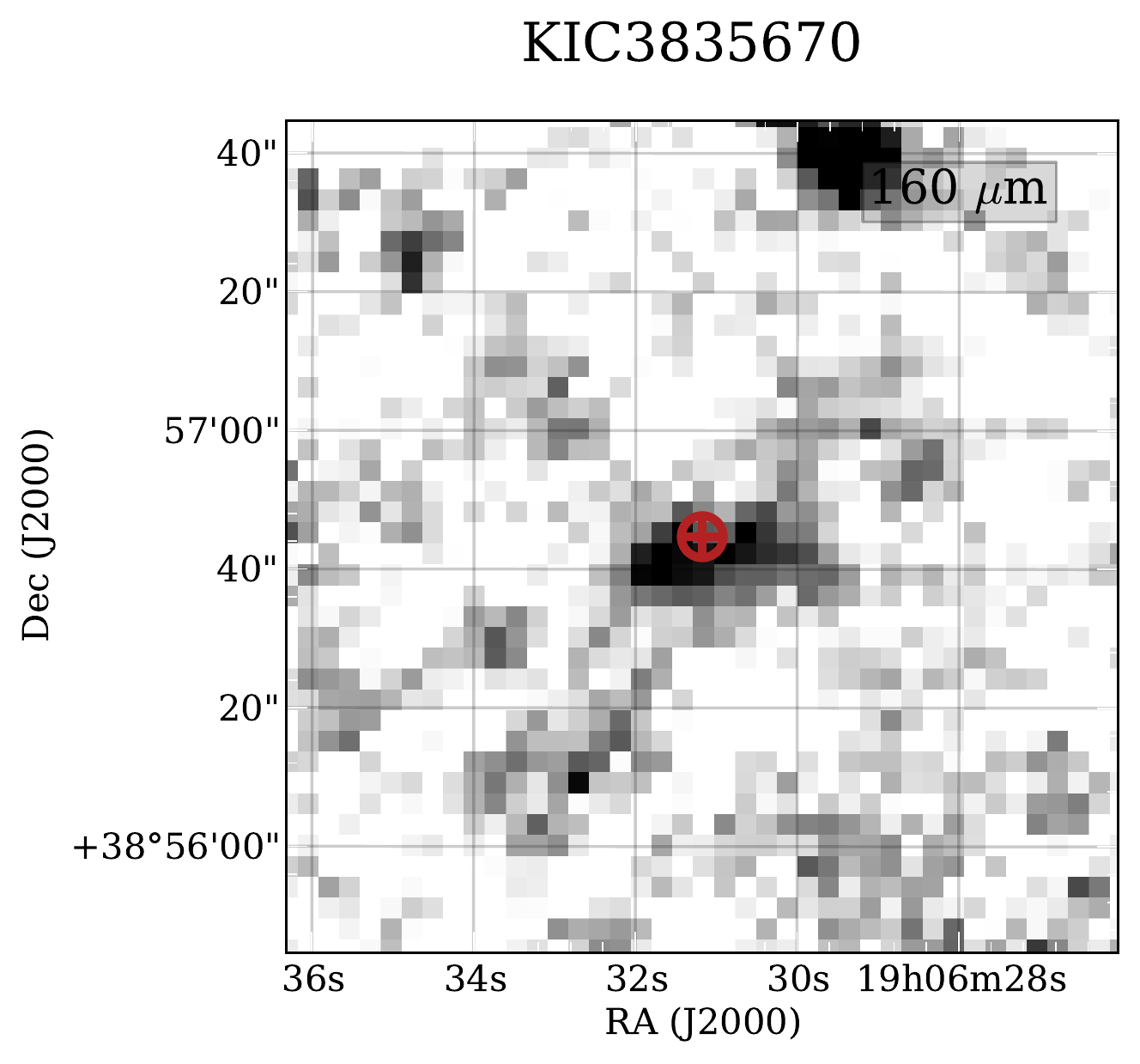}
\includegraphics[width=4.5cm]{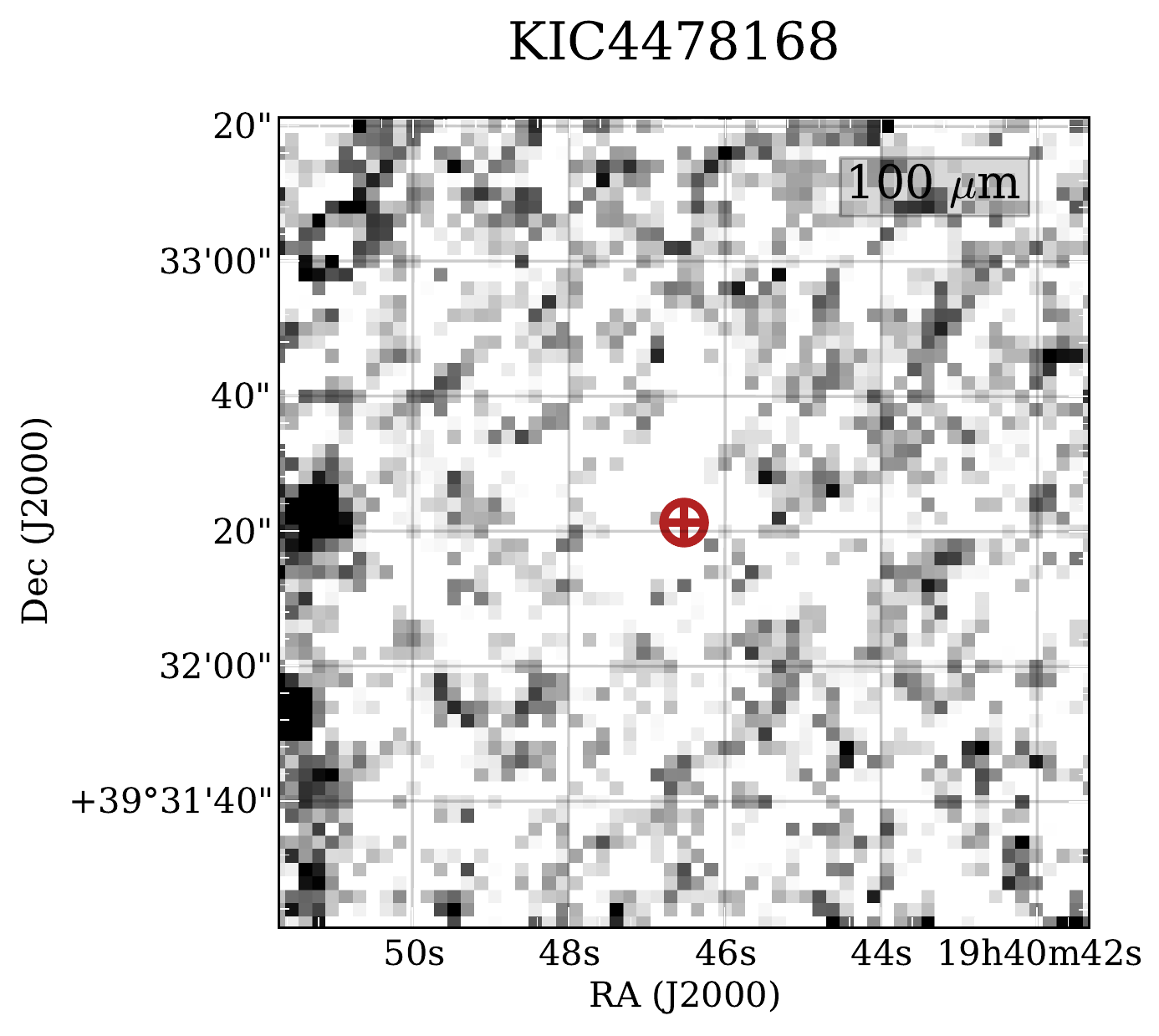}
\includegraphics[width=4.5cm]{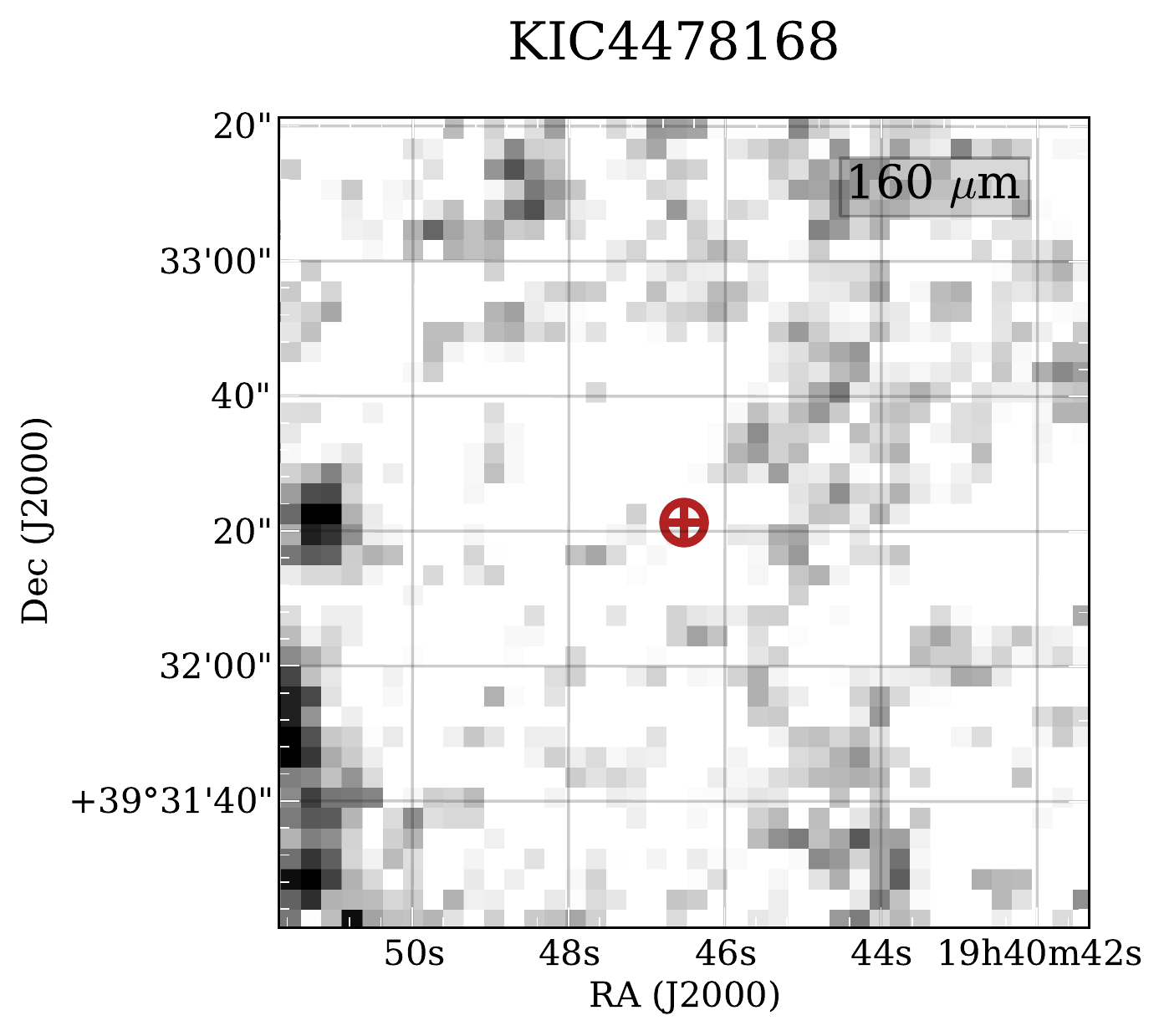}
\includegraphics[width=4.5cm]{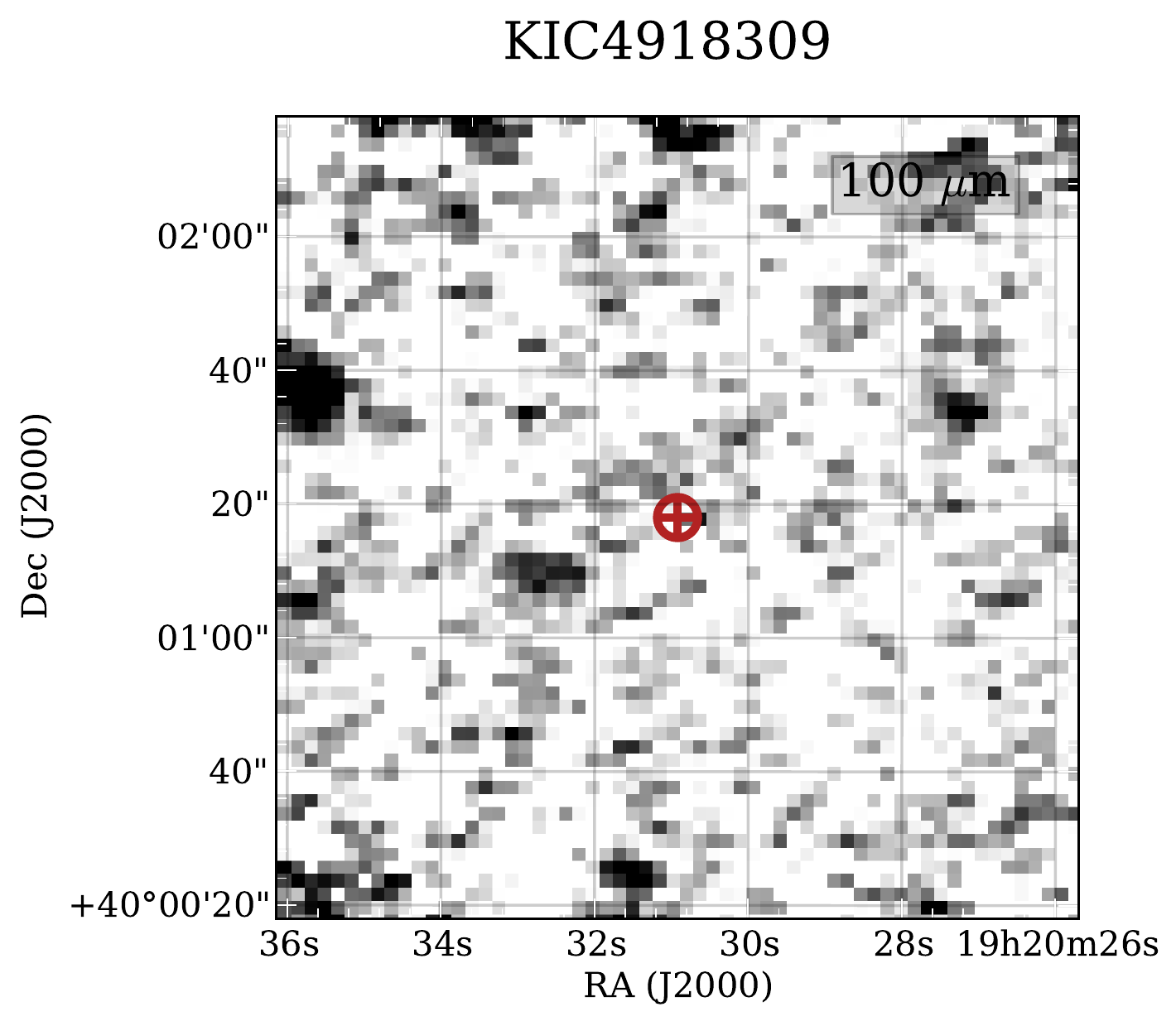}
\includegraphics[width=4.5cm]{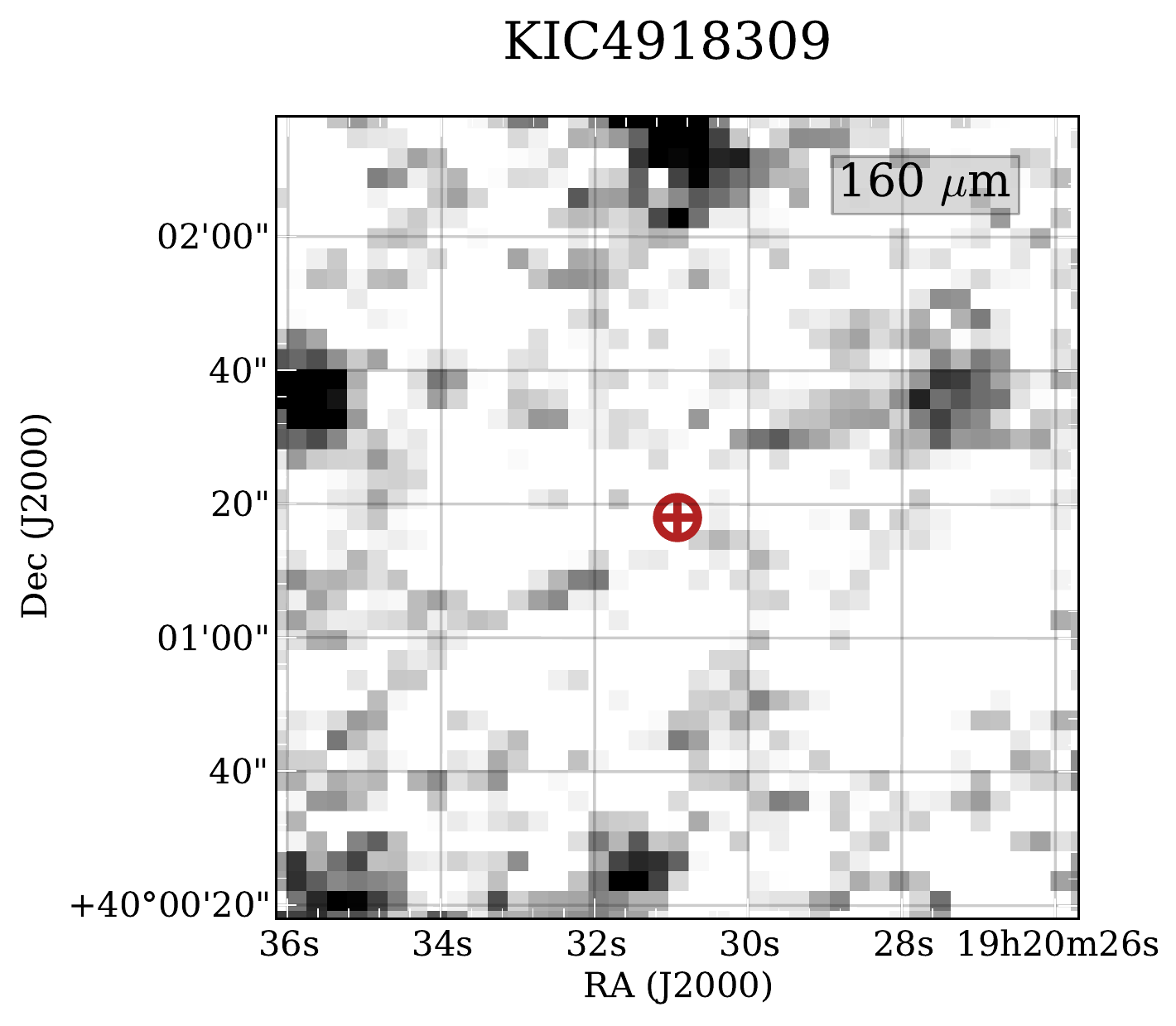}
\includegraphics[width=4.5cm]{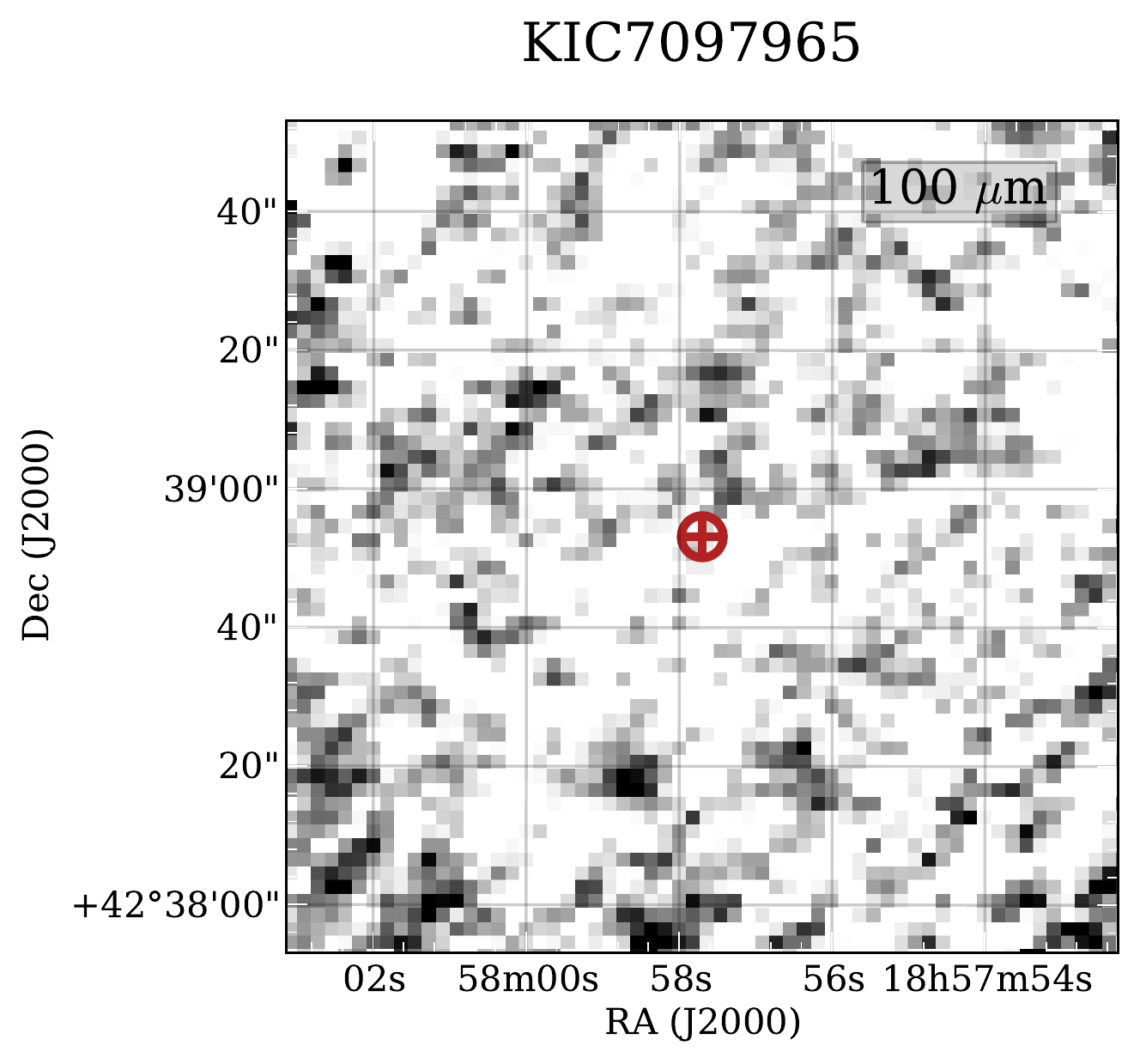}
\includegraphics[width=4.5cm]{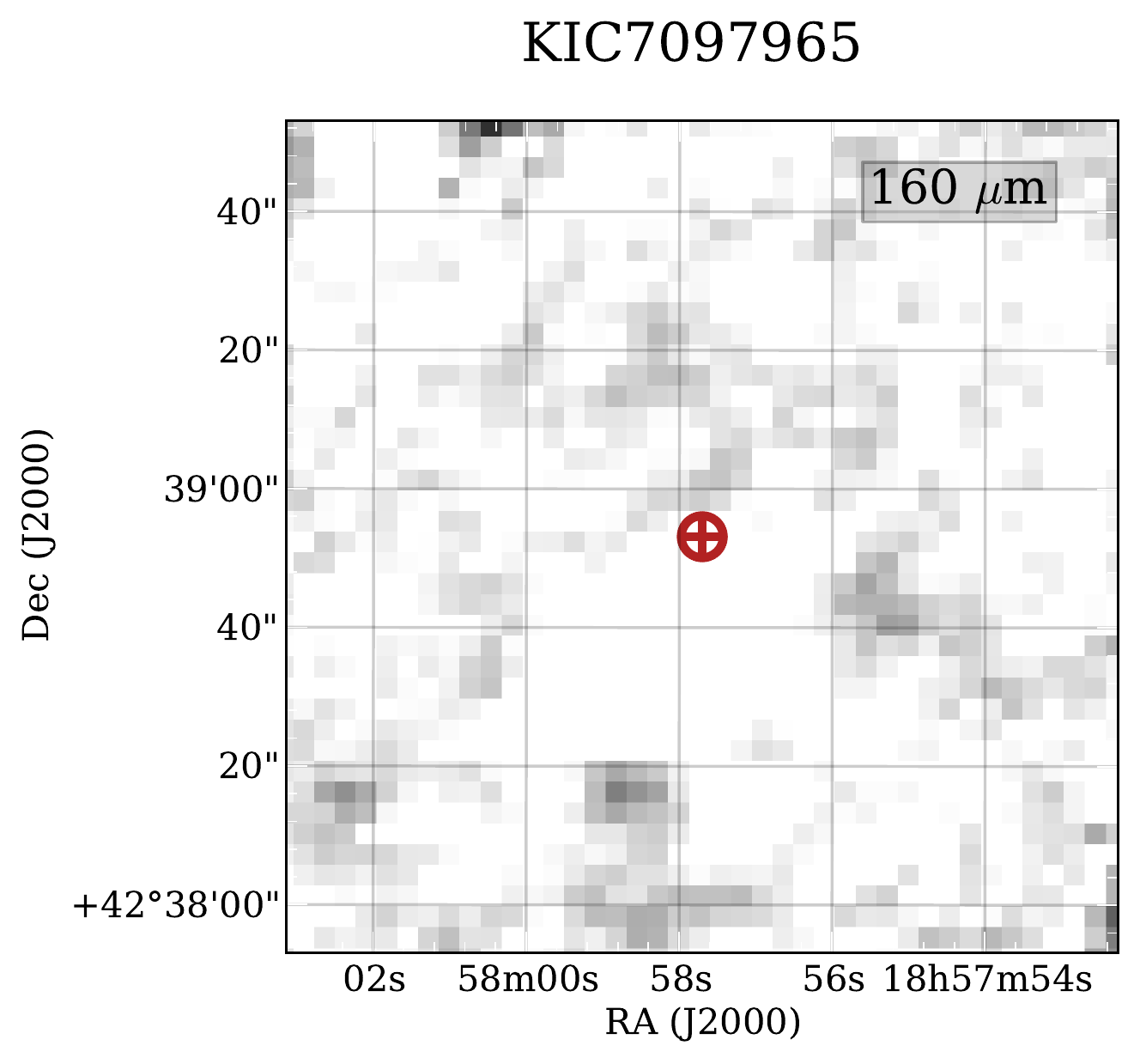}
\includegraphics[width=4.5cm]{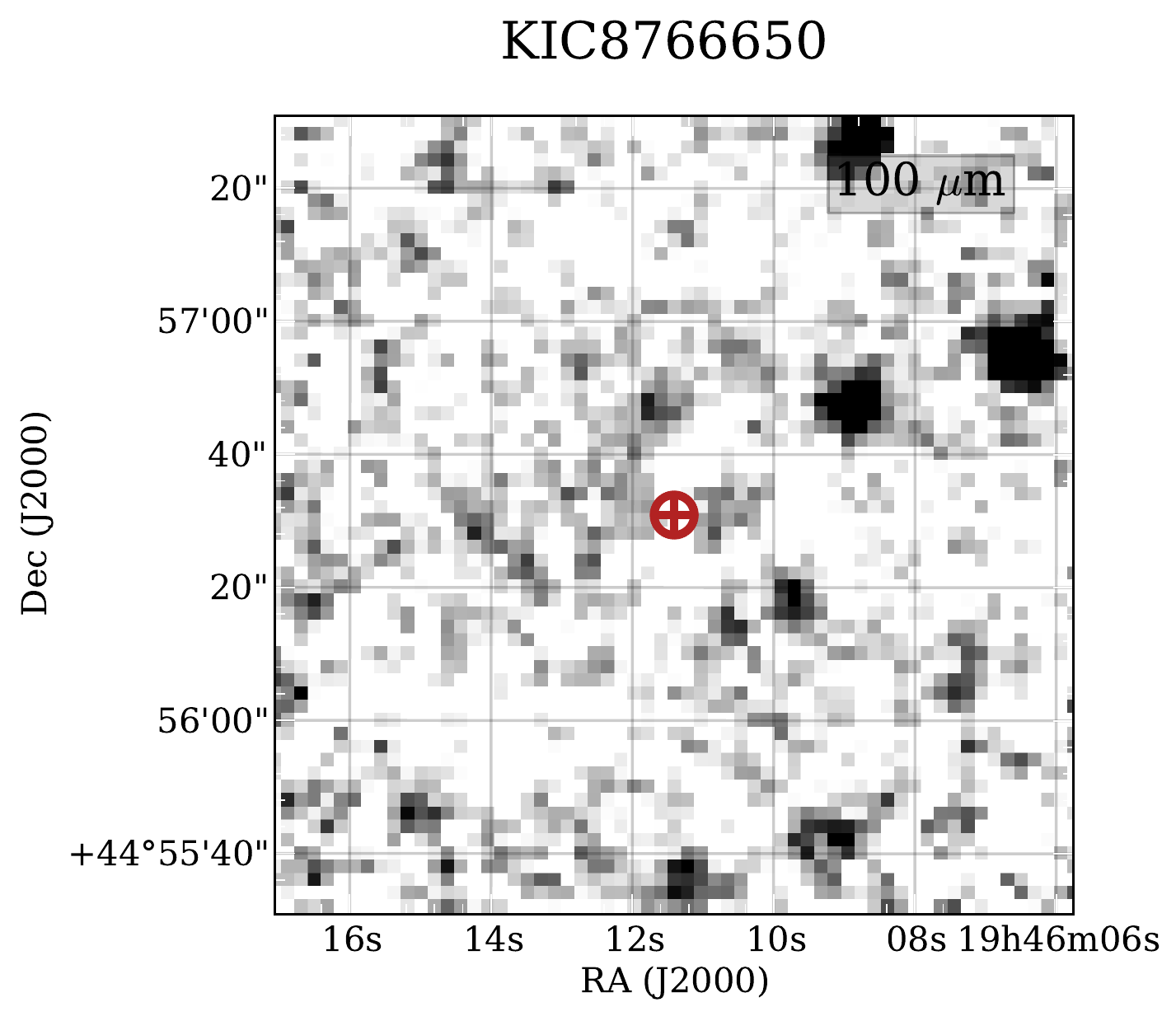}
\includegraphics[width=4.5cm]{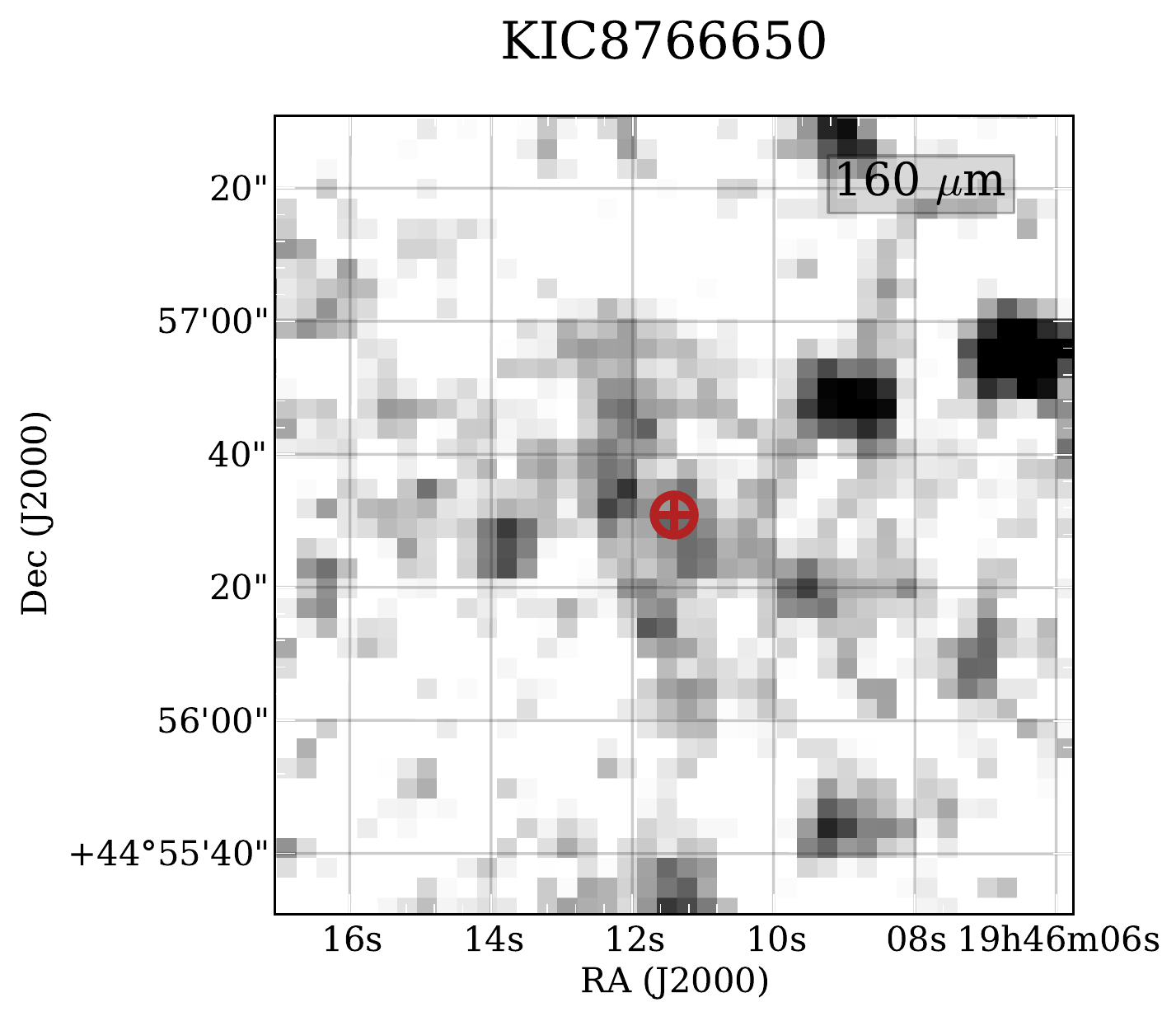}
\includegraphics[width=4.5cm]{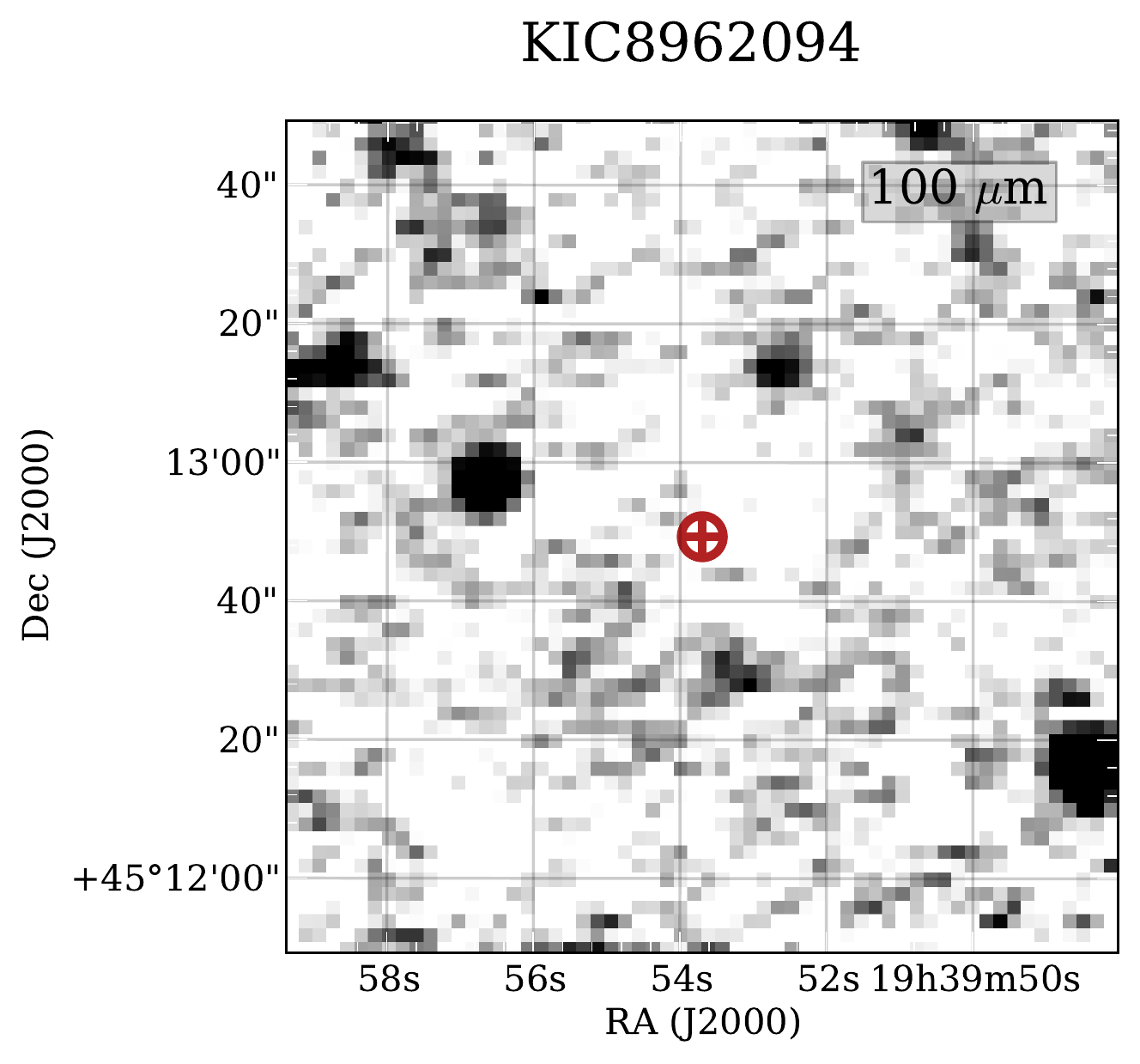}
\includegraphics[width=4.5cm]{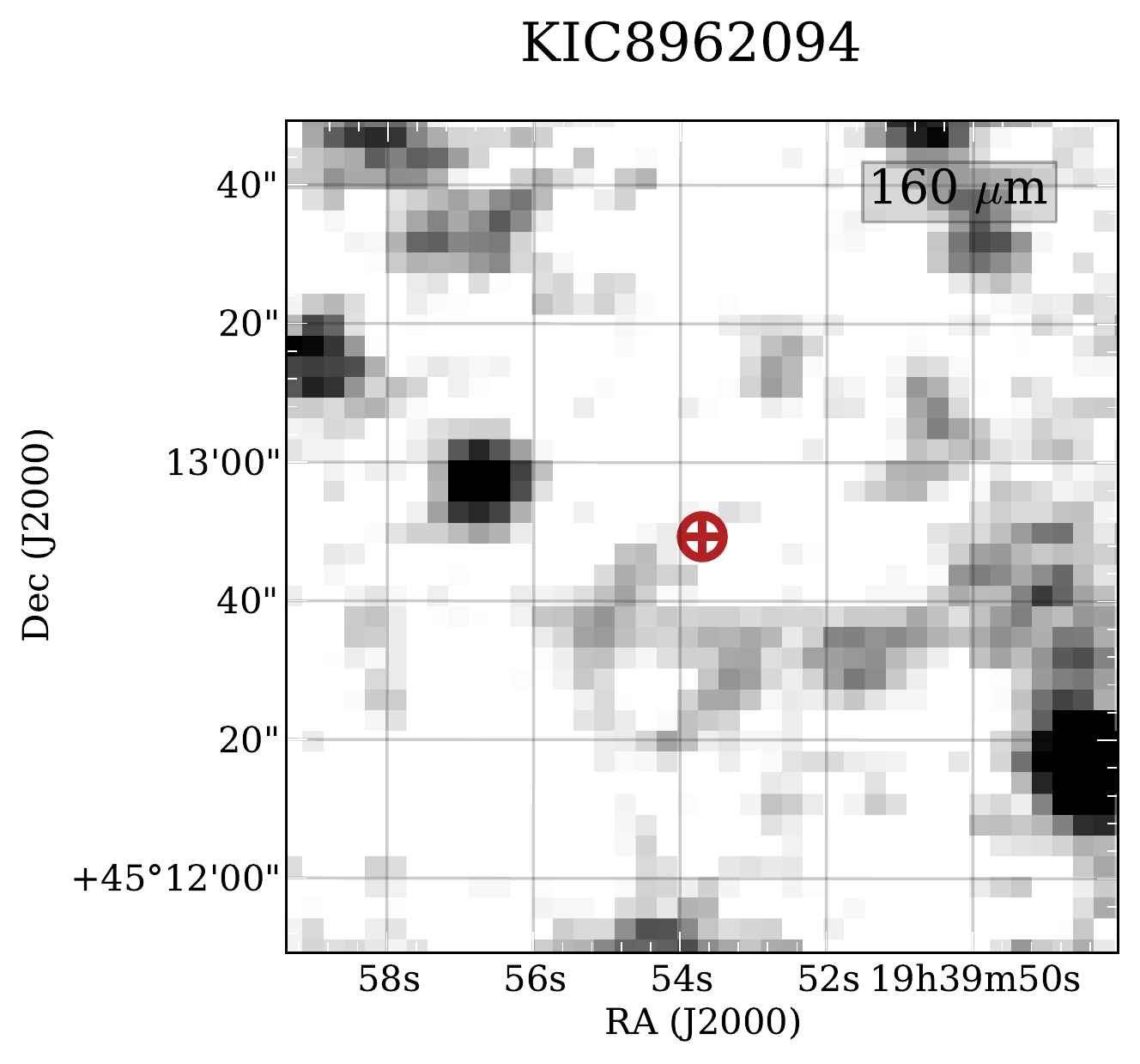}
\includegraphics[width=4.5cm]{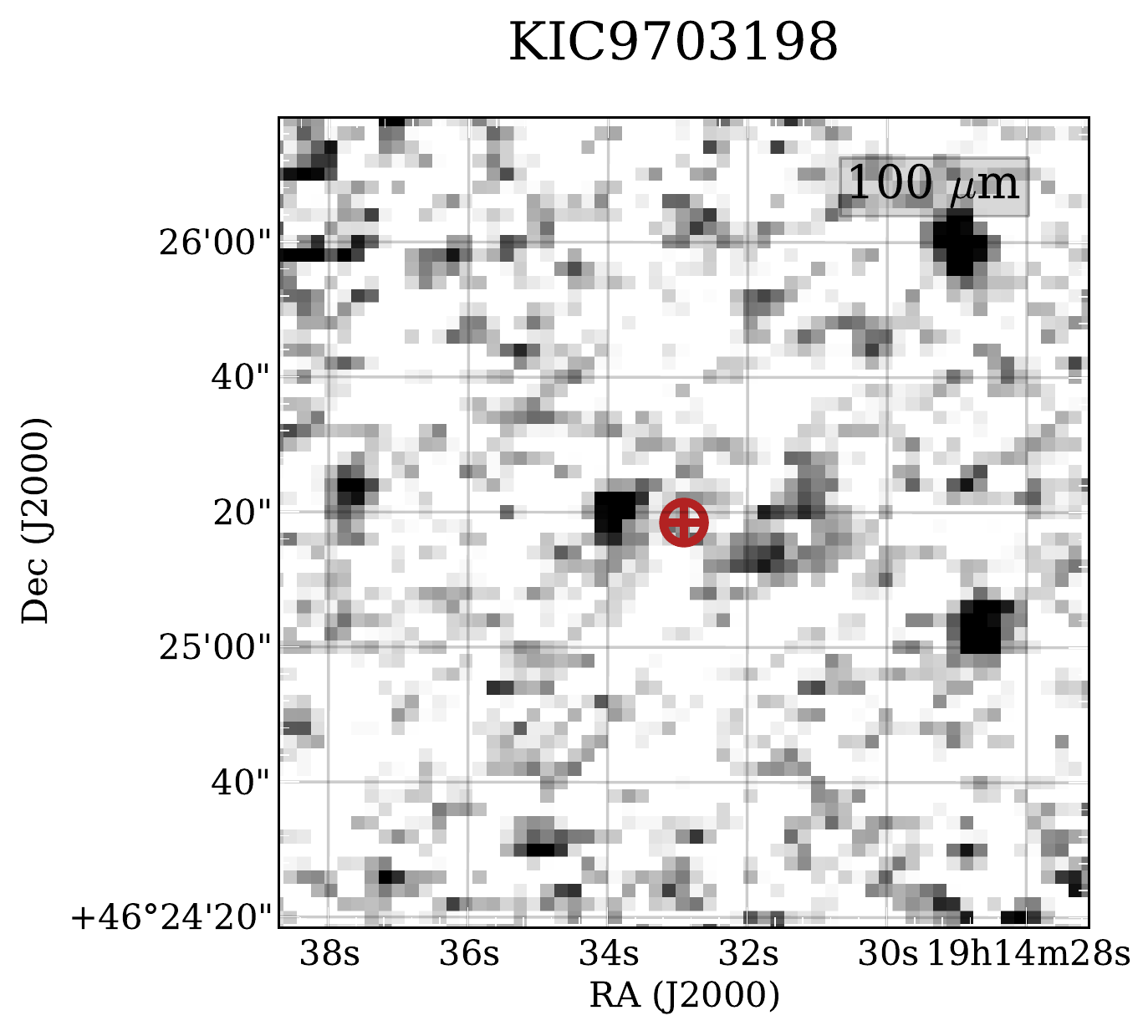}
\includegraphics[width=4.5cm]{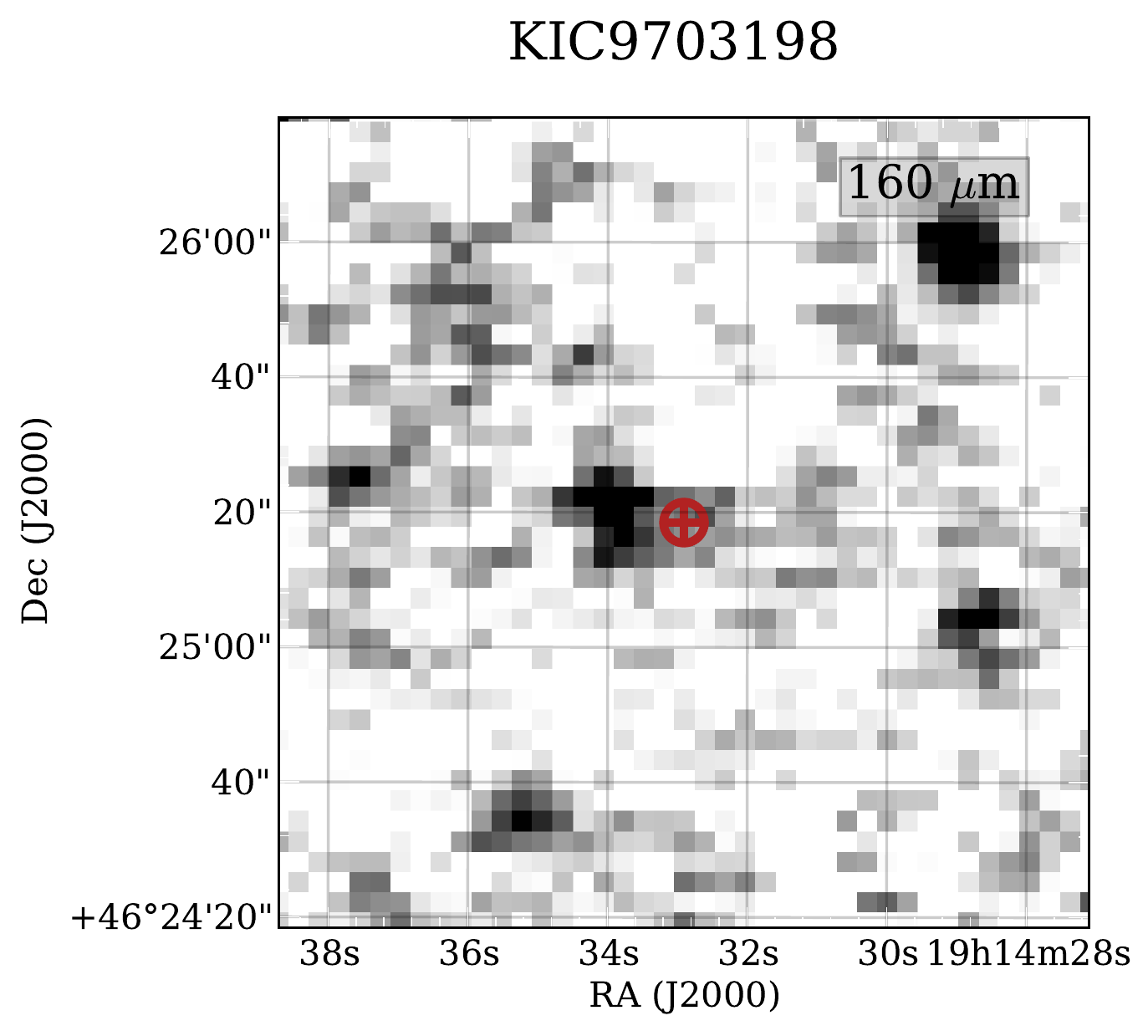}
\includegraphics[width=4.5cm]{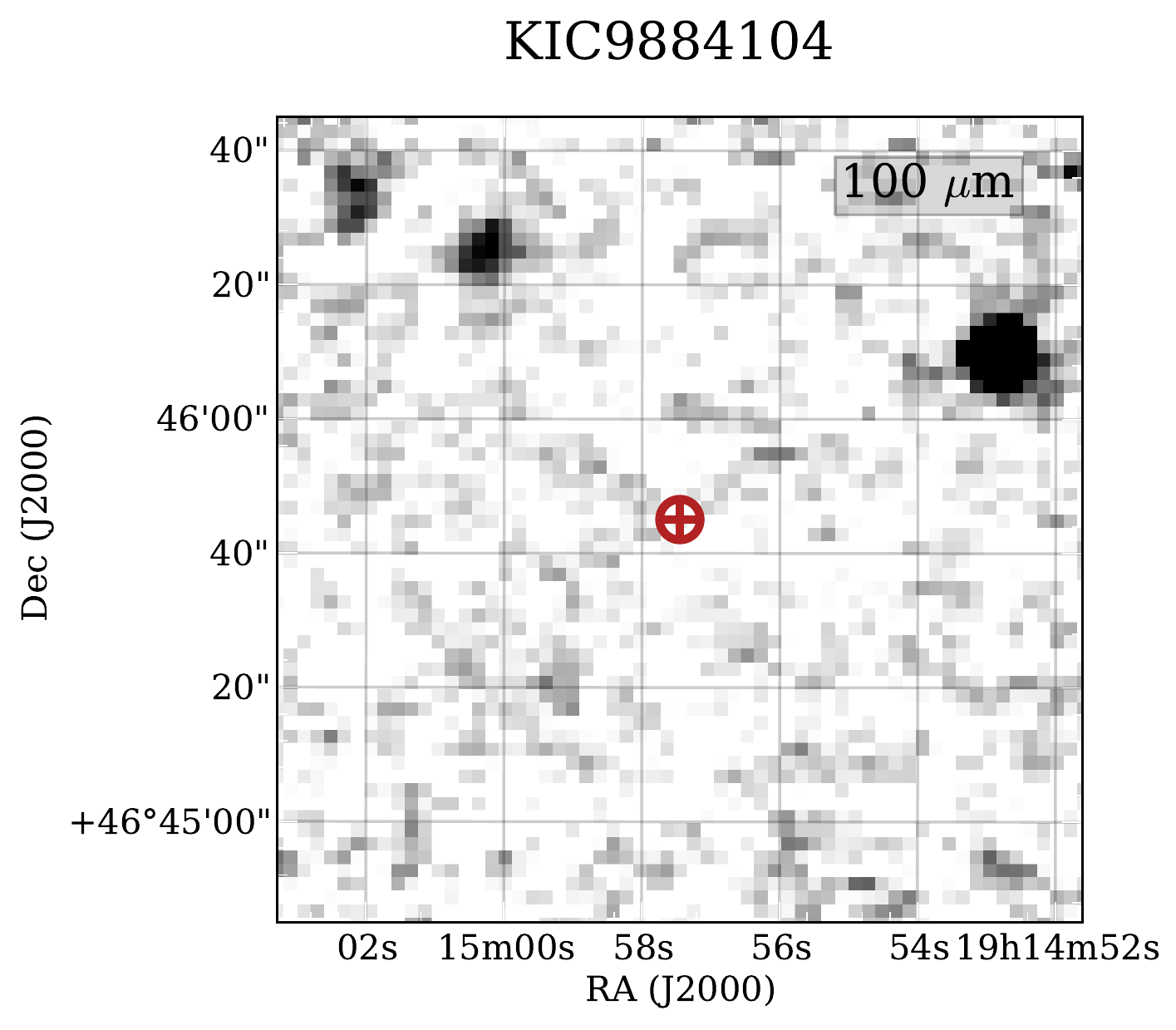}
\includegraphics[width=4.5cm]{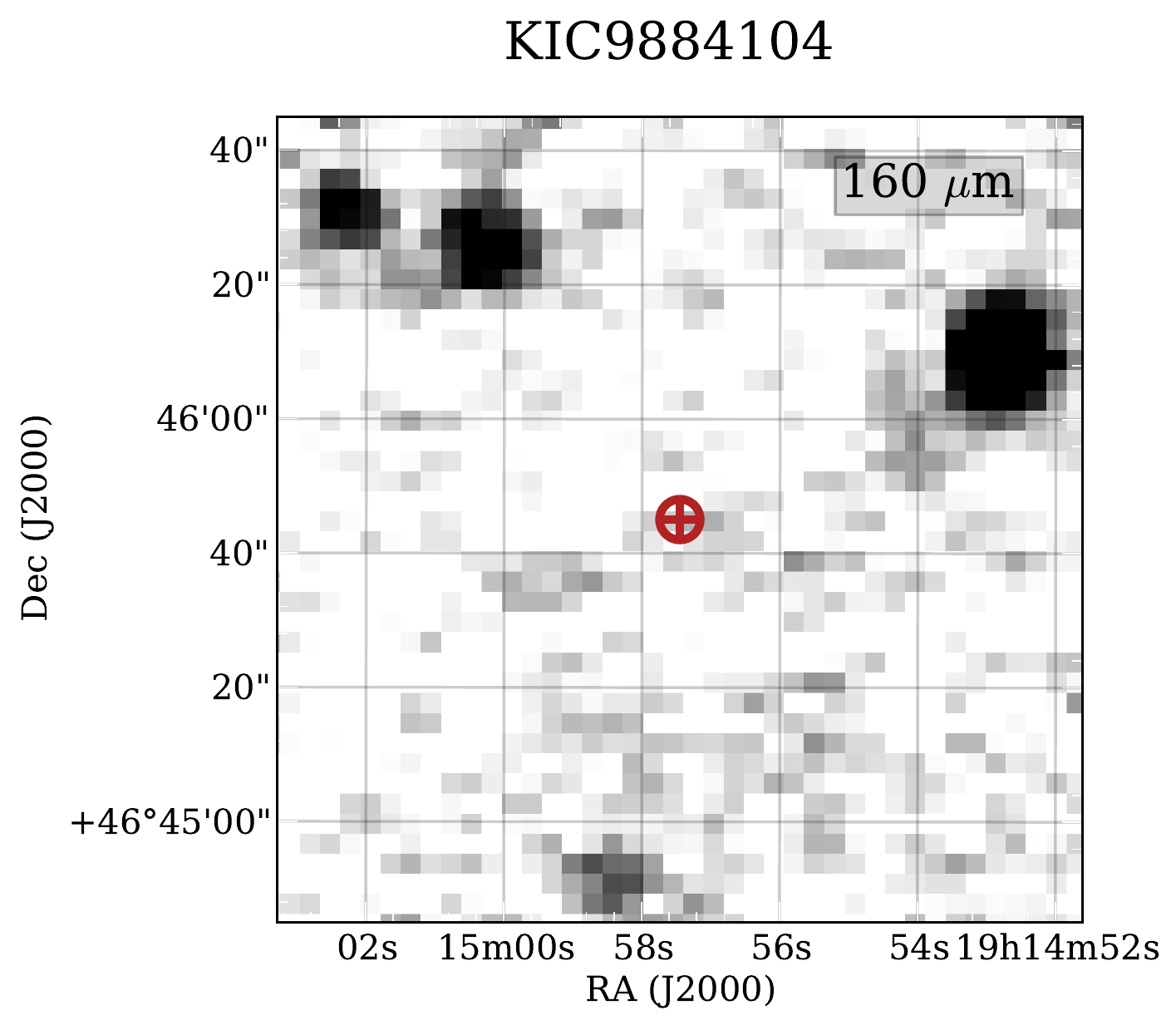}
  \caption{{\it Herschel}/PACS images of the planet host and host candidates in the sample. The red target symbol indicate the nominal position of the target in the KIC catalog or in the Exoplanet Catalog. The linear inverted greyscale goes from zero to 3$\sigma$, where the $\sigma$ is calculated as the standard deviation of the 50 x 50 central pixels in each image.}
  \label{fig:images1}
\end{figure*}

\begin{figure*}
  \centering
\includegraphics[width=4.5cm]{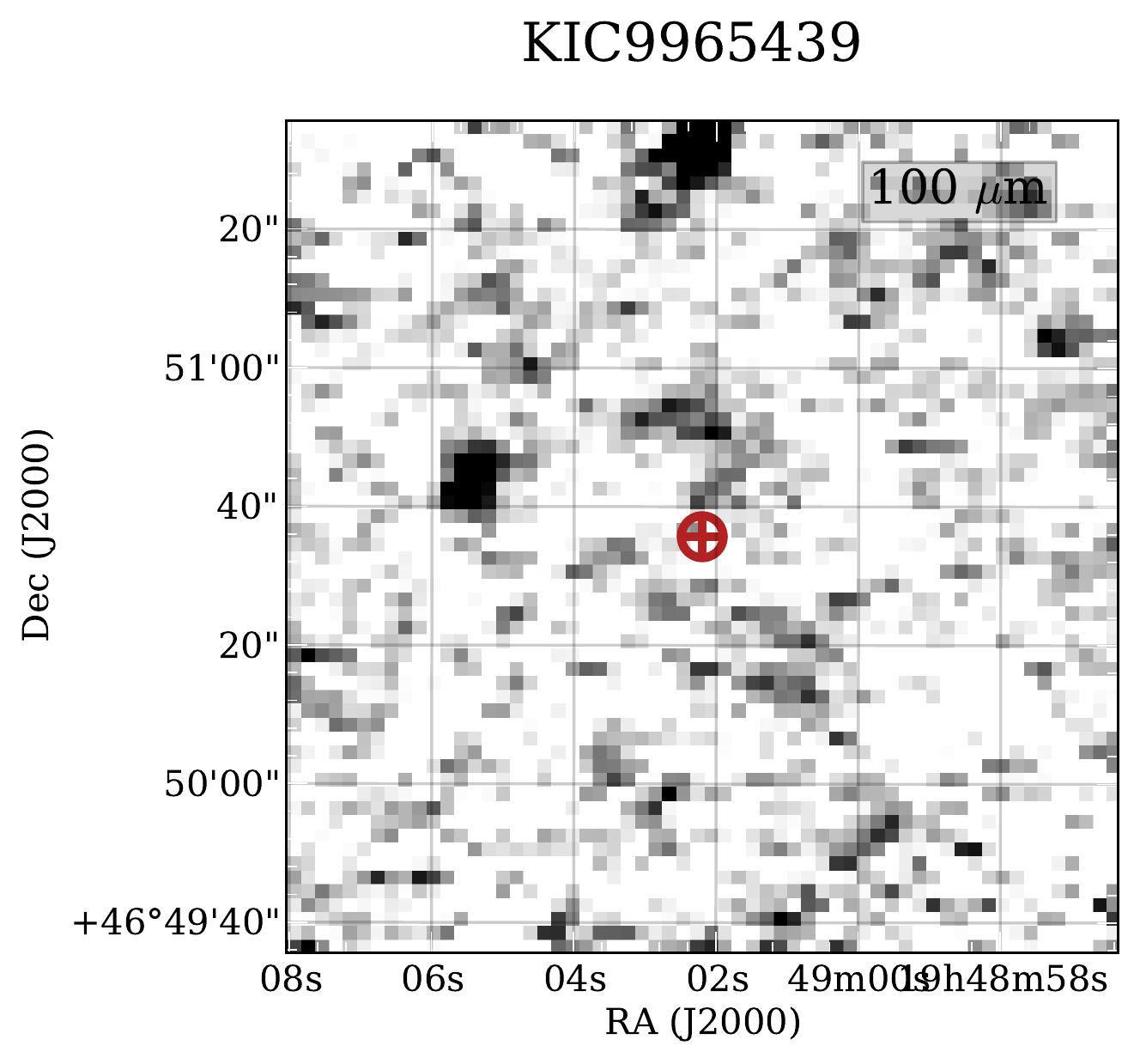}
\includegraphics[width=4.5cm]{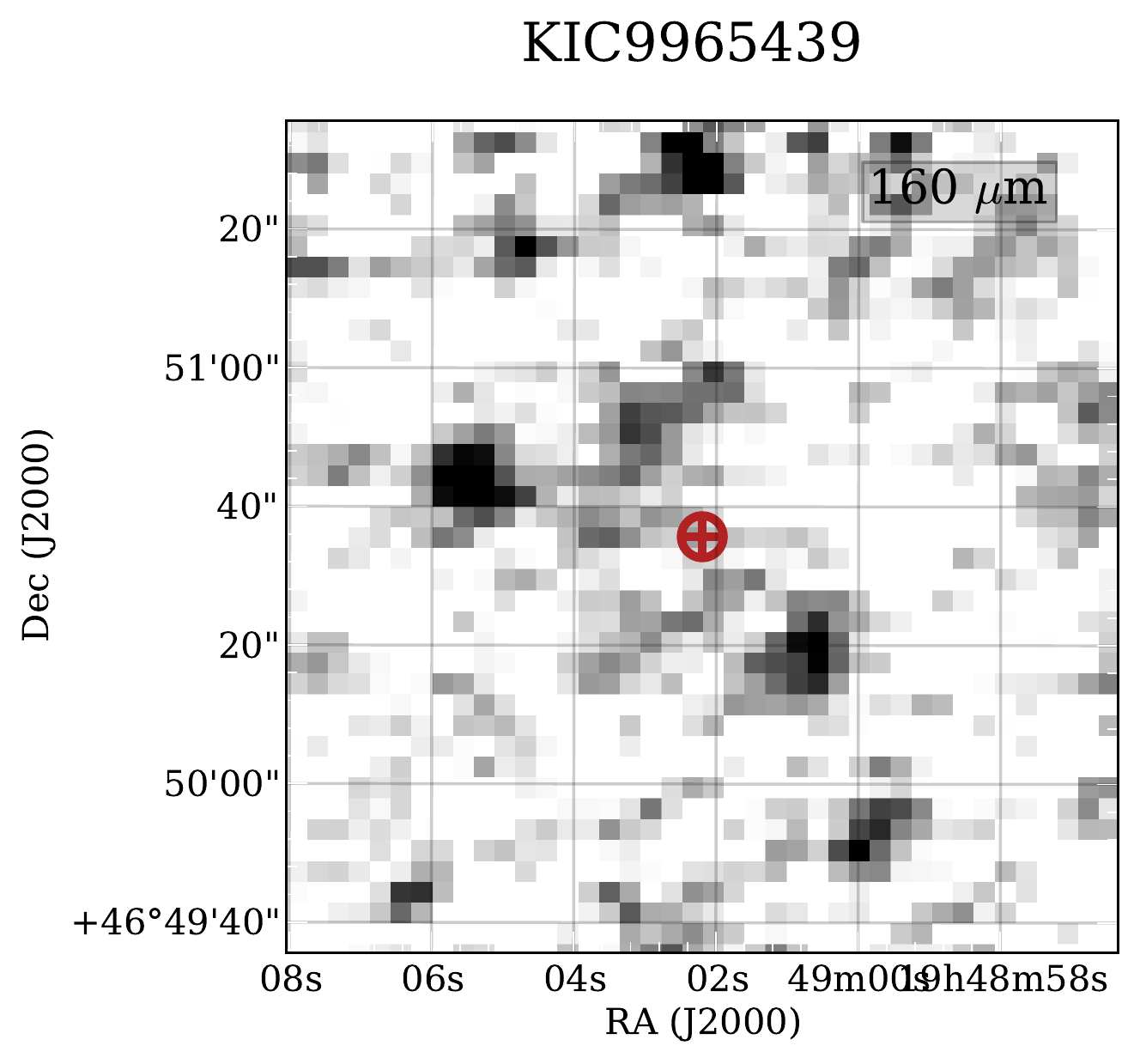}
\includegraphics[width=4.5cm]{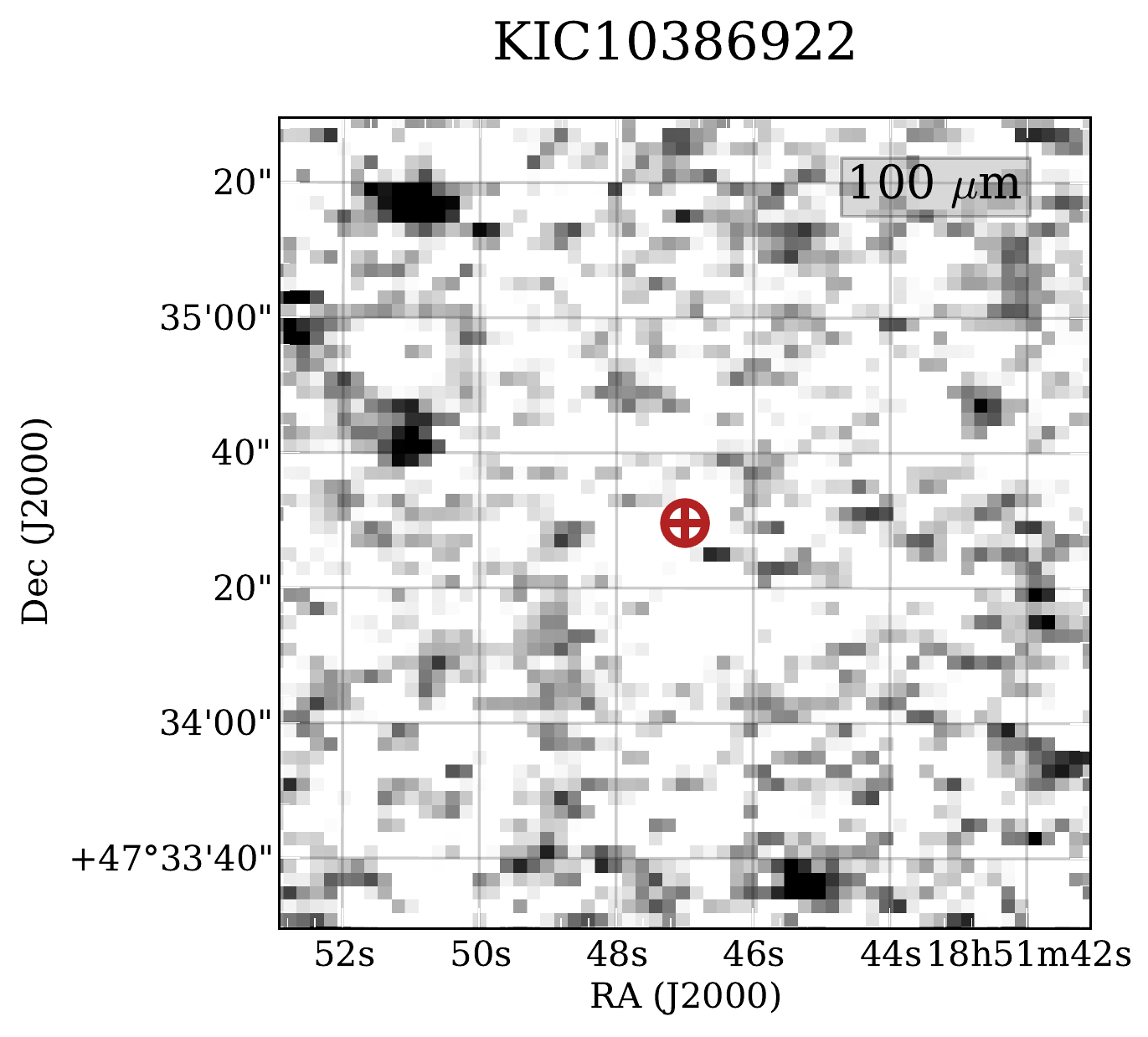}
\includegraphics[width=4.5cm]{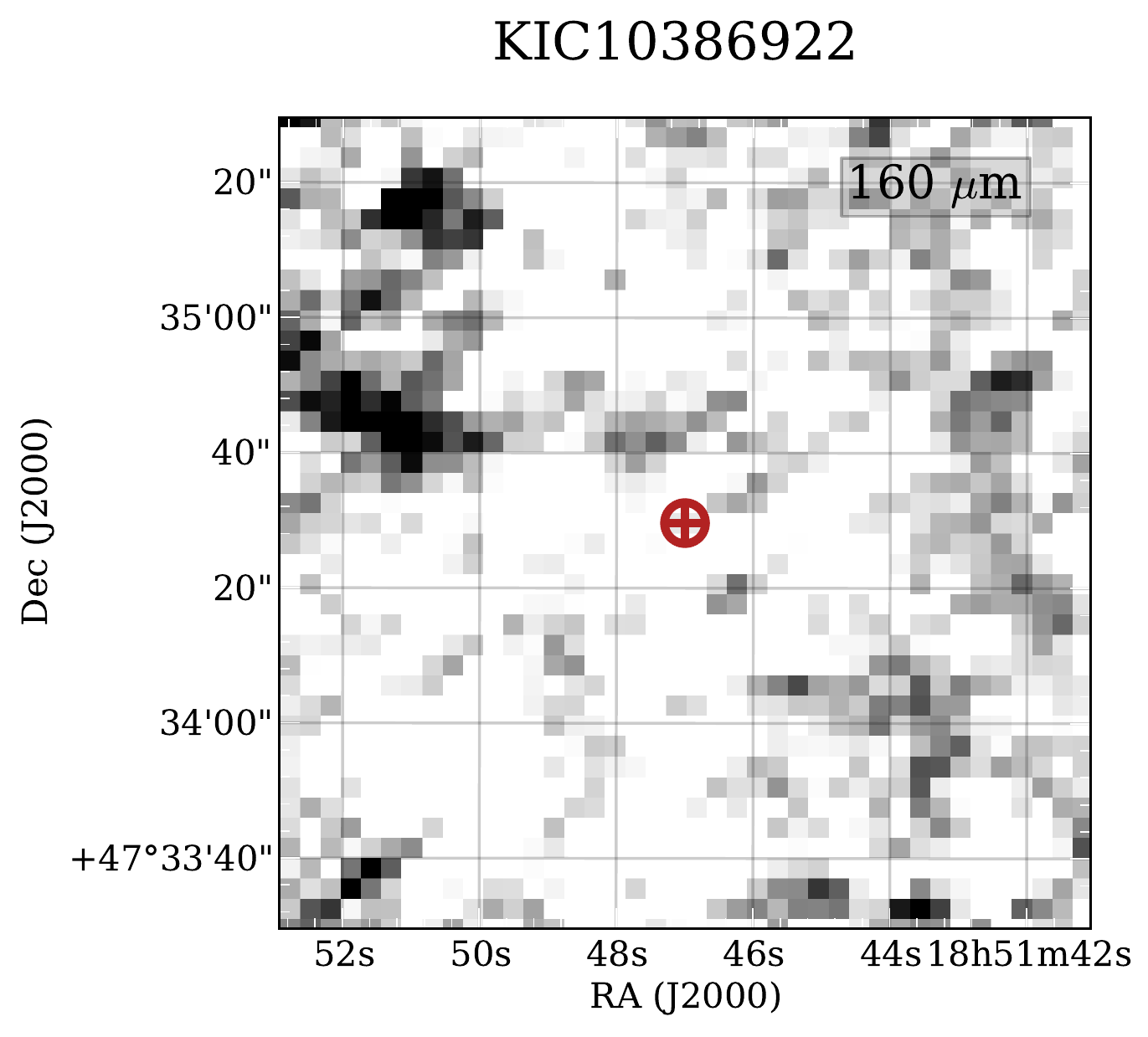}
\includegraphics[width=4.5cm]{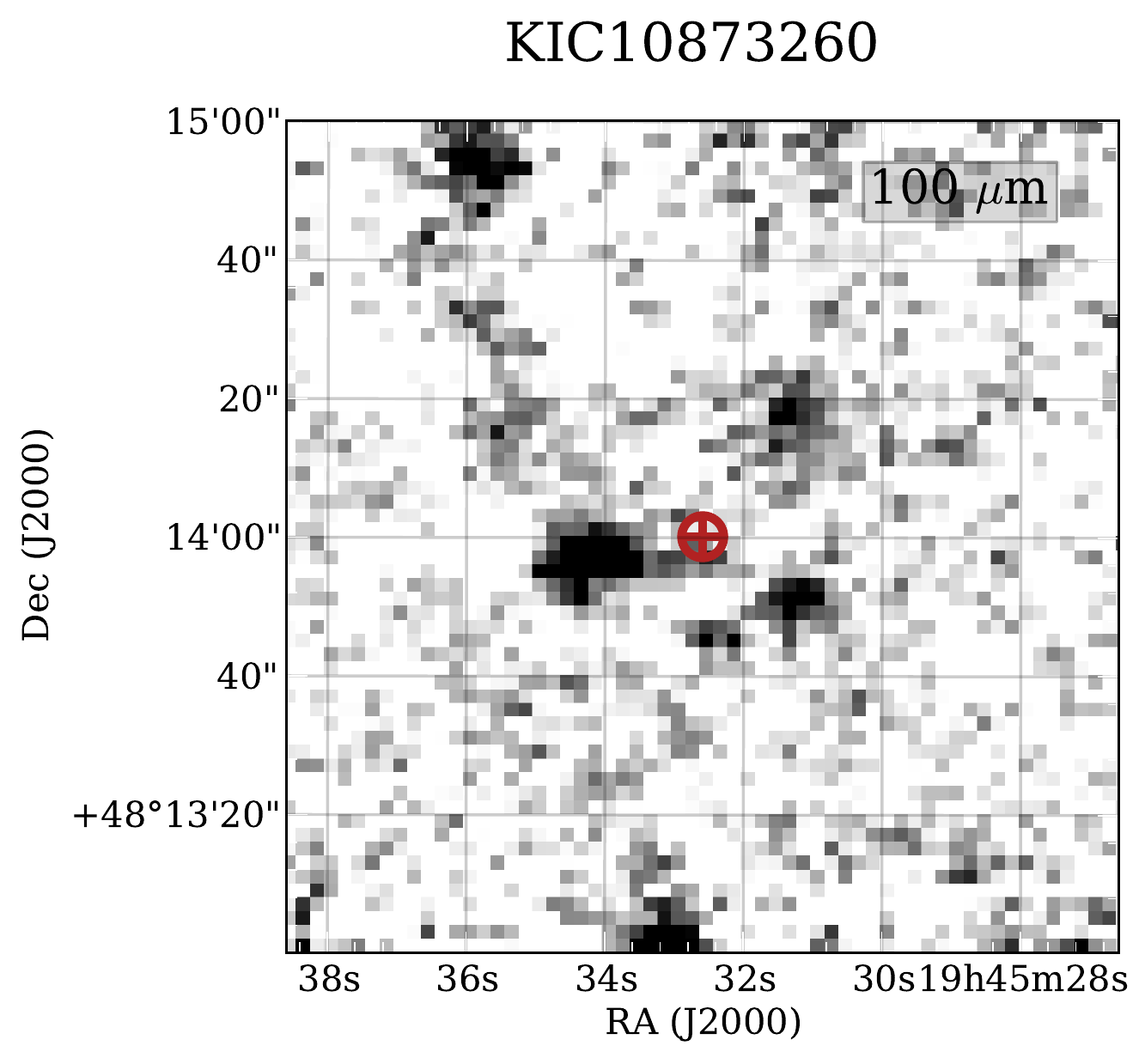}
\includegraphics[width=4.5cm]{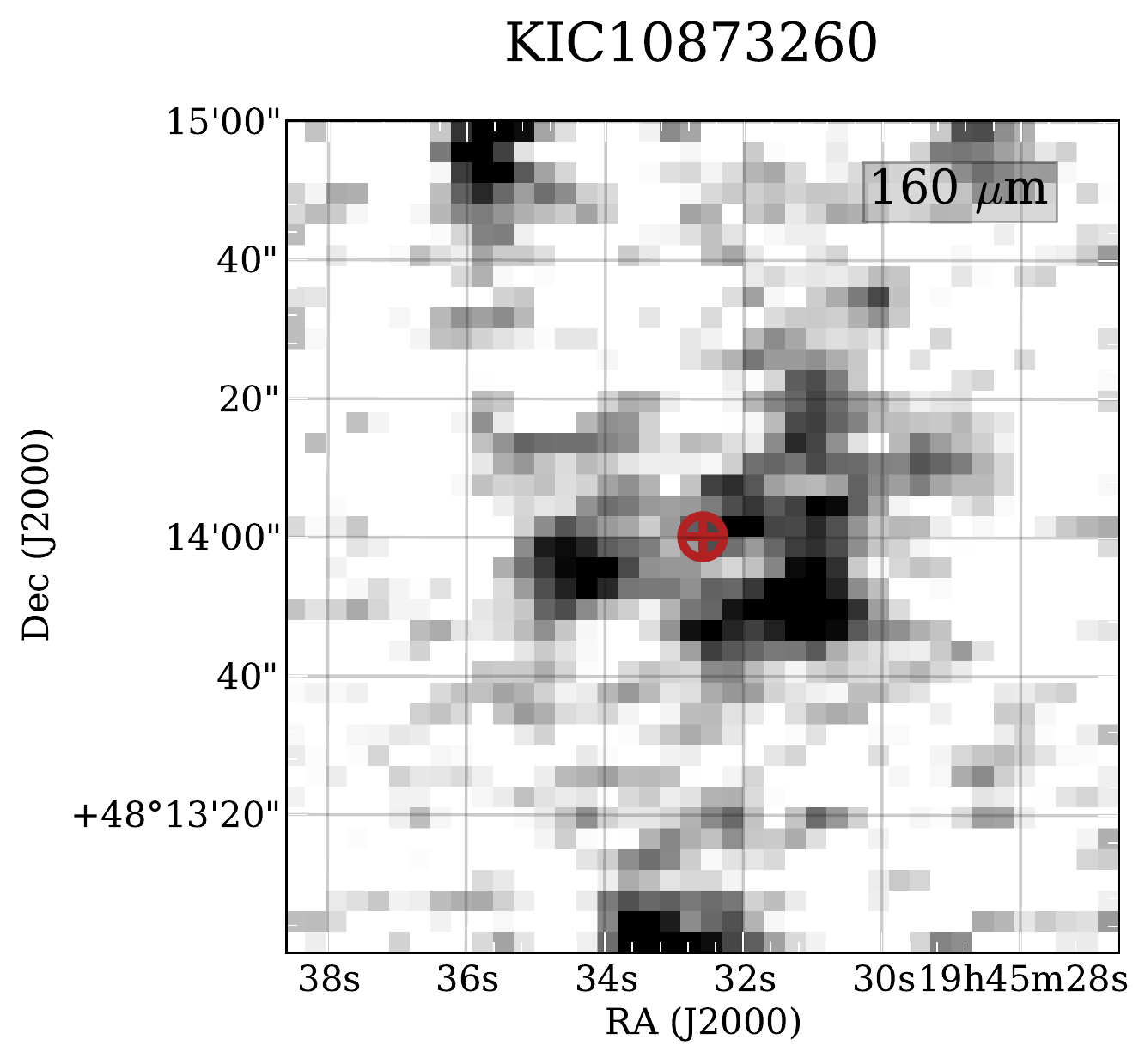}
\includegraphics[width=4.5cm]{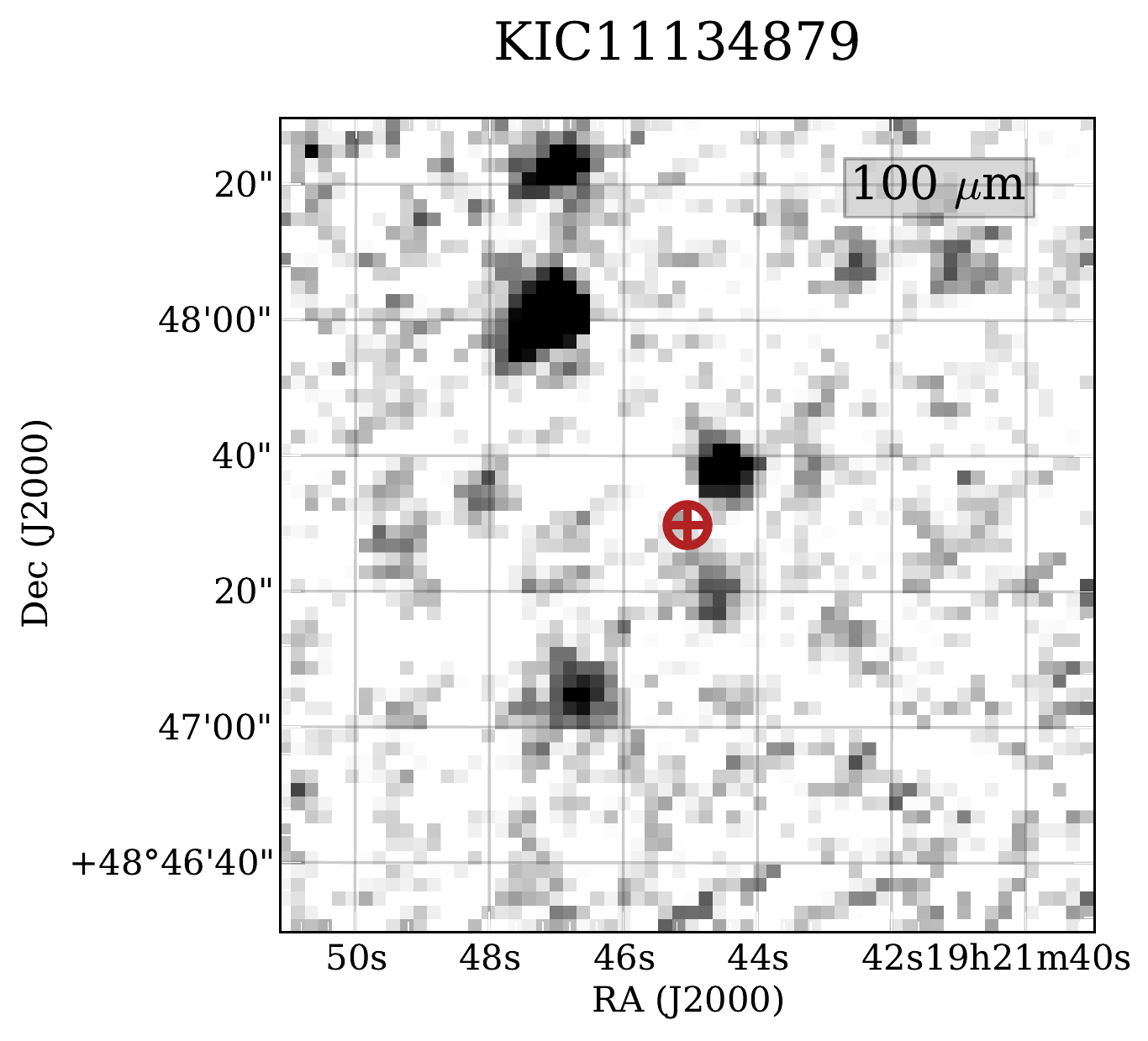}
\includegraphics[width=4.5cm]{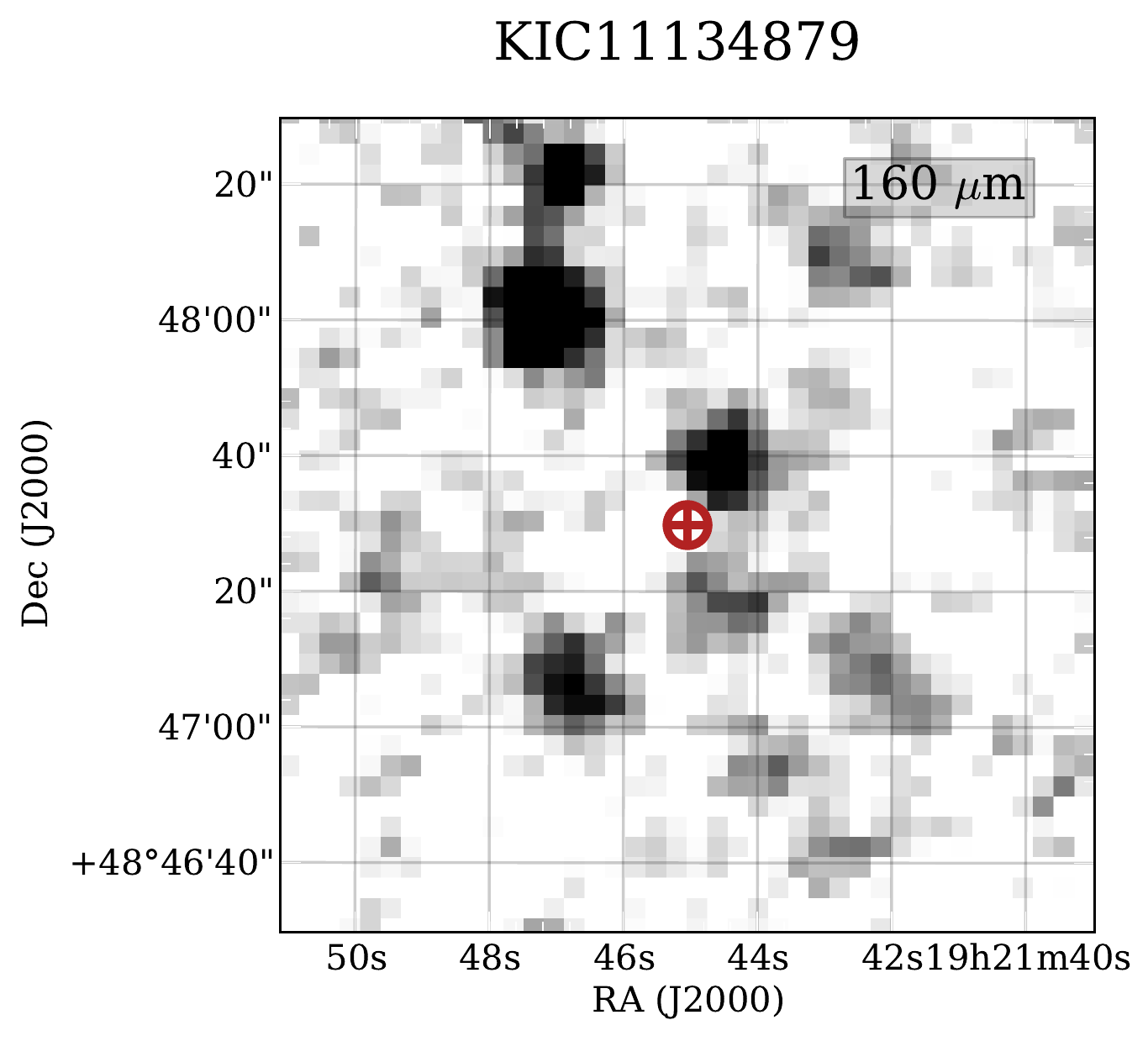}
\includegraphics[width=4.5cm]{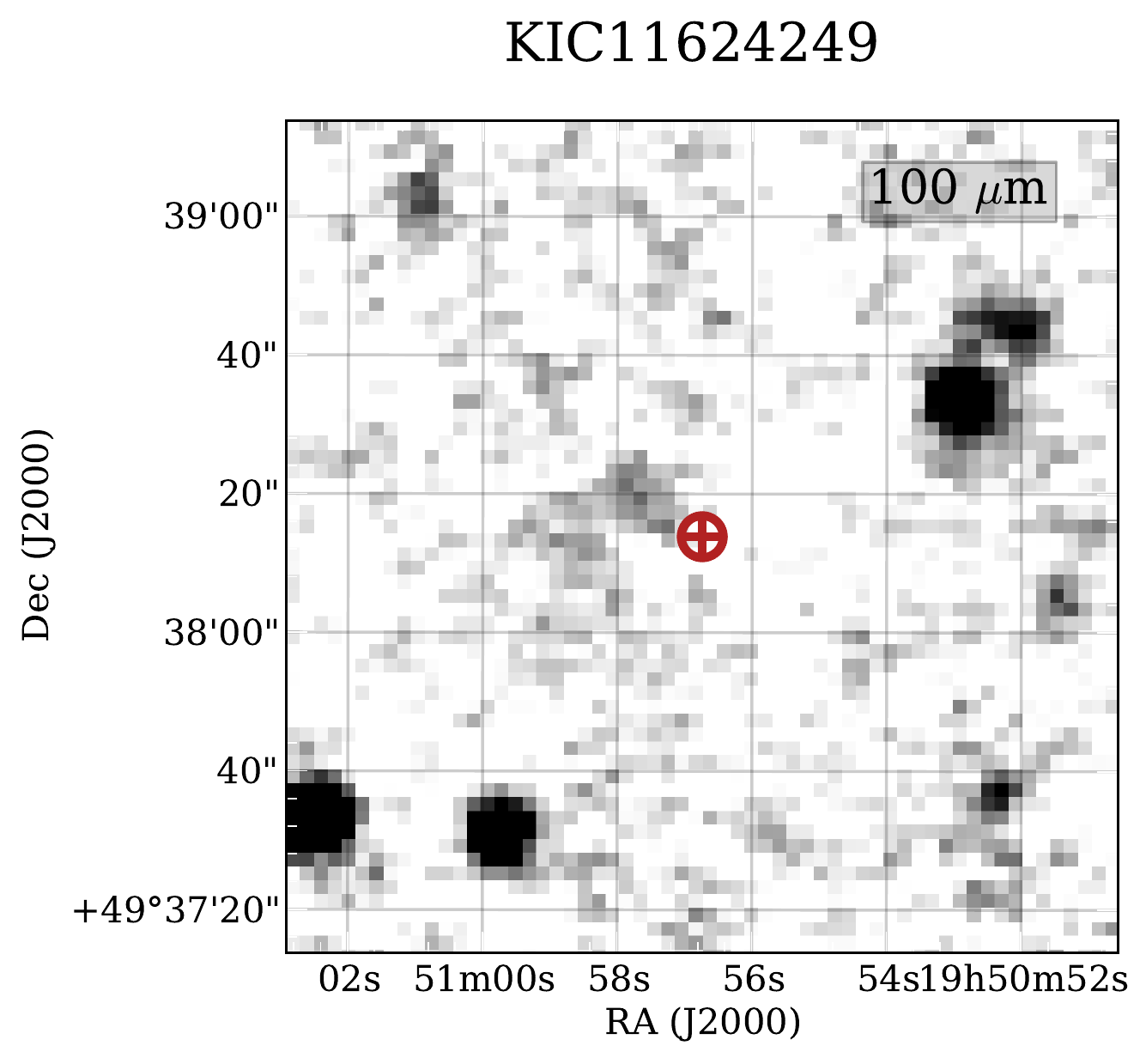}
\includegraphics[width=4.5cm]{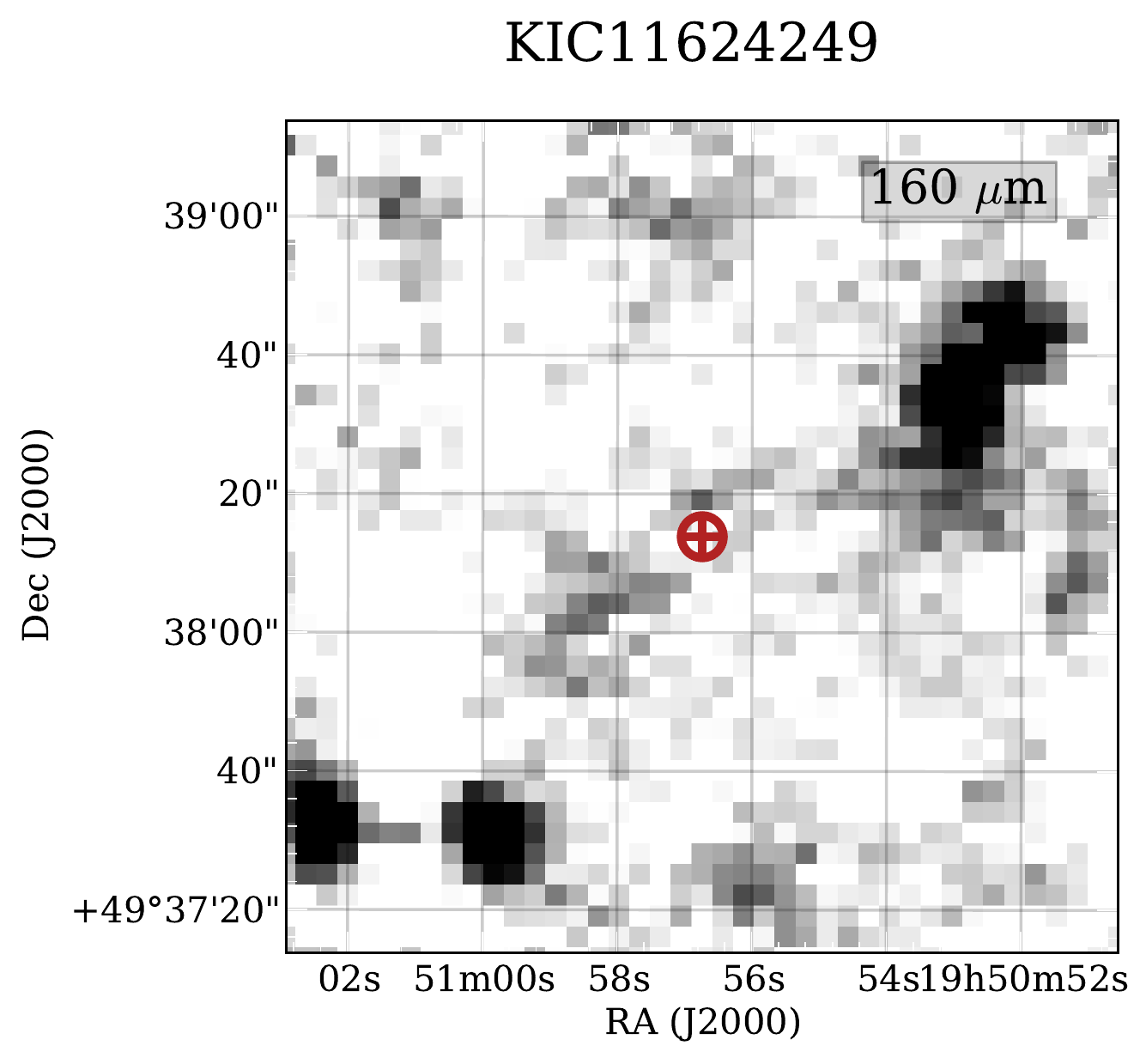}
\includegraphics[width=4.5cm]{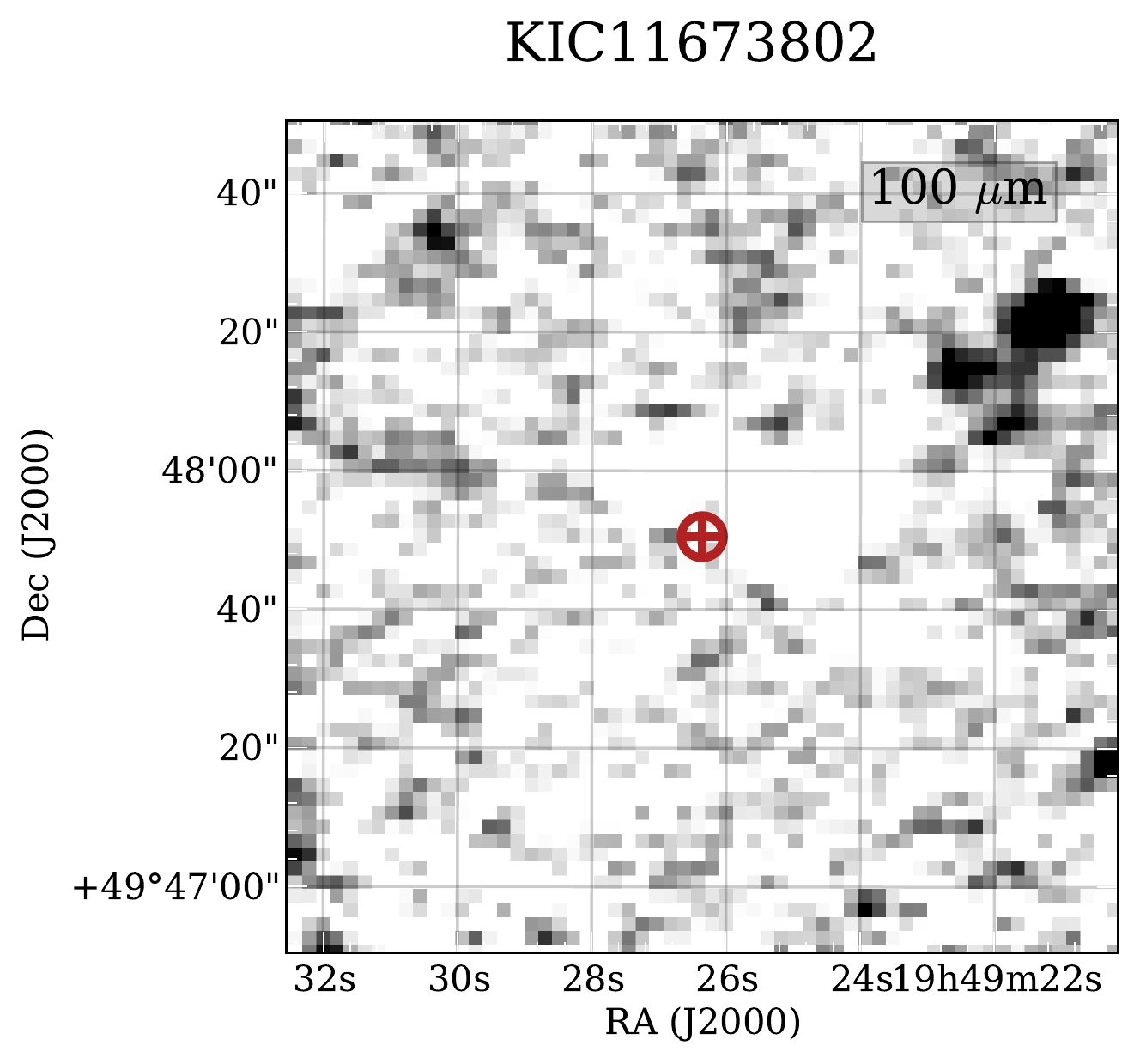}
\includegraphics[width=4.5cm]{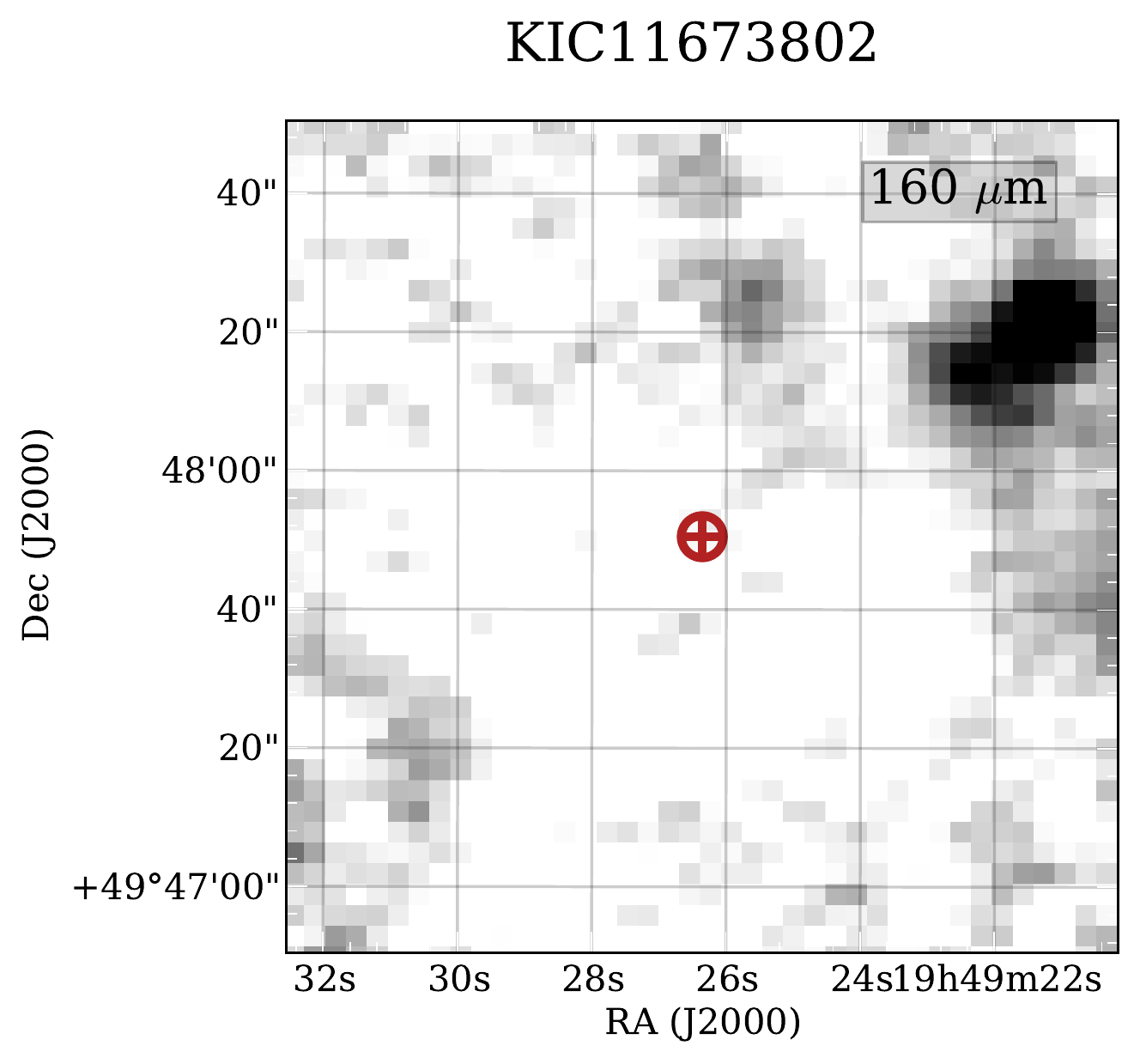}
\includegraphics[width=4.5cm]{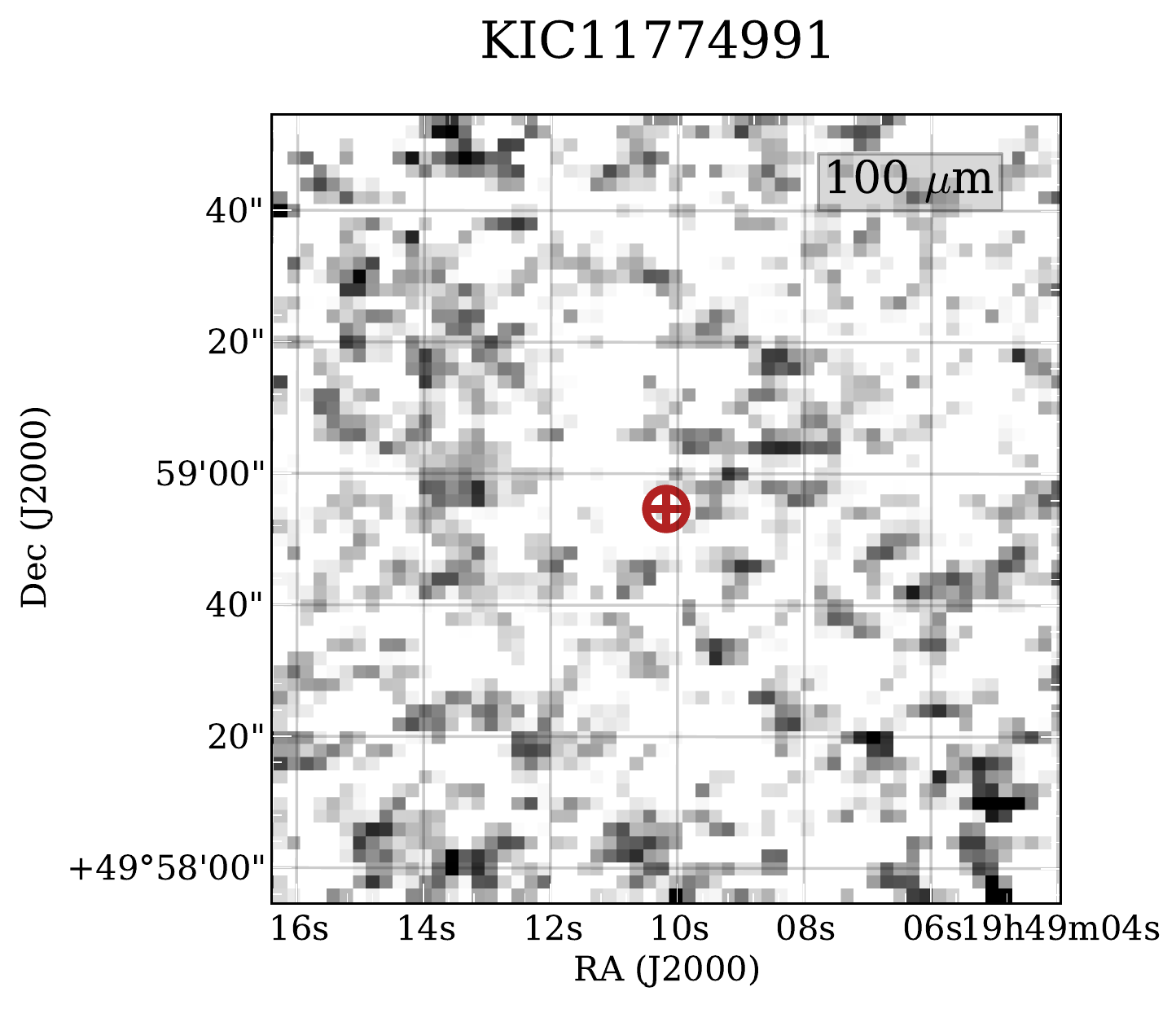}
\includegraphics[width=4.5cm]{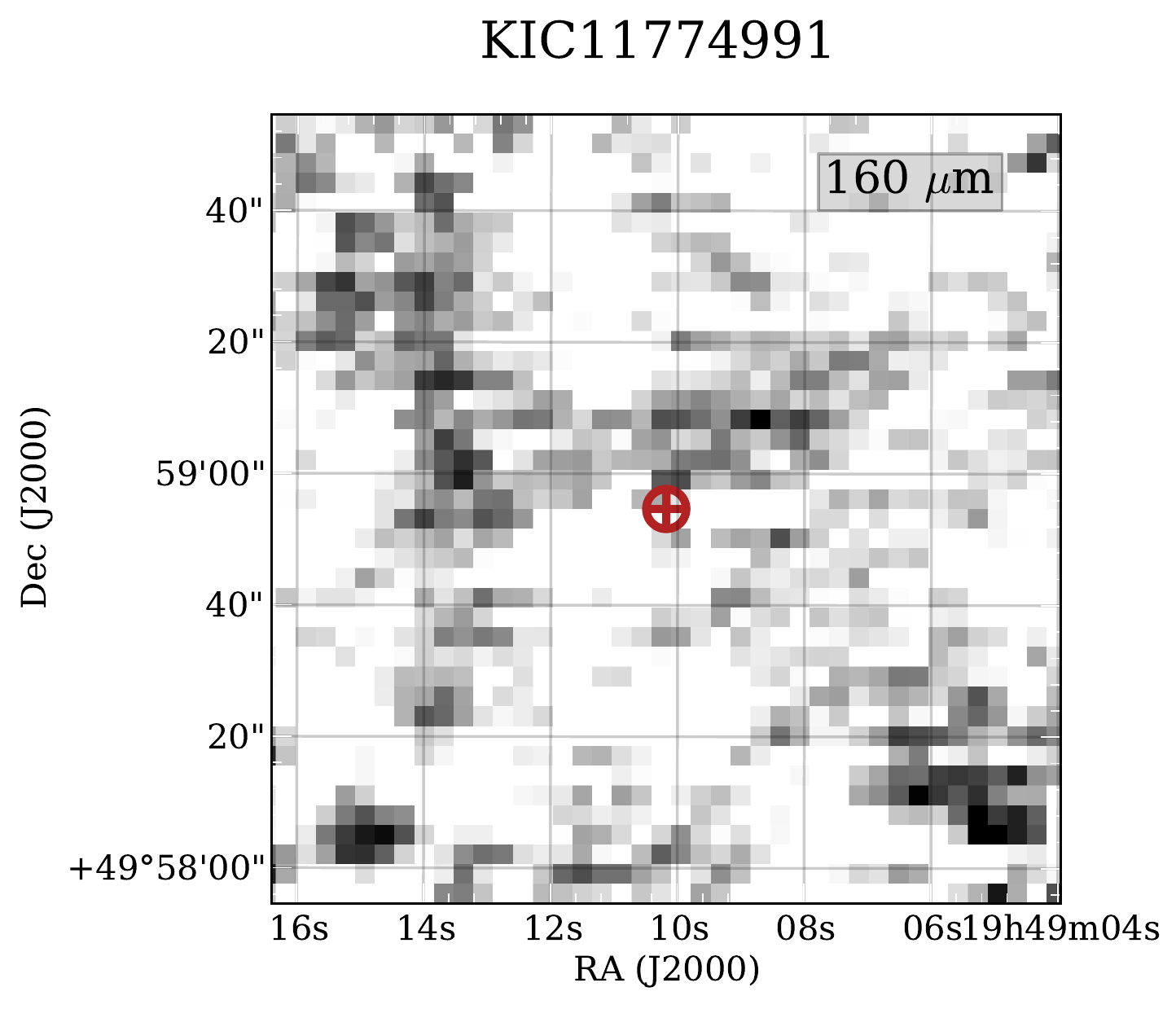}
  \caption{Continued from figure \ref{fig:images1}.}
  \label{fig:images2}
\end{figure*}

\subsection{RESULTS}
\label{results}

No clear point sources were detected at any of the target positions. The spacecraft pointing of the observations of object KIC3835670 was known to be affected by a reset of the spacecraft velocity vector (SVV) as identified during the quality control checks done at the Herschel Science Centre. This vector is used by the star tracker to compute the aberration correction of the coordinates of guide stars. To correct for these small offsets new pointing products were produced at the Herschel Science Centre with HCSS version 10.1 and applied to all affected observations from versions 11 onward \citep{SanchezPortal2014}. We also applied special checks to verify the pointing:  using HIPE, we retrieved all PACS photometer observations in each of the days when our observations were taken and calculated the offsets between intended (nominal RA and DEC, as in the observations metadata) and achieved astrometry by fitting 2D Gaussians to any source found within 5\arcsec x 5\arcsec around the intended central position of the images. This confirmed that the mean residual pointing offset on the days when our observations were taken was always smaller than 1.7$\arcsec$. This rules out the possibility that large offsets would still be present in any of these images, hence that some of the nearby point sources could indeed be associated to our target stars.



The distances in arc seconds to the closest point-source emissions in the images within 50\arcsec x 50\arcsec of the target positions were calculated by fitting 2D Gaussians on our images to the closest nearby point source and then calculating the distances between the 2D gaussian peak position and the target positions. These values are given in Table \ref{tab:sample} and range from 3 to 38 \arcsec. In particular, for a few targets, namely for HAT-P-28, KIC3835670, and KIC9703198,  point sources are found close to the targets positions. For KIC4918309 the source detection routine converged to a very nearby source as well, although probably has picked up a very low SN feature close to the target position. Doing blind aperture photometry with the recommended aperture of ten pixels in radius (2\arcsec\, for 100 $\mu$m and 30\arcsec\, for 160 $\mu$m) at those positions could produce incorrect fluxes for these objects due to emission from unrelated sources falling into the apertures \cite[see also, e.g.,][for more research on confused far-IR fluxes]{Matra2012,Lestrade2012}. We therefore only computed upper limits to the fluxes at the positions of our sources. For this, we used the recommended method (implemented in the user pipeline script in HIPE 11.0) of estimating the RMS of 100 circular apertures of five pixels in radius (10\arcsec\, for 100 $\mu$m and 15\arcsec\, and 160 $\mu$m) in a grid of 1.6\arcmin x 1.6\arcmin\, for 100 $\mu$m or 2.5\arcmin x 2.5\arcmin\, for 160 $\mu$m, centered around the target positions. The flux upper limit values obtained are given in Table \ref{tab:sample} and are in general consistent with the predictions from HSpot (v. 6.3.2) for the sensitivity of the PACS mini-scan AOT. 

While most fields appear clean, few of them show extended emission at distances of 3\arcsec to 6\arcsec from the target positions. Extended far-IR background emission is found to be more common at 160 $\mu$m, even at such high galactic latitude fields as these. In particular for KIC3835670, which has the strongest emission closest to the star's position, the extended emission has an extent of 29\arcsec. Assuming main sequence stellar radius and luminosity for the star and comparing its J-band magnitude with the absolute magnitude of a star from its spectral type, we compute a photometric distance to this object of $\sim$ 600 pc, which then implies that the extended emission seen at the source position in the PACS maps would have a physical extension of 17\,400 AU if it were associated to the star. This essentially rules out the possibility that this is a debris disk. There is a possibility that a 10 mJy debris disk would be hidden among this bright background extended emission but it is impossible to prove that with the current data. Only higher resolution mid-IR imaging can help sort out whether the extended emission could be related to these sources, although it is highly unlikely given the projected distances and sizes of the background sources. This result then gives support to the thesis by \citet{Kennedy2012} that all these objects are likely due to chance alignment with either background galaxies or interstellar medium.

\begin{figure*}
\includegraphics[width=4.5cm]{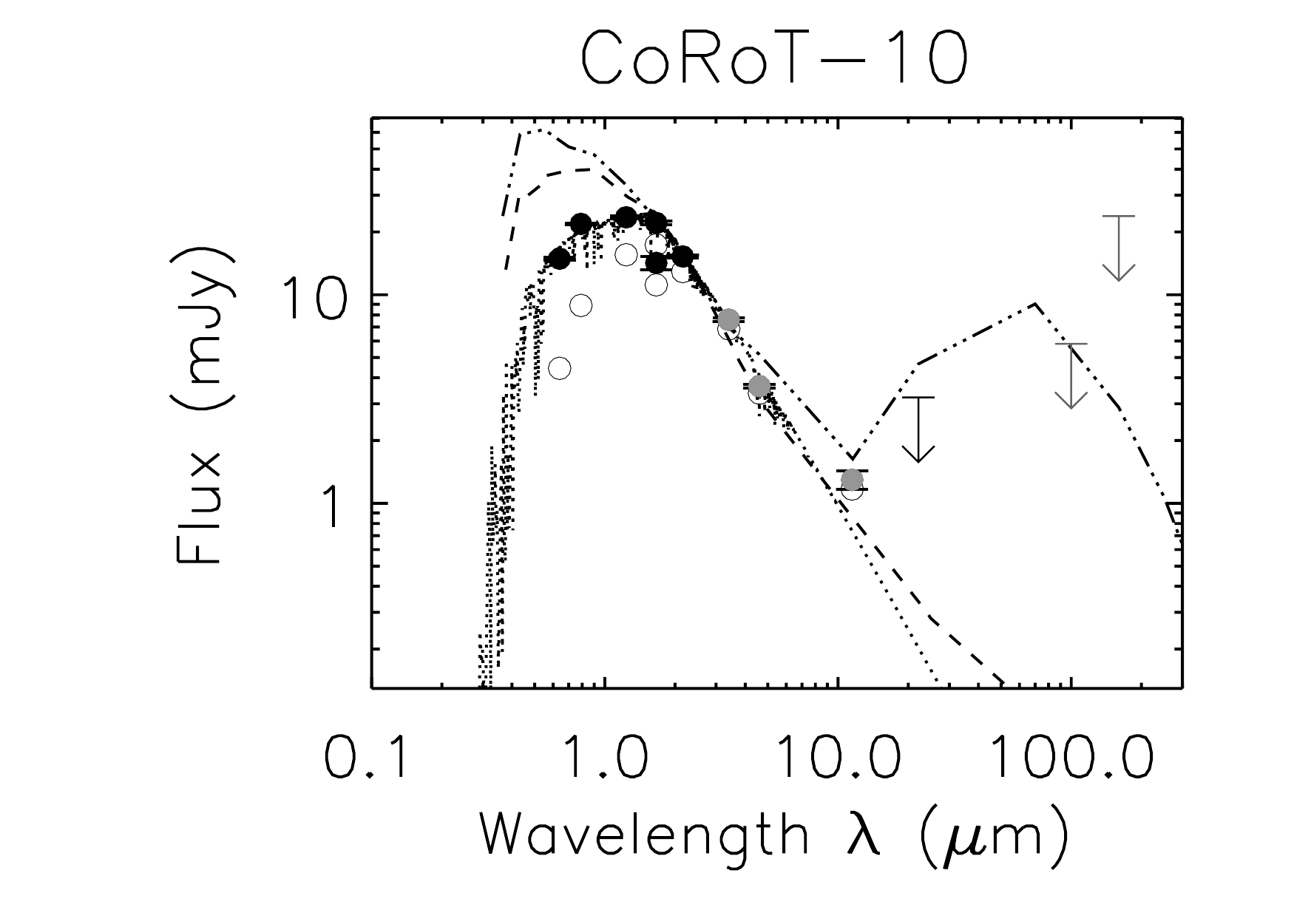}
\includegraphics[width=4.5cm]{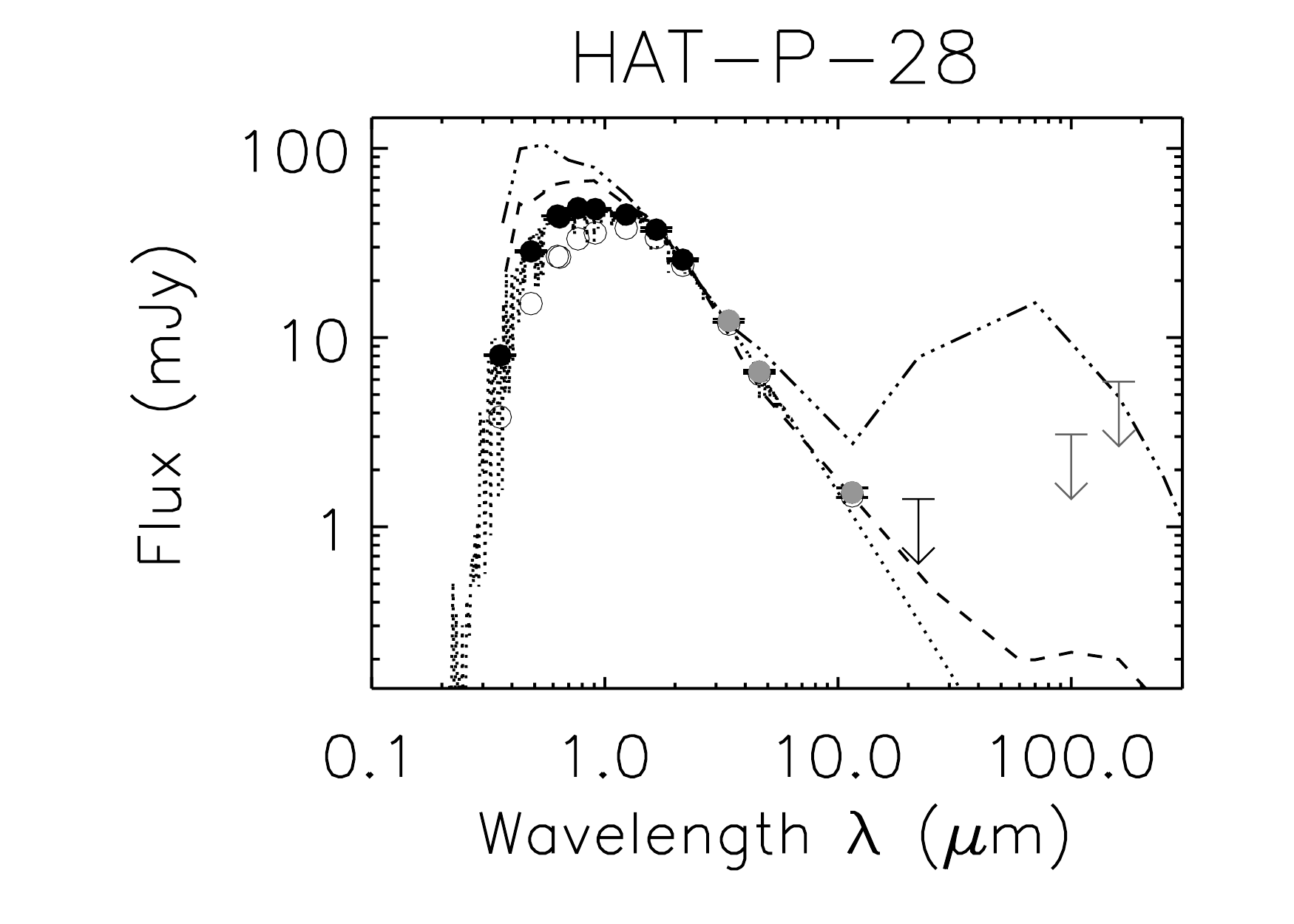}
\includegraphics[width=4.5cm]{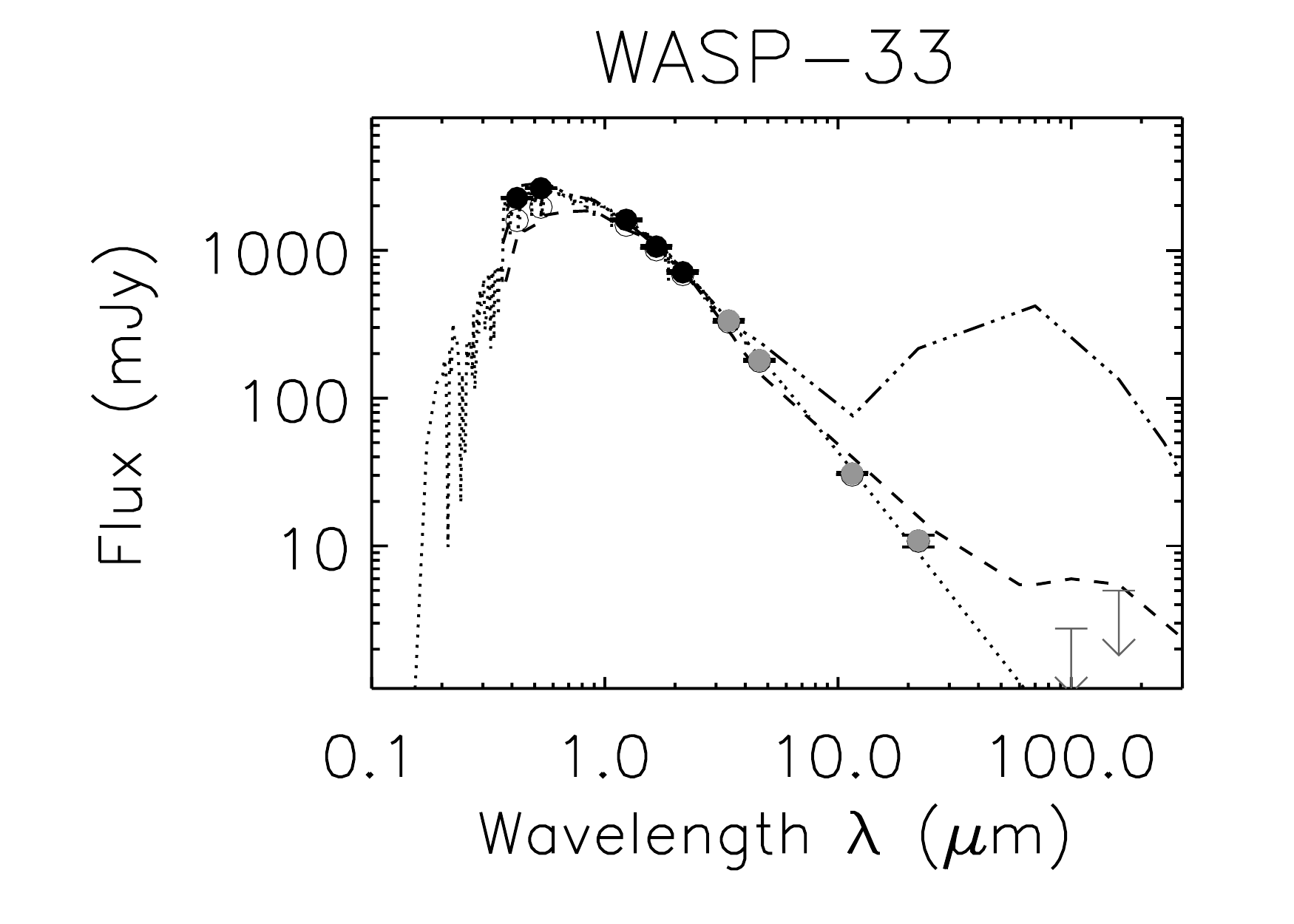}
\includegraphics[width=4.5cm]{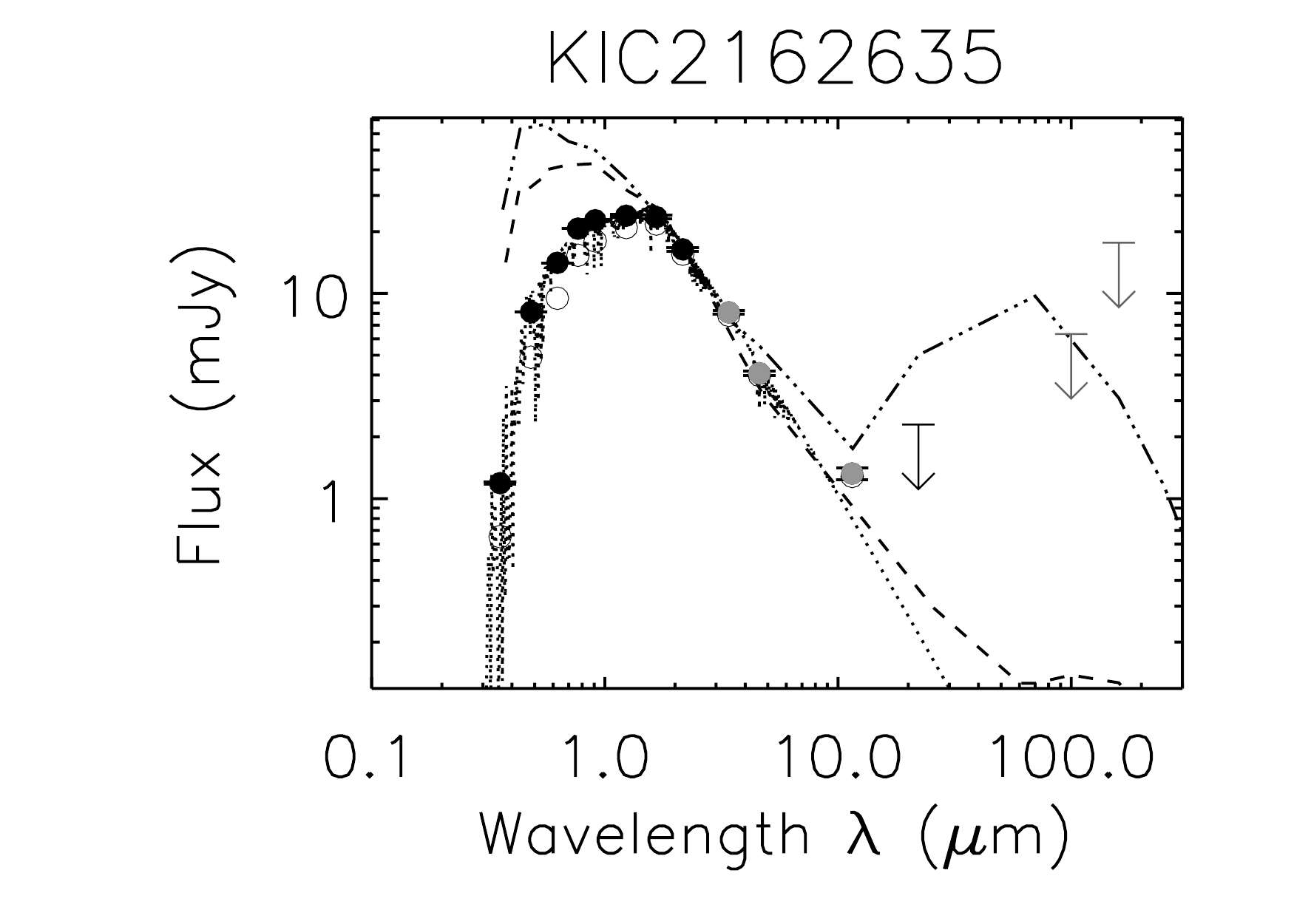}
\includegraphics[width=4.5cm]{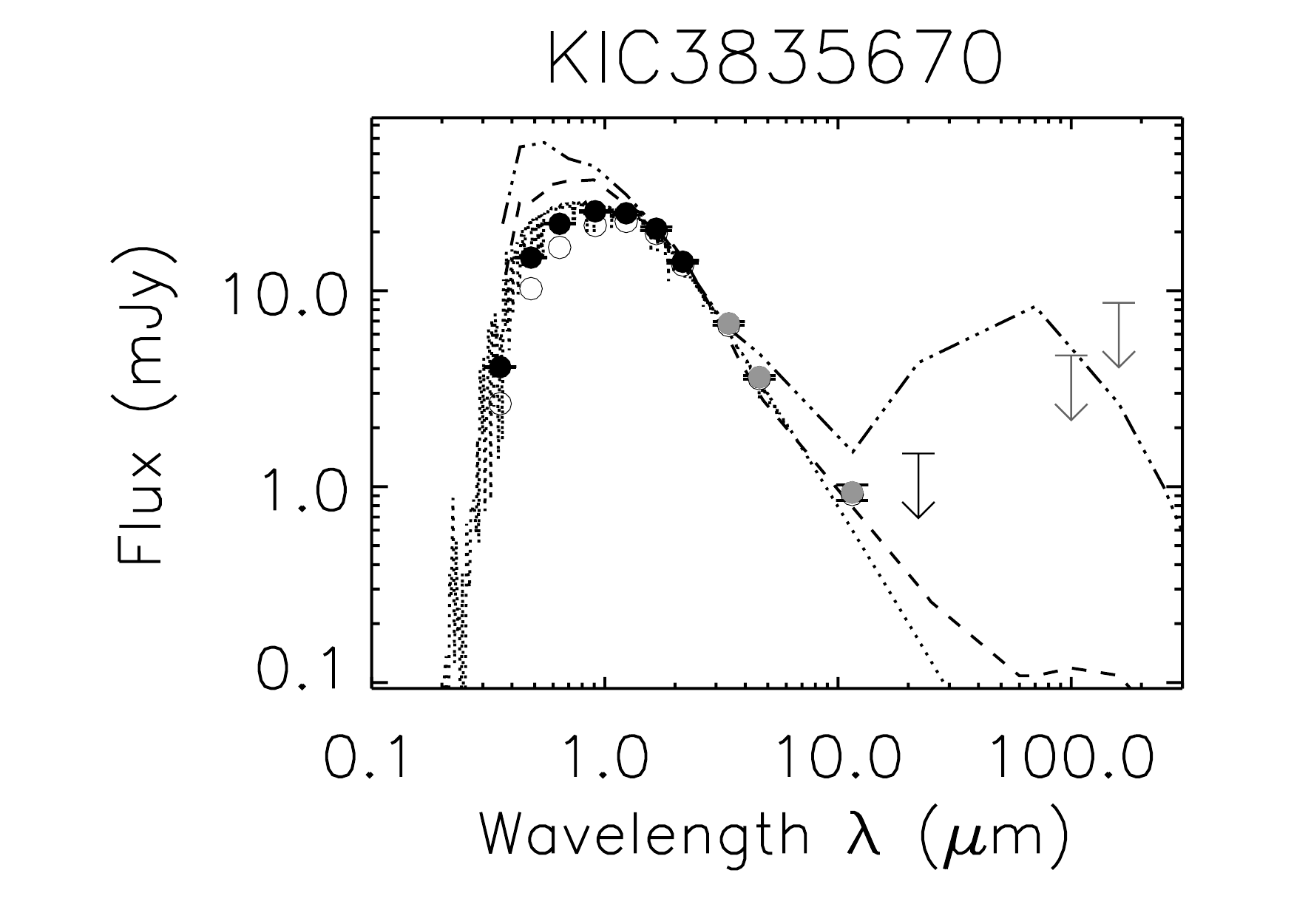}
\includegraphics[width=4.5cm]{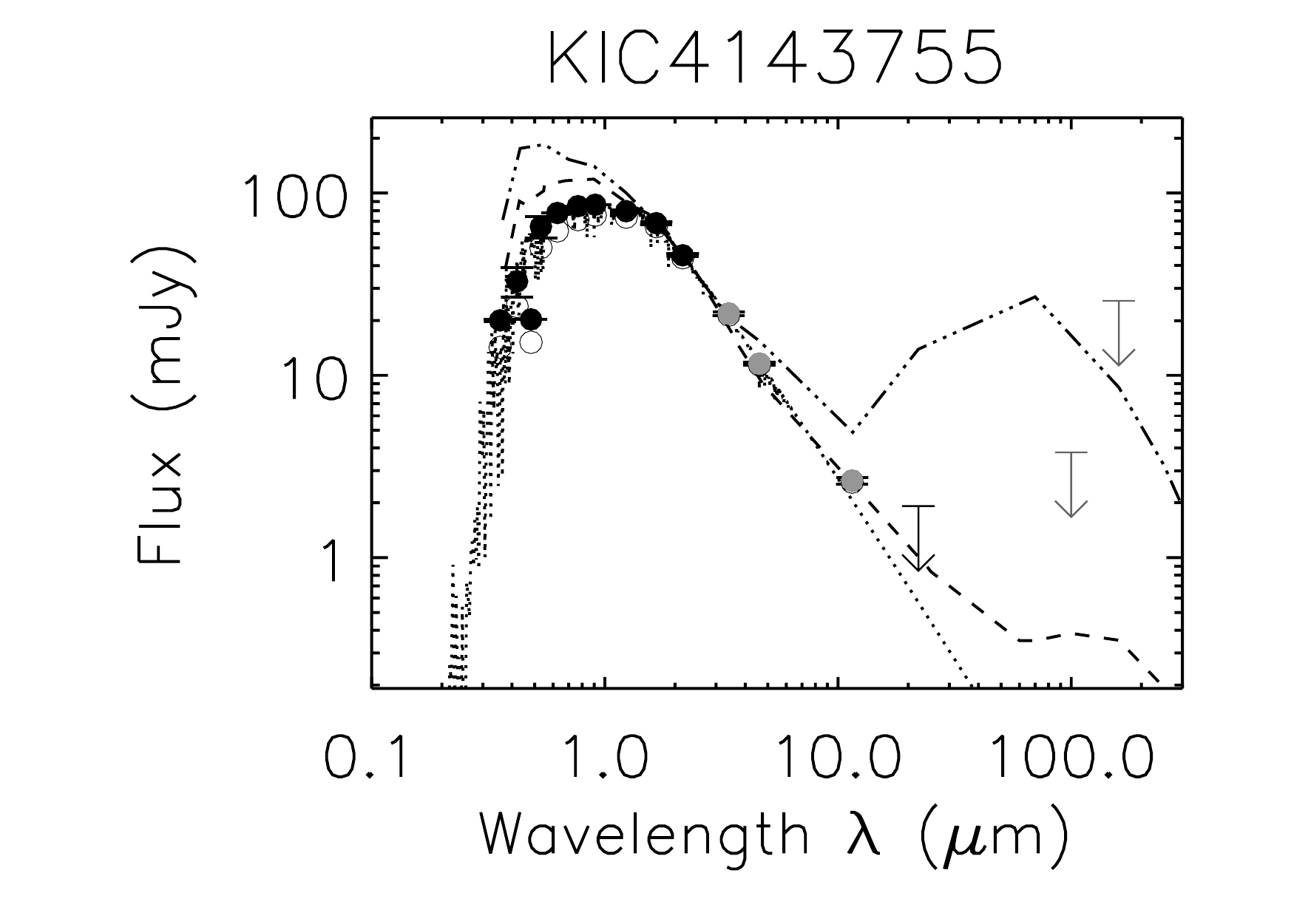}
\includegraphics[width=4.5cm]{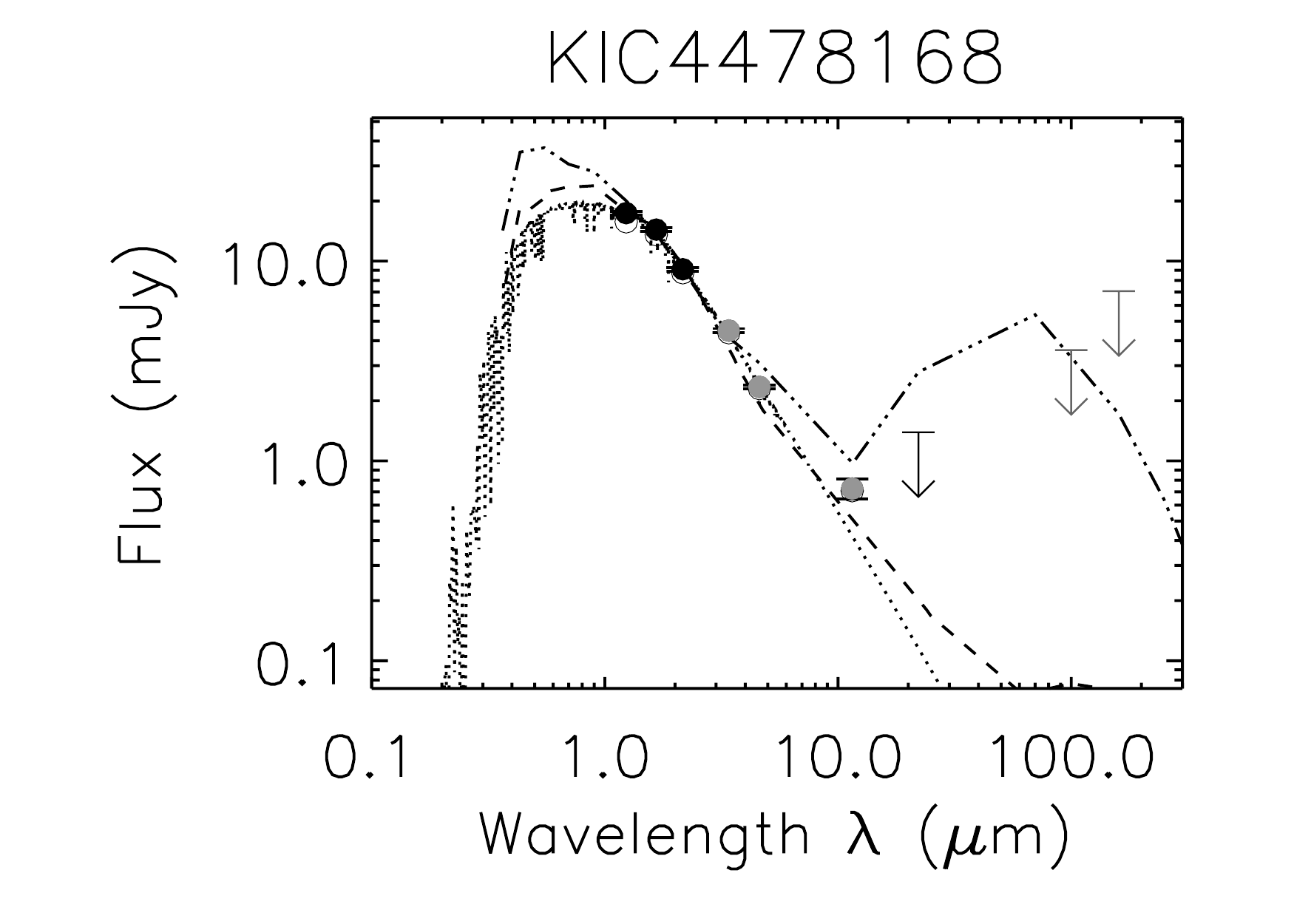}
\includegraphics[width=4.5cm]{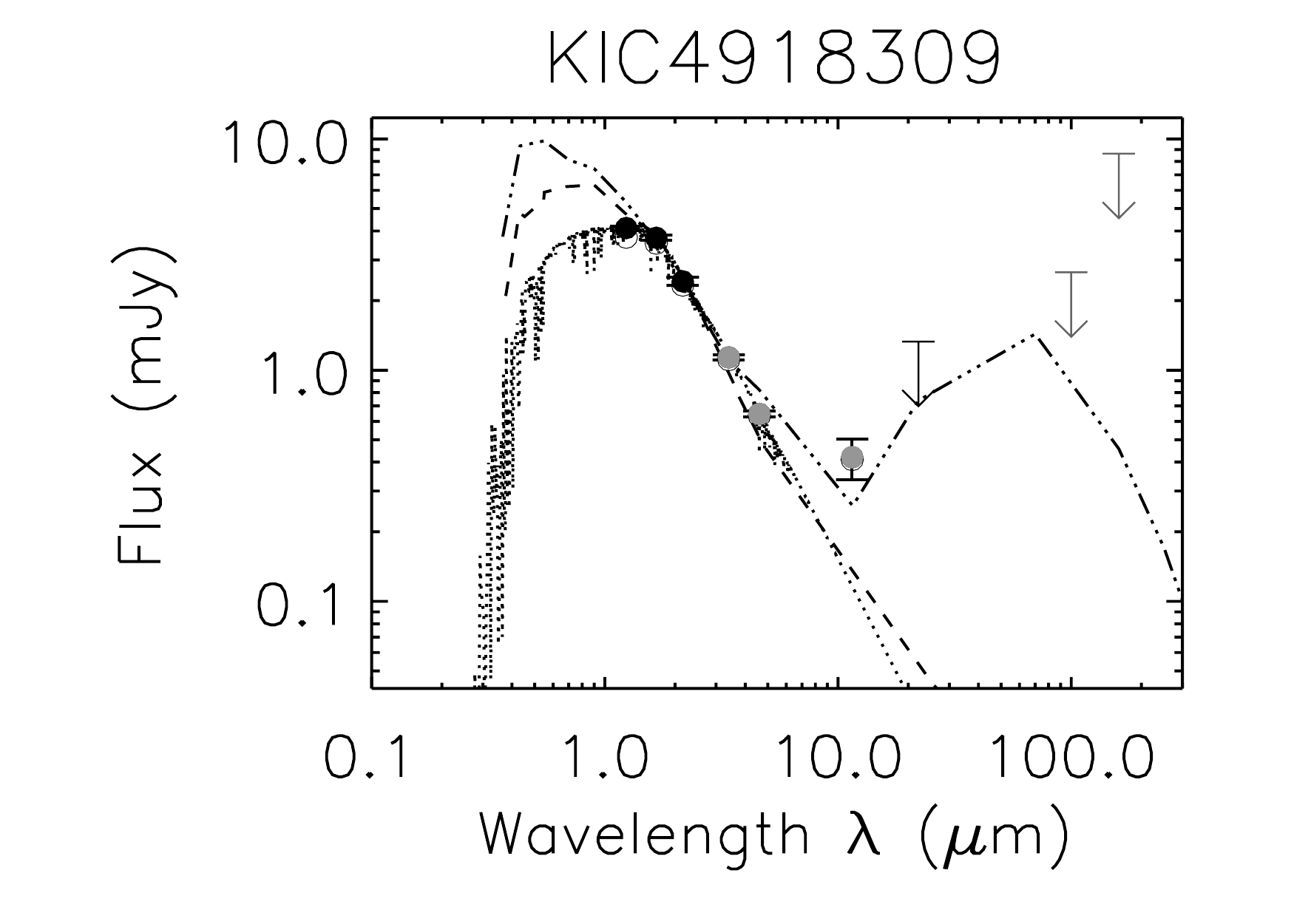}
\includegraphics[width=4.5cm]{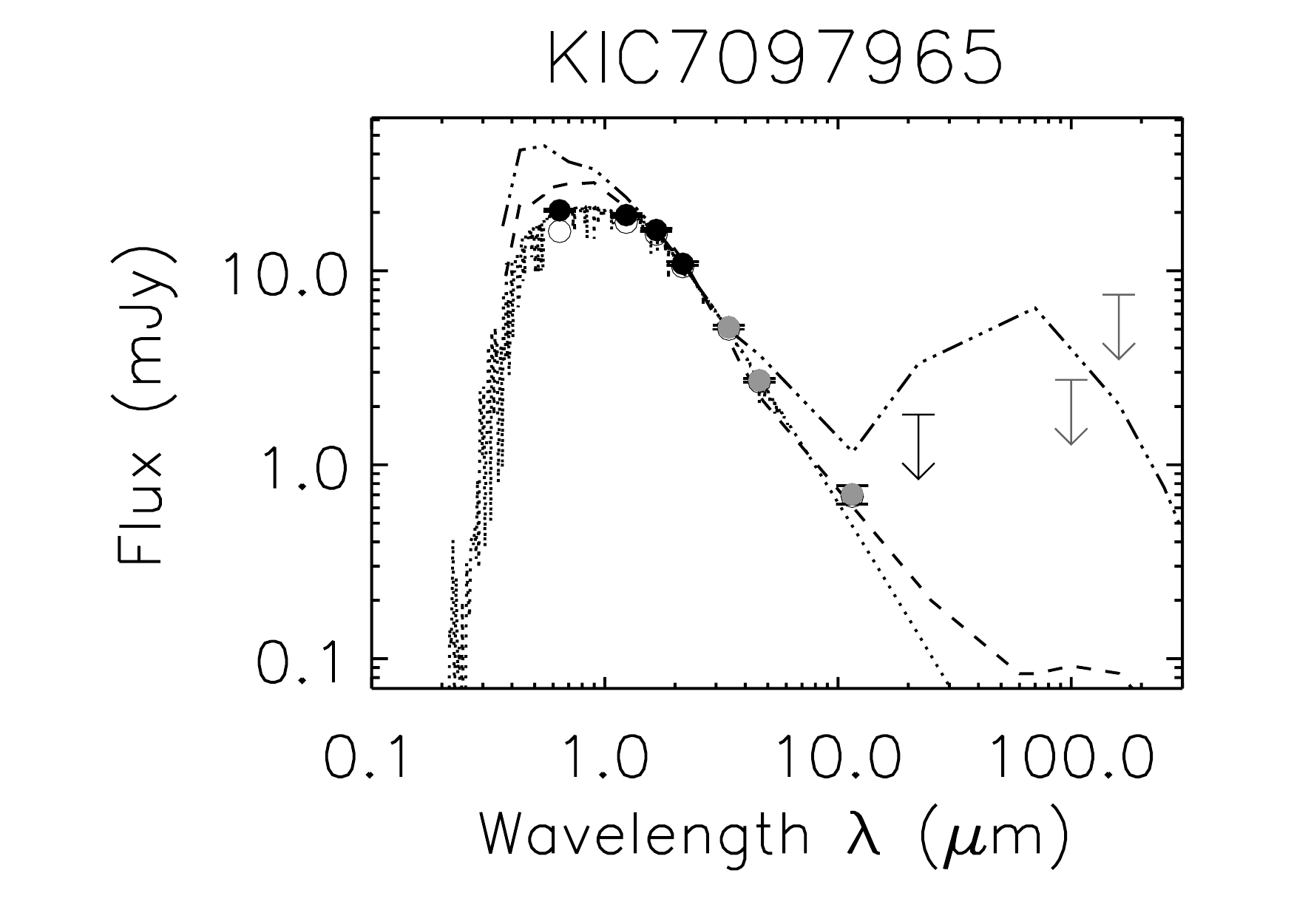}
\includegraphics[width=4.5cm]{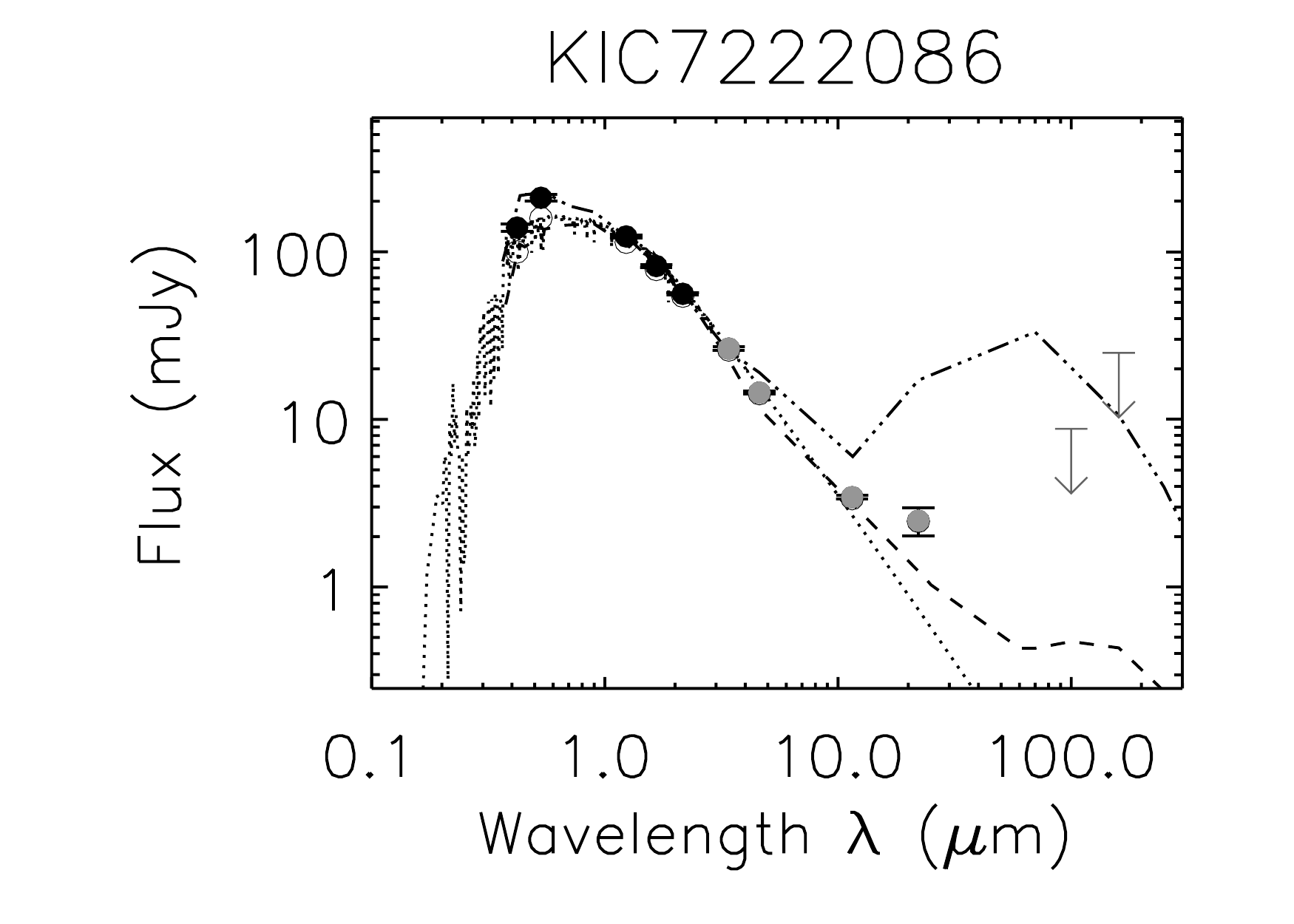}
\includegraphics[width=4.5cm]{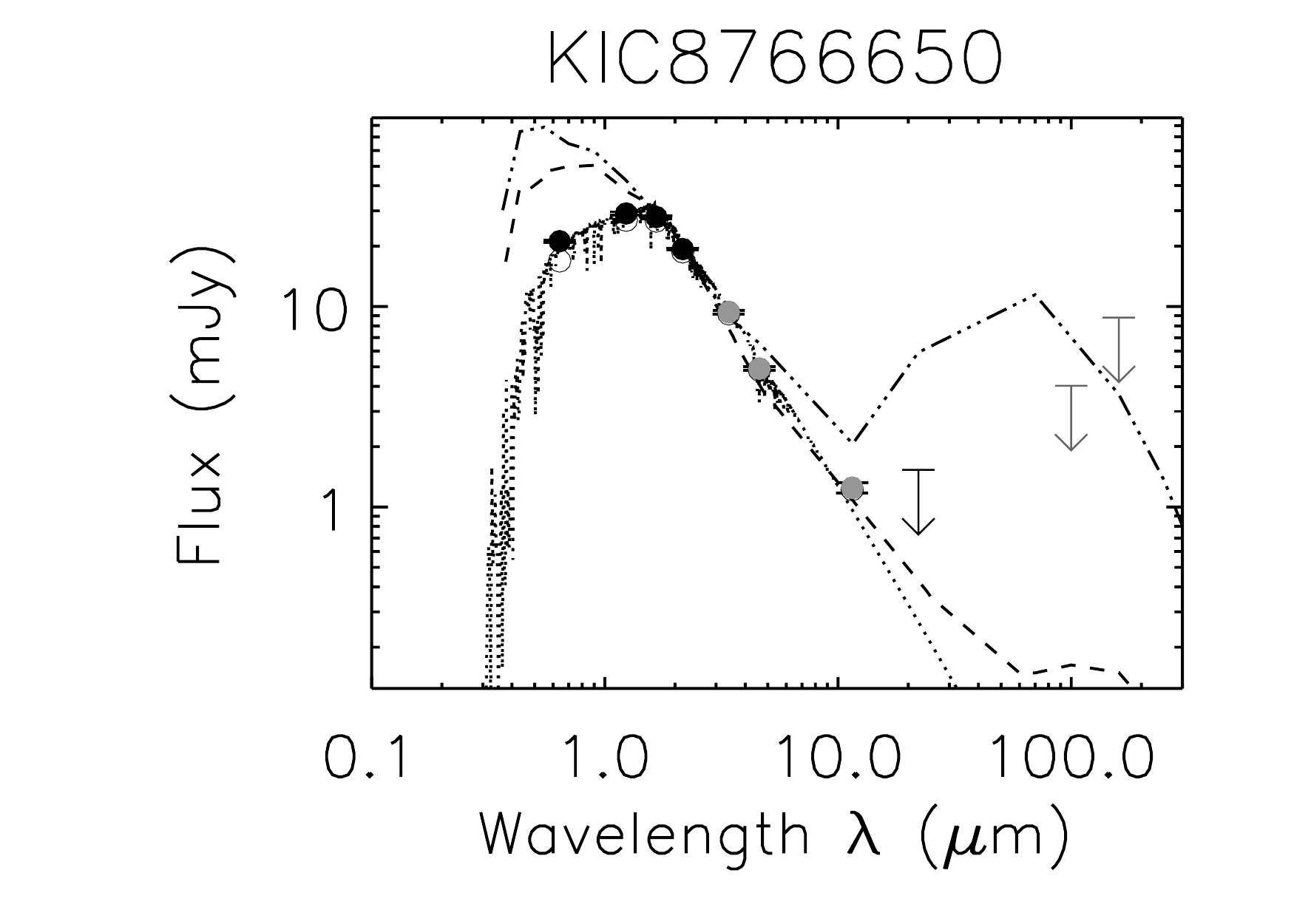}
\includegraphics[width=4.5cm]{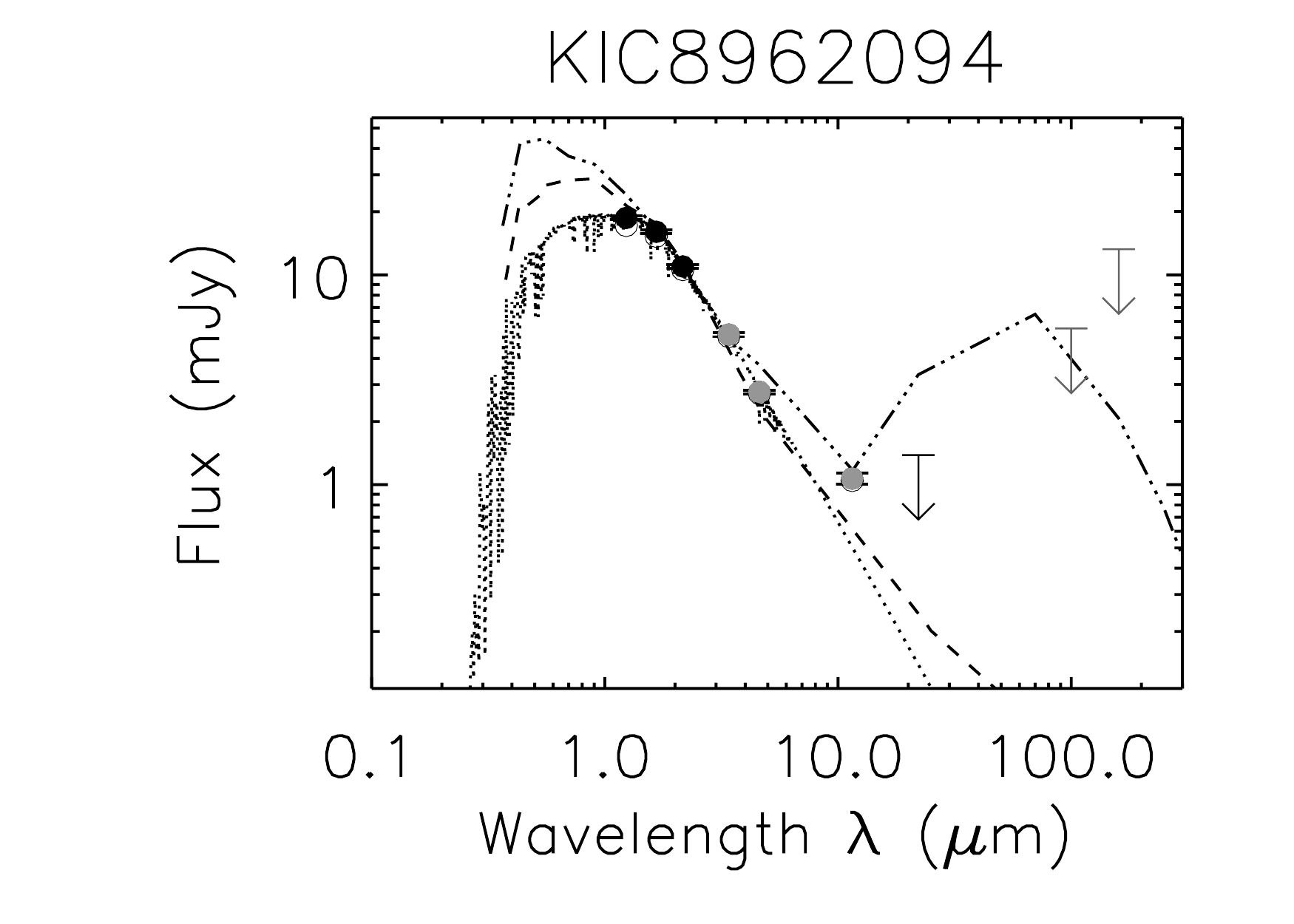}
\includegraphics[width=4.5cm]{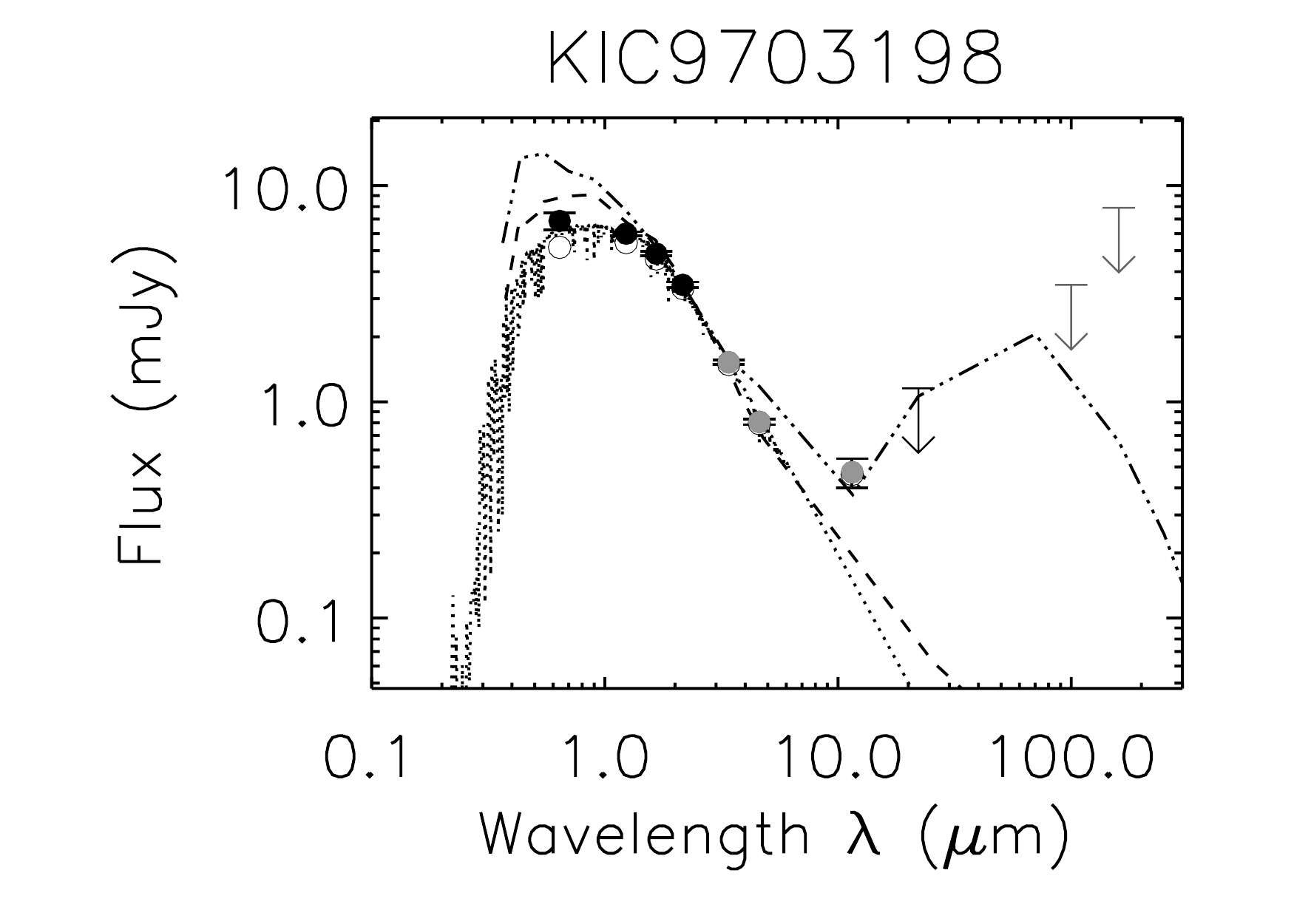}
\includegraphics[width=4.5cm]{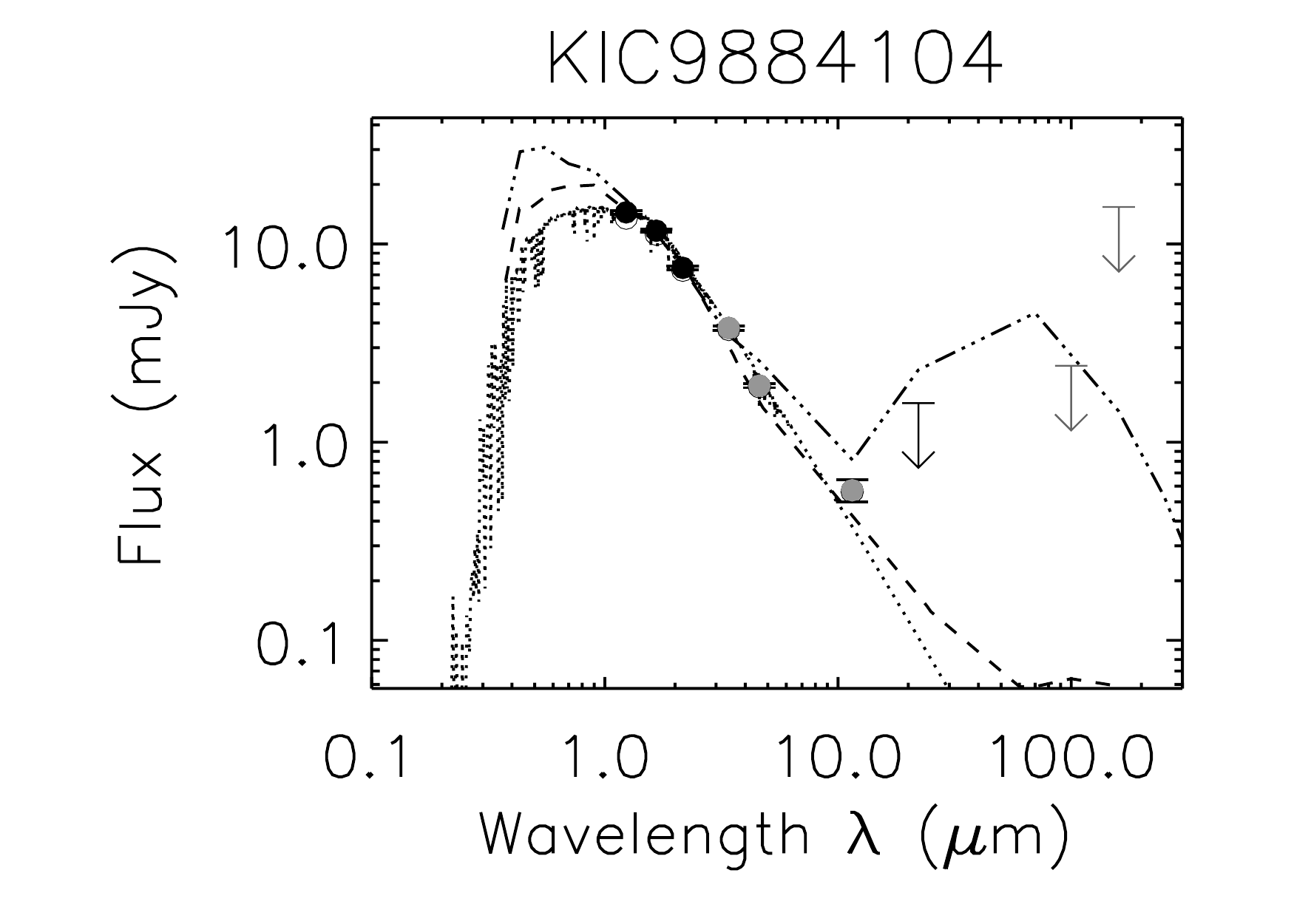}
\includegraphics[width=4.5cm]{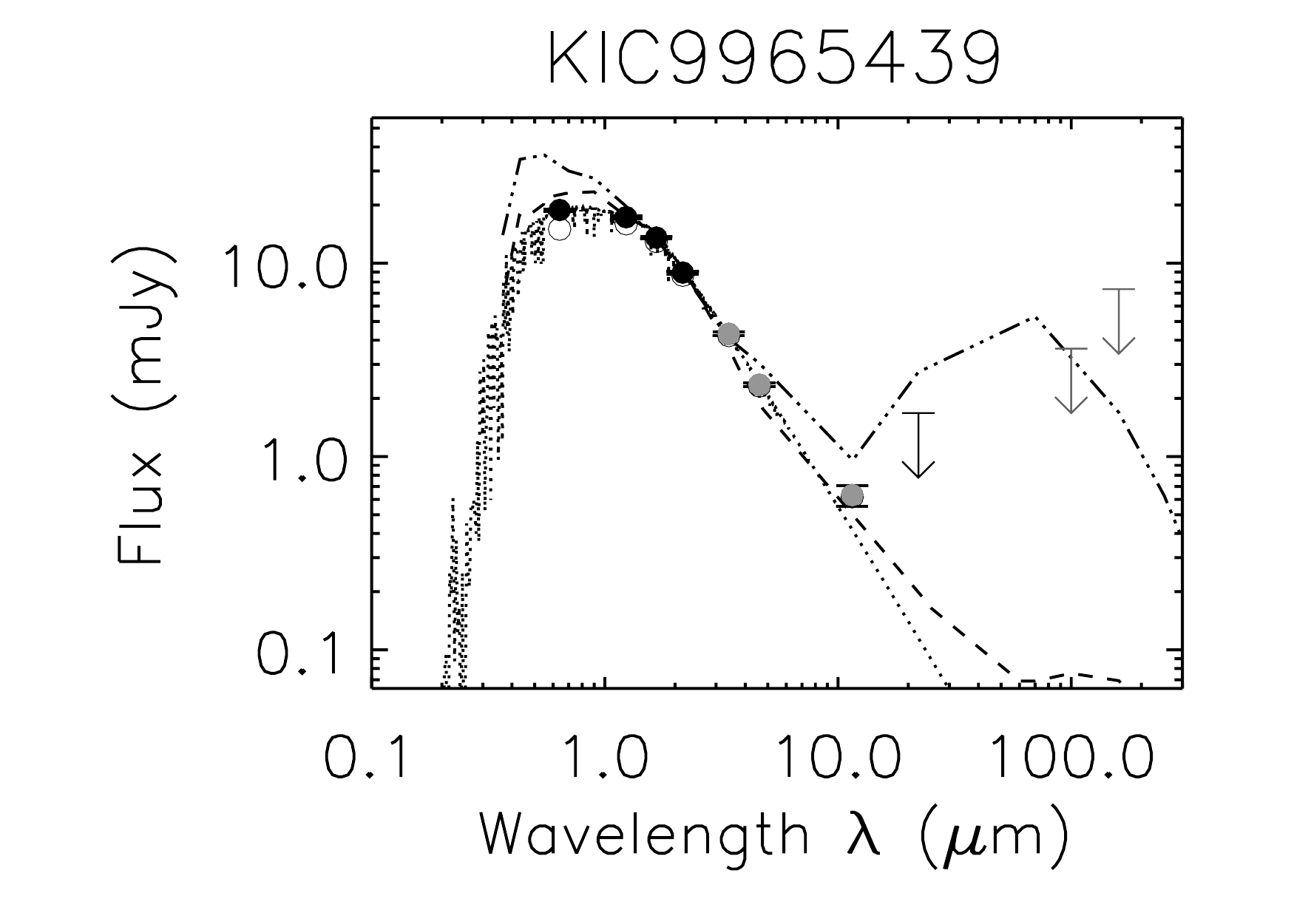}
\includegraphics[width=4.5cm]{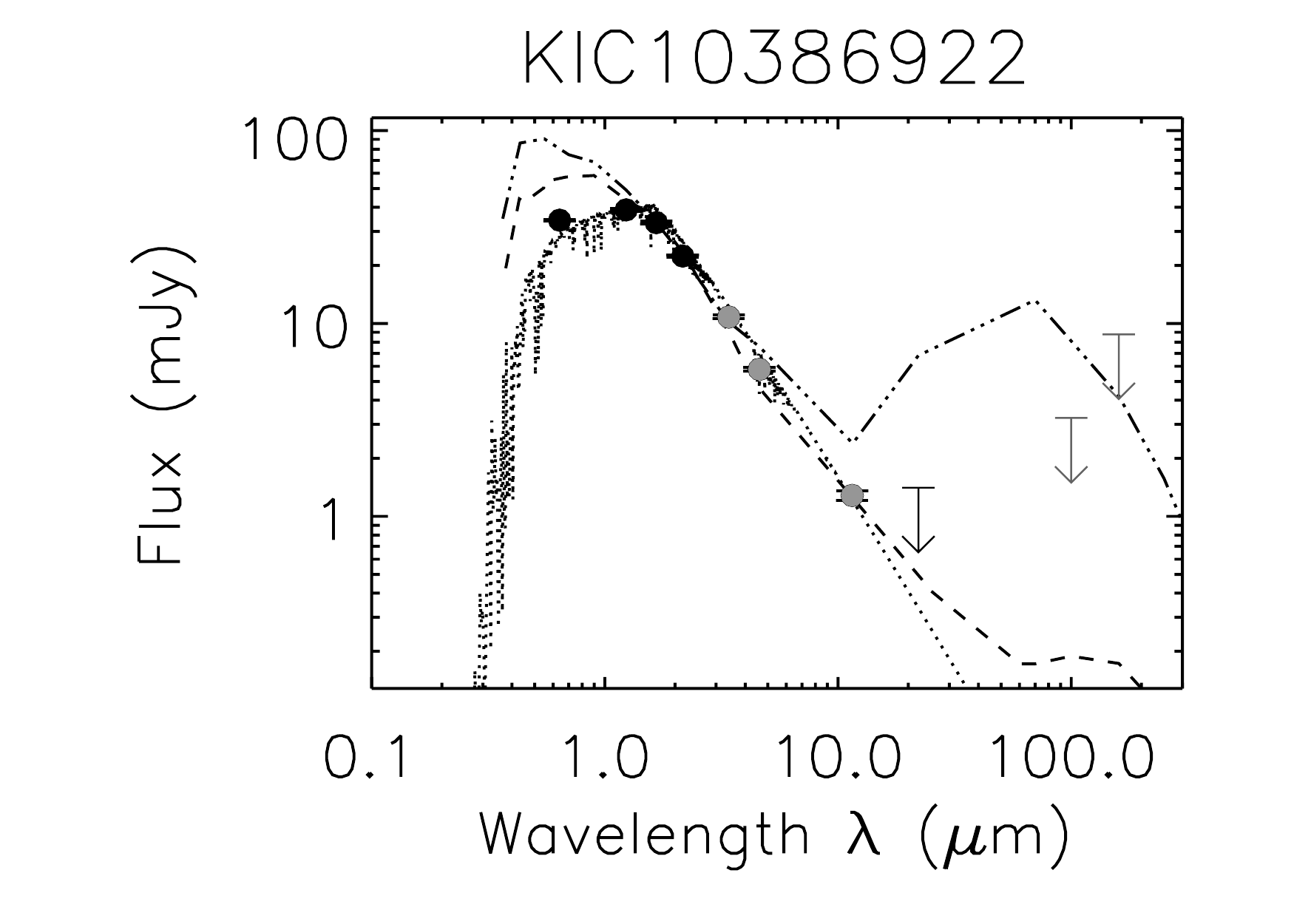}
\includegraphics[width=4.5cm]{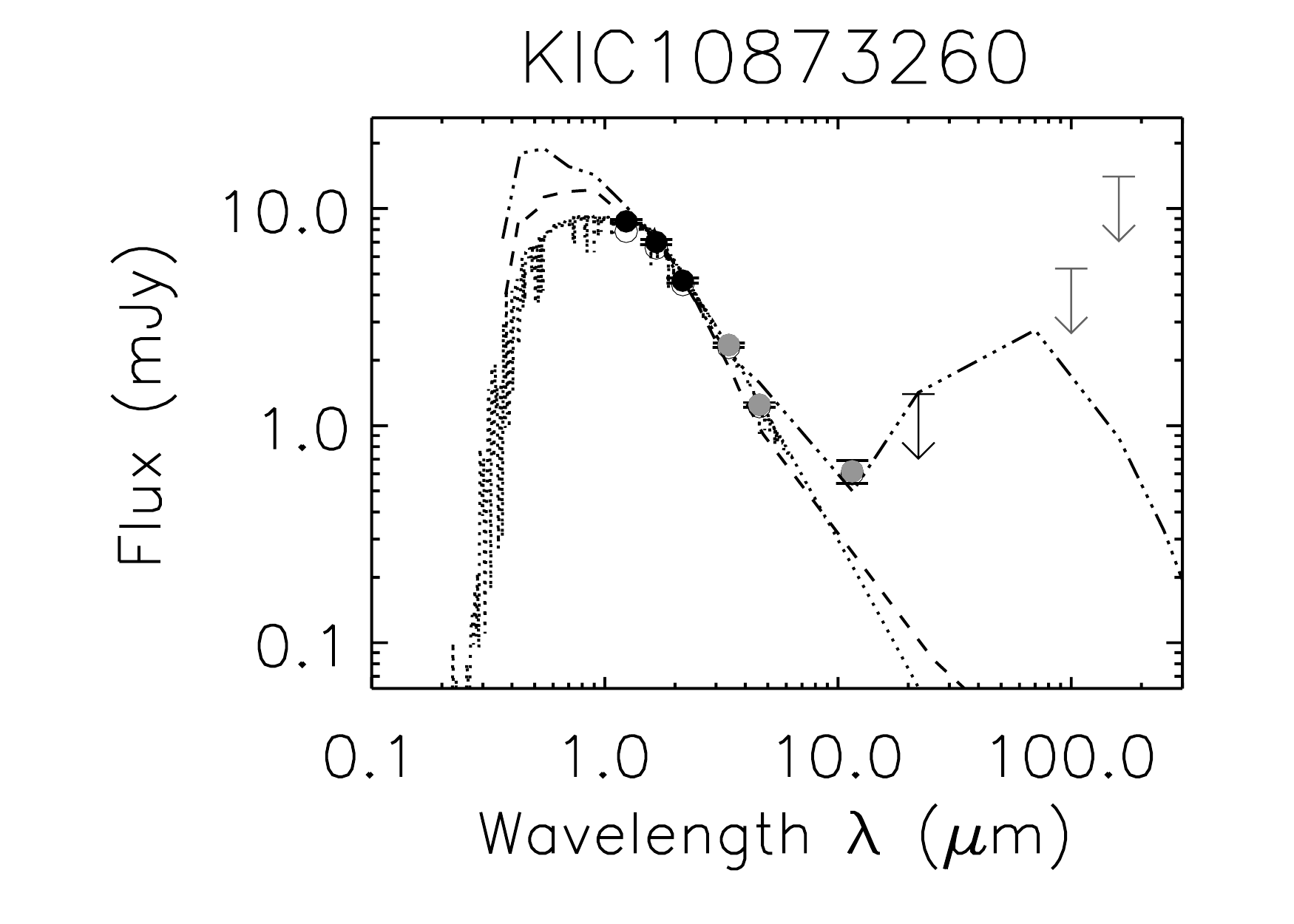}
\includegraphics[width=4.5cm]{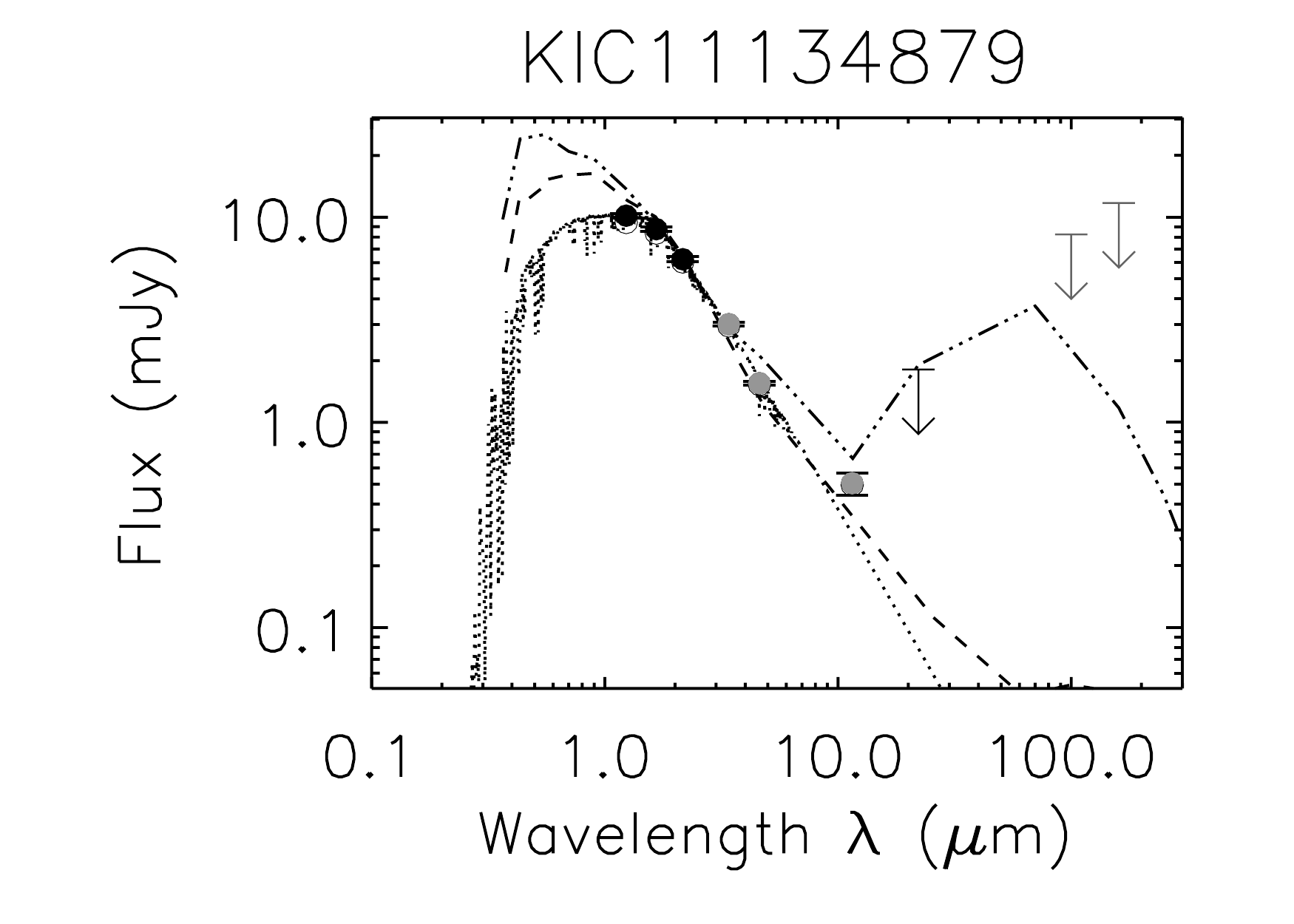}
\includegraphics[width=4.5cm]{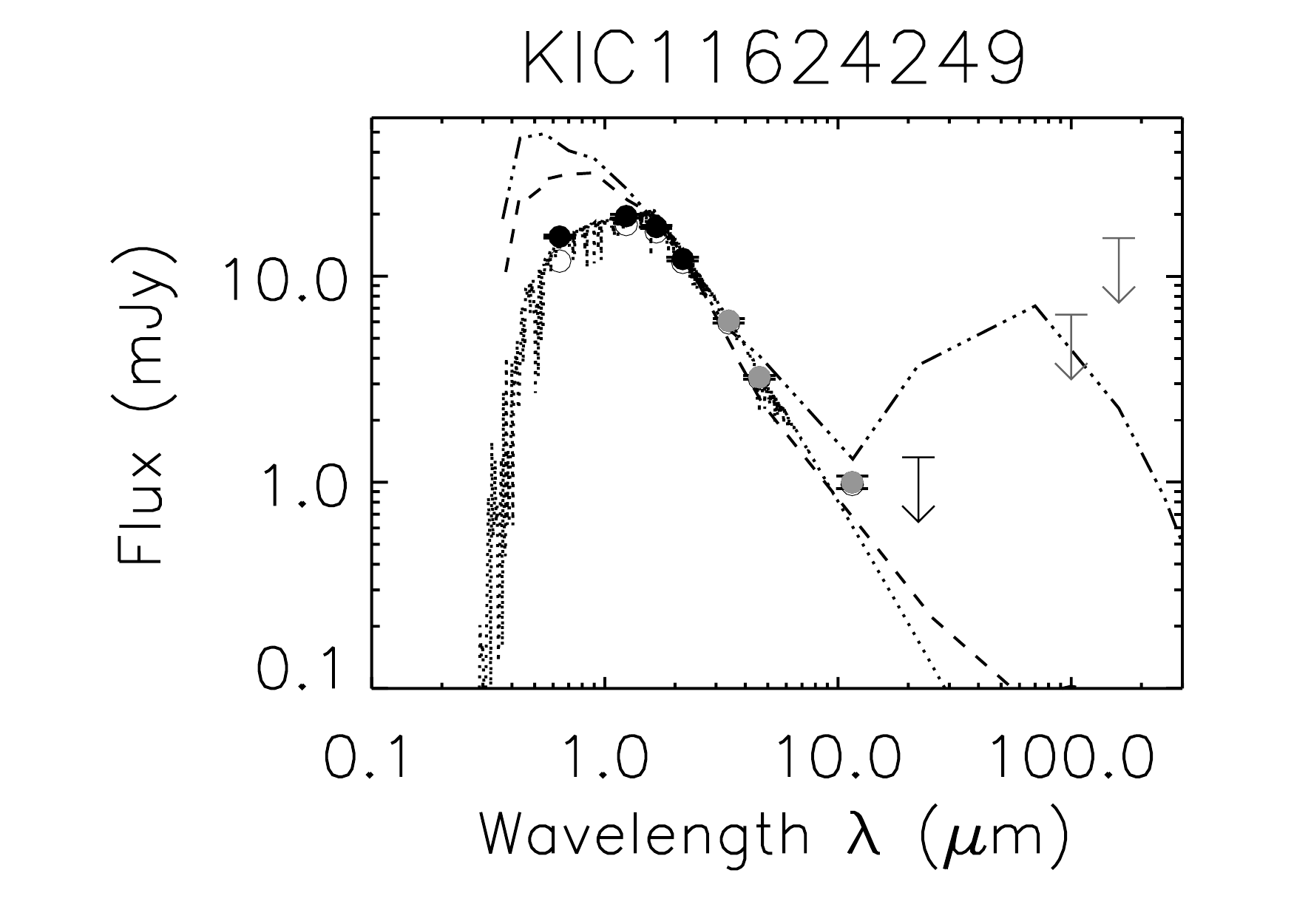}
\includegraphics[width=4.5cm]{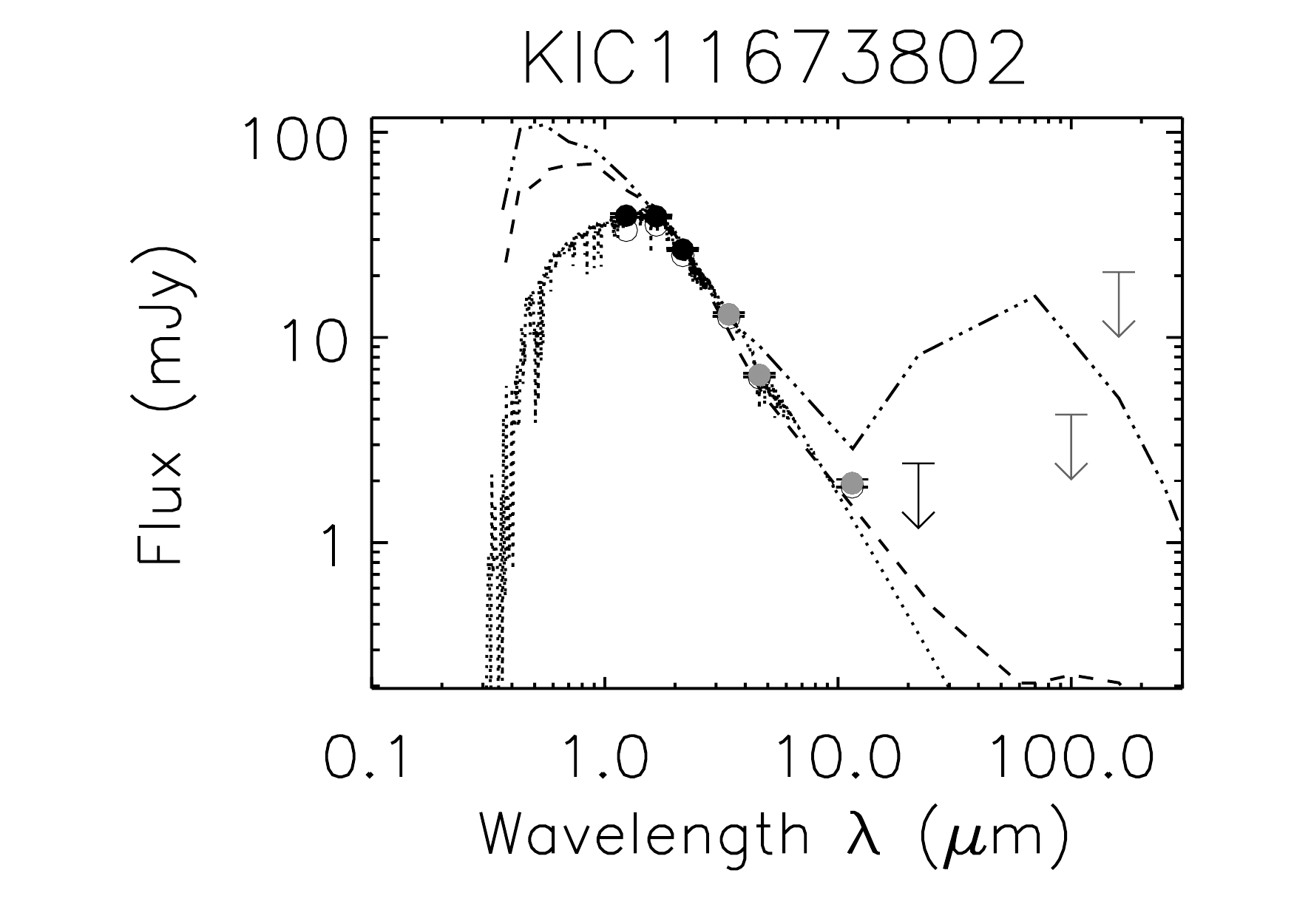}
\includegraphics[width=4.5cm]{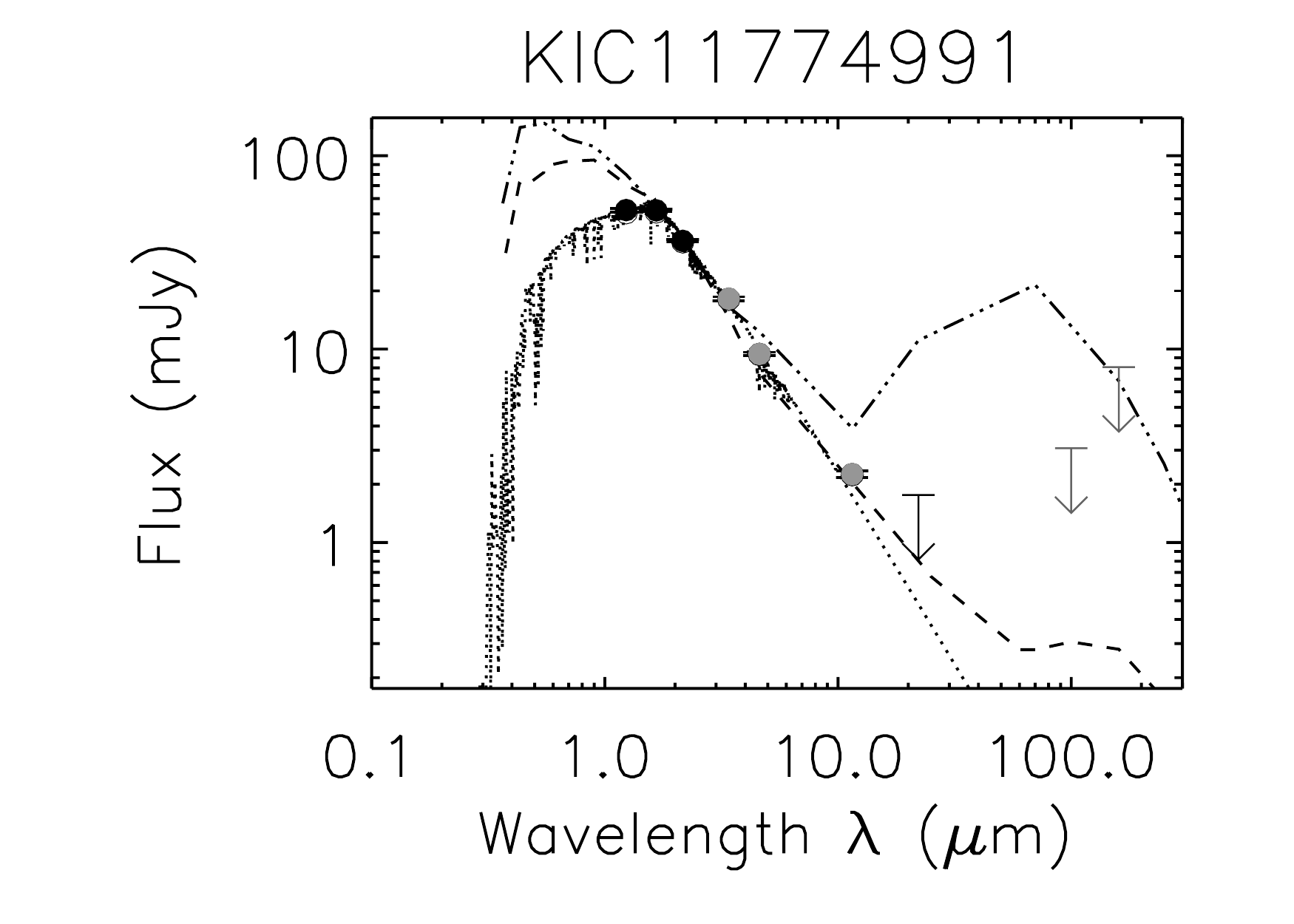}
  \caption{Spectral energy distributions (SEDs) of  planet hosts and planet host candidates. The dotted line represents the stellar photosphere model for the spectral type of the star scaled to the dereddened J-band flux. Open and solid dots are observed and dereddened fluxes, respectively. Symbols in gray are WISE fluxes. The two longest wavelength flux upper limits are the {\it Herschel}/PACS observations. For comparison, the SEDs of $\eta$ Crv \citep{Duchene2014} and $\beta$ Pic \citep{Vandenbussche2010} have been normalized to the dereddened J-band flux and added to each plot denoted by dashed and dot-dashed lines, respectively.}
  \label{fig:seds}
\end{figure*}

To constrain the possible dust content of the warm excess sources identified with WISE, we included the {\sl Herschel}/PACS upper limit flux measurements at 100 and 160 $\mu$m in the spectral energy distributions (SEDs) of the candidate mid-IR excess systems. The aim of this analysis is to quantify the relative constraints on the excess flux of these objects with the PACS far-IR wavelengths. To construct the SEDs we combined fluxes from the Tycho, SDSS, DENIS, 2MASS, and WISE for all targets and dereddened them with the \cite{Weingartner2001} interstellar extinction law using extinctions from the Kepler Input Catalog or computing them for the transiting planet systems with the help of optical plus 2MASS fluxes. We compared the dereddened fluxes with NEXTGEN stellar atmospheres \citep{NEXTGEN} with the same spectral types of the stars as obtained from the Kepler Input Catalog or from http://exoplanets.org in the case of the transiting planet systems.  The SEDs are shown in Figure \ref{fig:seds}. They include the normalized SEDs of $\eta$ Crv \citep[a prototypical warm excess debris disk,][]{Matthews2010,Duchene2014} and $\beta$ Pic \citep[a prototypical massive debris disk,][]{Vandenbussche2010}. The 12 to 24 $\mu$m excesses of these two prototypical sources are very different: while the 12 to 24 $\mu$m slope of $\eta$ Crv is rapidly declining, that of $\beta$ Pic rises. This indicates that only in the case of a bright massive debris disk, the rising 12 and 24 $\mu$m excesses mark the Wien tail of a colder excess emission. Therefore in the common case of having only detected excesses at 12 $\mu$m, it is impossible to distinguish between the two models. Higher SN data around 24 $\mu$m data for these objects is needed to achieve that.

On the other hand, while the large flux upper limits determined with the PACS data put weak constraints on the physical interpretation of the systems, these figures illustrate that conspicuous far-IR excesses like that in $\beta$ Pictoris \citep{Vandenbussche2010} would have been detected with these observations in most of the targets if they were present in these objects. For the nearby exoplanet-host star WASP-33, the new PACS data allows us to also discard weaker excesses at 100 $\mu$m and 160 $\mu$m like those in $\eta$ Crv \citep{Matthews2010,Duchene2014}. Therefore, our new PACS data allows us to rule out massive debris disks in most of these objects but cannot be used to rule out smaller excesses or warm-only debris disks.

\section{Discussion and conclusions}
\label{discussion}

\citet{Kennedy2012} argue that warm excess debris disk candidates identified with all-sky surveys like WISE do suffer from serious mid-IR background contamination and that the resulting numbers match the expected values of extragalactic counts  suggesting that these warm-excess candidates might result from chance alignment with background galaxies. However, although reasonable, this is a statistical argument so it cannot rule out the presence of some such objects, such as in the case of the now famous $\eta$ Crv warm debris disk \citep{Matthews2010,Duchene2014}. 

An alternative {\sl direct} check of this hypothesis, beyond high-resolution mid-IR imaging of the candidates, is to search for far-IR excesses on these same targets, which one would expect if the candidate excesses found in these objects were indeed associated to the objects and the disks were not extremely warm.

Having no result in this program confirms that the planet-hosting star WASP-33 does not have a debris disk comparable to that of $\eta$ Crv, that most other candidate mid-IR  excess sources do not have a debris disk like that of $\beta$ Pic, and places a loose constraint on the presence of such rare warm disks around the Kepler planet-hosting candidates, given the sensitivity of the observations at the distance to the {\it Kepler} sample. However, this result and the nature of the emissions found at longer wavelengths support the hypothesis by \citet{Kennedy2012} that most of the WISE-identified mid-IR candidate excesses around these systems do indeed stem from chance alignment from either background IR-bright galaxies and/or interstellar emission.

\begin{acknowledgements}

  
  This work has been possible thanks to the support from the ESAC Trainee Program, ESAC Space Science Faculty, and of
  the Herschel Science Centre. AR acknowledges support from the ESA SRE-OA Research Funding via contract SC 1300016149. Support for this work was provided by NASA through an award issued by JPL/Caltech. This publication is based on observations made by the Herschel Space Telescope and the PACS instrument. PACS was developed by a consortium of institutes led by MPE (Germany) and including UVIE (Austria); KU Leuven, CSL, IMEC (Belgium); CEA, LAM (France); MPIA (Germany); INAF-IFSI, OAA, OAP, OAT, LENS, SISSA (Italy); IAC (Spain). This development has been supported by the funding agencies BMVIT (Austria), ESA-PRODEX (Belgium), CEA/CNES (France), DLR (Germany), ASI, INAF (Italy), and CICYT/MCYT (Spain). HCSS, HSpot and HIPE are joint developments by the Herschel Science Ground Segment Consortium, consisting of ESA, the NASA Herschel Science Center, and the HIFI, PACS and SPIRE consortia. This publication is also based on
  observations made with the Kepler Spacecraft. Funding for this
  mission is provided by National Aeronautics and Space
  Administration's Science Mission Directorate (NASA). This study also
  makes use of data products from Herschel Space Observatory and the Wide-field Infrared Survey
  Explorer, a joint project of the University of California, Los
  Angeles, and the Jet Propulsion Laboratory (JPL) / California
  Institute of Technology (Caltech); the NASA Infrared Processing and
  Analysis Center (IPAC) Science Archive and the NASA/IPAC/NExScI Star
  and Exoplanet Database, operated by JPL, Caltech, and funded by NASA;
  the SIMBAD database and the Vizier service, operated at the Centre
  de Donnes astronomiques de Strasbourg, France; the data products
  from the Two Micron All Sky Survey (2MASS), a joint project of the
  University of Massachusetts and IPAC at Caltech, funded by NASA and the
  National Science Foundation; the Multimission Archive at the Space
  Telescope Science Institute (MAST). STScI is operated by the
  Association of Universities for Research in Astronomy, Inc., under
  NASA contract NAS5-26555. Support for MAST for non-HST data is
  provided by the NASA Office of Space Science via grant NNX09AF08G
  and by other grants and contracts.
\end{acknowledgements}

\bibliographystyle{aa}
\bibliography{mybiblio}

\end{document}